\documentclass[showpacs,11pt,
preprint,preprintnumbers,nofootinbib,
groupedaddress,superscriptaddress]{revtex4}
\usepackage{amsmath,amssymb}	
\usepackage[dvipdfmx]{graphicx}
\usepackage{cancel}
\usepackage[hypertex]{hyperref}

\usepackage{color}

\begin{document}
\title{Fingerprinting non-minimal Higgs sectors } 
\preprint{KUNS-2500}
\preprint{UT-HET-095}
\pacs{12.60.Fr, 
      14.80.Cp  
}
\author{Shinya Kanemura}
\email{kanemu@sci.u-toyama.ac.jp}
\affiliation{Department of Physics, University of Toyama, Toyama
930-8555, Japan} 
\author{Koji Tsumura}
\email{ko2@gauge.scphys.kyoto-u.ac.jp}
\affiliation{Division of Physics and Astronomy, Kyoto University, Kyoto 606-8502,
Japan} 
\author{Kei Yagyu}
\email{keiyagyu@ncu.edu.tw}
\affiliation{Department of Physics, National Central University, 
Chungli 32001, Taiwan} 
\author{Hiroshi Yokoya}
\email{hyokoya@sci.u-toyama.ac.jp}
\affiliation{Department of Physics, University of Toyama, Toyama
930-8555, Japan} 


\begin{abstract}

After the discovery of the standard-model-like Higgs boson at the LHC, 
the structure of the Higgs sector remains unknown. 
We discuss how it can be determined by the combination of 
direct and indirect searches for additional Higgs bosons
at future collider experiments. 
First of all, we evaluate expected excluded regions for the 
mass of additional neutral Higgs bosons from direct searches at the LHC with the 14 TeV collision energy
in the two Higgs doublet models with a softly-broken $Z_2$ symmetry. 
Second, precision measurements of the Higgs boson couplings at future experiments
can be used for the indirect search of extended Higgs sectors 
if measured coupling constant with the gauge boson slightly deviates from the standard model value.
In particular, in the two Higgs doublet model with the softly-broken discrete symmetry, 
there are four types of Yukawa interactions, so that they can be discriminated by measuring the pattern of deviations
in Yukawa coupling constants. 
Furthermore, we can fingerprint various extended Higgs sectors with future precision data
by detecting the pattern of deviations in the coupling constants of the standard-model-like Higgs boson.
We demonstrate how the pattern of deviations can be different 
among various Higgs sectors which predict the electroweak rho parameter to be unity; such as 
models with additional an isospin singlet, a doublet, triplets or a septet. 
We conclude that as long as the gauge coupling constant of the Higgs boson slightly differs from 
the standard model prediction but is enough to be detected at the LHC and its high-luminosity run or at the 
International Linear Collider, 
we can identify the non-minimal Higgs sector even without direct discovery of additional Higgs bosons at the LHC.

\end{abstract}
\maketitle

\section{Introduction}

The new particle discovered at the LHC in 2012 was identified as a Higgs boson~\cite{Ref:Higgs_Dicovery}. 
With the current LHC data, 
its measured properties are consistent with those of the Higgs boson in the Standard Model (SM)~\cite{Ref:Higgs_Dicovery2,Ref:spin-parity,Ref:H2ZZ}. 
On the other hand, so far, no evidence for new physics beyond the SM has been found directly at the LHC. 
Therefore, our standard picture for high energy particle phenomena, which is based on 
the gauge theory with spontaneous symmetry breaking, seems successful. 

However, it has been well known that the Higgs sector in the SM is problematic from the theoretical viewpoint. 
First of all, the existence of the scalar boson causes the hierarchy problem~\cite{Weinberg_prd13,Gildener,Susskind}, so that many 
physicists try to understand the essence of the Higgs boson, e.g., elementary or composite particle. 
In order to solve the hierarchy problem, several scenarios for the new paradigm have been introduced such as 
supersymmetry, dynamical symmetry breaking and extra dimensions. 
Each of them gives a different answer for the question of essence of the Higgs boson. 
Second, there is no principle for the structure of the Higgs sector, so
that the minimal Higgs sector adopted in the SM is just an assumption. 
There are many possibilities of non-minimal Higgs sectors with additional scalar fields such as singlets, doublets and triplets.	
A model based on the above-discussed paradigms can predict the specific structure and property of the Higgs sector.  
For example, the minimal supersymmetric SM (MSSM) predicts the Higgs sector with two isospin doublet scalar fields~\cite{Ref:SUSY_Higgs,Ref:HHG}. 
In addition, 
non-minimal Higgs sectors can also be introduced in new physics models to explain the origin of neutrino masses, the existence of 
dark matter and baryon asymmetry of the Universe, {\it et cetera}, which cannot be explained in the SM. 
Each new physics model predicts a characteristic structure for the Higgs sector.   
Therefore, if the Higgs sector is determined by experiments in future, 
the new physics scenario can be selected from many candidates. 
 
To probe the extended Higgs sectors, the simplest way is to directly search for additional Higgs bosons such as the second scalar boson. 
By measuring its properties; e.g., the mass, the electric charge, the spin and the parity, important information to 
reconstruct the Higgs sector can be extracted. 
Non-observation of the second Higgs boson gives the experimental constraint on the parameter space in the Higgs sector.  
Current bounds from searches for additional Higgs bosons at the LHC with the collision energy of 7 and 8 TeV can be
found in Refs.~\cite{Aad:2012cfr, CMS:2013hja, Ref:A2bb, ATLAS:2013zla, CMS:2013eua, Ref:H2ZZ,
Ref:Hp_to_TauNu,ATLAS_charged_constraints,Ref:Hp_to_CS,Ref:Hpp,Ref:invisible}. 

In addition to the direct search results, precision measurements of various low energy observables
can be an indirect search for extended Higgs sectors, since the existence of
additional Higgs multiplets can affect them. 
The electroweak rho parameter defined in terms of the masses of the $W$ boson $m_W$ and the $Z$ boson $m_Z$, and the weak mixing angle $\theta_W$ by 
\begin{align} 
\rho =\frac{m_W^2}{m_Z^2\cos^2\theta_W},
\end{align} 
is one of the most important tools to constrain the structure of the Higgs sector, 
whose experimental value is very close to unity; i.e., $\rho_{\text{exp}}^{}=1.0004^{+0.0003}_{-0.0004}$~\cite{PDG}.
In the SM, the rho parameter is predicted to be unity at the tree level. 
In general, in extended Higgs sectors, predicted values for the rho parameter can deviate from unity.  
In the Higgs sector with arbitrary number of scalar fields $\phi_i$ (a hypercharge $Y_i^{}$ and an isospin $T_i^{}$) with vacuum expectation values (VEVs) 
$v_i$, the rho parameter is calculated at the tree level by~\cite{Ref:HHG}
\begin{align}
\rho_\text{tree}^{}
= \frac{\sum_i \big[ T_i^{}(T_i^{}+1)-Y_i^2 \big]\, v_i^2}
{2\,\sum_i Y_i^2\, v_i^2}.  \label{Eq:rho_tree}
\end{align}
From Eq.~(\ref{Eq:rho_tree}), an additional VEV of a Higgs field satisfying with $T_i^{}\,(T_i^{}+1)-3Y_i^2=0$
does not change the value of the rho parameter from the SM\footnote{Although there are infinite number of solutions for the above equation, 
larger isospin representation fields cause violation of perturbative unitarity~\cite{Earl:2013jsa}. 
Therefore, only the three possibilities can be substantially considered; i.e., 
isospin singlets with $Y_i=0$, doublets with $Y_i=1/2$ and septets with $Y_i=2$. 
The next possibility to the septet representation  
is isospin 26-plet with $Y_i=15/2$. }. 
A VEV of a Higgs multiplet without satisfying the above equation; e.g., 
a triplet Higgs field, deviates $\rho_\text{tree}^{}$ from unity, so that we need finetuning for such a VEV
to avoid the constraint from $\rho_{\text{exp}}^{}$.
However, even if the above equation is not satisfied, by allowing an alignment among VEVs, we can keep $\rho_\text{tree}=1$. 
The simplest realization is known as the Georgi-Machacek (GM) model~\cite{Ref:GM} whose Higgs sector is composed of additional
real and complex triplet fields with $Y_i=0$ and $Y_i=1$, respectively.  

The above discussion is very useful to discriminate extended Higgs sectors. 
However, the rho parameter has been measured quite precisely, so that 
we need to take into account quantum effects to the rho parameter.   
Let us discuss how the rho parameter is modified at the one-loop level. 
The deviation of the rho parameter from unity measures the violation of the custodial $SU(2)_V$ symmetry~\cite{Weinberg_prd13,Susskind,custodial} in the sector 
of particles in the loop\footnote{
In models with $\rho_{\text{tree}}\neq 1$, we need a different prescription for the calculation of the radiative corrections to the rho parameter 
from that in models with $\rho_{\text{tree}}= 1$, because 
of an additional input parameter in the electroweak sector. 
In the Higgs sector with a real triplet Higgs field with $Y=0$, one-loop corrections to the rho parameter have been calculated in Refs.~\cite{Blank_Hollik,rho_real}. 
That has been applied to the Higgs sector with a complex triplet Higgs field with $Y=1$ in Refs.~\cite{KY,AKKY}. 
In the GM model, although $\rho_\text{tree}=1$ can be satisfied, 
similar prescription in models with $\rho_{\text{tree}}\neq 1$ is necessary due to the 
VEV alignment~\cite{GVW_rho}. }. 
For example, in the Yukawa Lagrangian, the custodial symmetry is broken by the mass splitting between the top and bottom quarks.  
As a result, the deviation of the rho parameter from unity $\Delta\rho \equiv \rho-1$ due to the loop contribution of the top and bottom quarks
takes a form of $\Delta\rho \propto(m_t-m_b)^2$.
In fact, since $m_t\gg m_b$, there remains $m_t^2$ dependence in $\Delta\rho$. 
On the other hand, the Higgs potential in the SM respects the custodial symmetry, 
so that the Higgs boson loop contribution is at most the logarithmic dependence of the Higgs boson mass through the hypercharge gauge interaction. 
However, in general, the custodial symmetry is broken in extended Higgs sectors. 
For example, in two Higgs doublet models (THDMs), 
the mass splitting between singly-charged Higgs bosons and a CP-odd Higgs boson gives 
a quadratic mass dependence similarly to the top and bottom quark contributions in $\Delta\rho$~\cite{Ref:rho-2hdm,Ref:rho5-2hdm,Gerard1,Gerard2,Ref:rho6-2hdm}. 
Therefore, a sizable amount of the mass difference has already been excluded~\cite{Ref:Peskin_Wells,Ref:KOTT}.
In the above way, we can take bounds on various physical parameters by comparing 
precisely measured observables with theory predictions with radiative corrections.  

Experimental data for flavor changing neutral current (FCNC) processes such as $K_L^0\to \mu^+\mu^-$ and the $B^0$-$\bar{B}^0$ mixing strongly constrain
extended Higgs sectors with multi-doublet structures.  
The way to avoid such dangerous FCNC processes at the tree level is to assign a different quantum number for 
each Higgs doublet. 
Consequently, each quark or lepton can obtain its mass from only one Higgs doublet just like in the SM, and therefore the model escapes
FCNC processes at the tree level.  
In the THDM, for example, 
this can be achieved by imposing a discrete $Z_2$ symmetry to the model~\cite{Ref:GW} as the simplest way, which can be softly-broken in the potential. 
There are four independent types of Yukawa interactions under the $Z_2$ symmetry~\cite{4types_barger,4types_grossman,Akeroyd}, which are called as 
Type-I, Type-II, Type-X and Type-Y~\cite{Ref:AKTY}\footnote{The Type-X (Type-Y) THDM is referred to as the Type-IV (Type-III) THDM in~\cite{4types_barger}, 
Type-I' (Type-II') THDM in~\cite{4types_grossman,Akeroyd} and 
the lepton-specific (flipped) THDM in~\cite{Logan_MacLennan_X,Logan_MacLennan_Y,Ref:2HDM_Rev}. 
Because the term ``Type-III" is sometimes used for the THDM with tree-level FCNCs~\cite{Type-III}, 
we adopt the terms ``Type-X" and ``Type-Y" to avoid confusion. }
\footnote{If we introduce right-handed neutrinos, four more types of Yukawa interactions can be defined. 
In particular, if one of the two doublets gives Dirac neutrino masses, and another one gives masses of all the other fermion, it is known 
as the  neutrino-philic THDM~\cite{nuphilic}.  }. 

How can we explore extended Higgs sectors?
It is important to understand that in general, 
a new scale $M$ is introduced in extended Higgs sectors, which is irrelevant to the VEV of the Higgs boson. 
When $M$ is much larger than the TeV scale, the mass of the second Higgs boson is approximately given by $M$. 
In this case, the second Higgs boson is too heavy to be discovered directly at the LHC. 
In addition, the indirect effect of new particles decouples from the low energy observables~\cite{decoupling_theorem} such as the coupling constants of the 
discovered Higgs boson. 
However, if $M$ is as high as the TeV scale, there can be two possibilities in searches for additional Higgs bosons. 
The first possibility is that the second Higgs boson can be discovered directly at the LHC. 
In this case, the properties can be directly measured at the LHC, 
and the useful information to determine the structure of the Higgs sector can also be obtained at the High Luminosity (HL)-LHC~\cite{HLLHC,HLLHC2,Ref:A2hZ-HL-LHC}. 
The second possibility is that it cannot be discovered directly, but its indirect effect on the Higgs couplings 
can be significant and thus detectable by precision measurements at the HL-LHC and at the International Linear Collider (ILC)~\cite{Ref:ILC_TDR}. 
It goes without saying that in order to realize the second possibility a small but detectable mixing between the SM-like Higgs boson and an additional Higgs boson is required. 
In this case, in addition to obtaining information on the mass of the second Higgs boson, 
the structure of the Higgs sector could be determined without the direct discovery by 
finding the pattern in deviations in various Higgs boson couplings~\cite{Ref:ILC_White,Ref:Snowmass}.  
On the other hand, if $M$ stays at the electroweak scale, 
a large mixing between the SM-like Higgs boson and an additional Higgs boson can occur, and 
the Higgs boson couplings can deviate significantly from the SM values. 
If the last scenario is realized, both the direct search and the indirect search are possible to determine the Higgs sector.
The direct search for additional Higgs bosons in THDMs at the LHC 
has been discussed in Refs.~\cite{Craig:2013hca,Baglio:2014nea,Ref:Chiang-Yagyu,Su,Ref:THDM-LHC,Ref:Ferreira} after the discovery of 
the Higgs boson. 
The complementarity of additional Higgs boson searches at the LHC and at the ILC is recently discussed in Ref.~\cite{KYZ}. 

In this paper, we discuss how the structure of the Higgs sector can be determined 
at the LHC and at the ILC. 
In particular, we shed light on complementarity of direct
searches of additional Higgs bosons at the upcoming 13 TeV or 14 TeV run of the LHC and 
precision measurements of the coupling constants of the discovered Higgs boson at future collider experiments. 
We consider extended Higgs sectors which satisfy $\rho_{\text{tree}}=1$ without predicting FCNCs at the tree level; i.e., 
the THDM with the softly-broken $Z_2$ symmetry, 
the Doublet-Singlet model~\cite{Ref:singlet}, the GM model~\cite{Ref:GM} and the Doublet-Septet model~\cite{septet_ellis,Ref:septet,KKY}.
For the THDM, we discuss the four types of Yukawa interaction.  
We at first give a detailed explanation for properties in the THDMs such as 
the decay branching ratio, perturbative unitarity and vacuum stability. 
We then analyze the direct search for additional neutral Higgs bosons at the LHC. 
The expected excluded regions on the mass of extra neutral Higgs bosons are shown 
assuming the 14 TeV energy at the LHC\footnote{In Fig.~1.20 in the ILC Higgs White Paper~\cite{Ref:ILC_White}, 
we have shown the expected excluded parameter space in the Type-II and Type-X THDMs at the LHC.
We update this analysis with more detailed explanations.}. 
Next, as the indirect search, we show various patterns of deviations 
in the gauge interaction $hVV$ and the Yukawa interactions $hf\bar{f}$ of the SM-like Higgs boson $h$ from the SM predictions. 
We show the deviation in the $hf\bar{f}$ couplings in the THDMs. 
For the rest models, we also show those in the $hf\bar{f}$ and $hVV$ couplings, where these models predict universal modifications 
for the $hf\bar{f}$ couplings. 
We use the latest results of allowed values of the Higgs boson couplings
which have been obtained from the global fit to all Higgs data~\cite{Strumia} in order to compare the various
prediction of deviations in the Higgs boson couplings\footnote{In Figs.~1.17 and 1.18 in the ILC Higgs White Paper~\cite{Ref:ILC_White}, 
we have shown the deviation in the $hf\bar{f}$ and $hVV$ couplings in the THDMs and in the models with universal modification of the $hf\bar{f}$ couplings. 
We update the plots for the $hf\bar{f}$ and $hVV$ couplings with more detailed explanations by using the latest data~\cite{Strumia}. }.

This paper is organized as follows.
In Section II, we define the Higgs potential and Yukawa Lagrangian in the THDM with the softly-broken $Z_2$ symmetry. 
After we derive the Yukawa couplings in the four types, we discuss the decay branching ratios of the Higgs bosons. 
The bounds from unitarity and vacuum stability are also discussed. 
In Section III, we study the direct search for the additional Higgs bosons at the LHC. 
In Section IV, we present expected accuracy of the precise measurement of the Higgs boson couplings at the ILC, and then
we discuss the deviation in the SM-like Higgs boson couplings are calculated in the THDMs and 
models with universal Yukawa couplings. 
Complementarity between the direct search and the indirect search at the LHC and at the ILC is discussed in Section IV. 
Conclusion is summarized in Section V.


\section{The two Higgs doublet model}

\subsection{Lagrangian}

The Higgs potential of the THDM under the softly-broken $Z_2$ symmetry to avoid FCNC at the tree level
and the CP invariance is given by~\cite{Ref:HHG,Gunion:2002zf,KOSY,Ref:2HDM_Rev}
\begin{align}
V_\text{THDM} &=m_1^2|\Phi_1|^2+m_2^2|\Phi_2|^2-m_3^2(\Phi_1^\dagger \Phi_2 +\text{h.c.})\notag\\
& +\frac{1}{2}\lambda_1|\Phi_1|^4+\frac{1}{2}\lambda_2|\Phi_2|^4+\lambda_3|\Phi_1|^2|\Phi_2|^2+\lambda_4|\Phi_1^\dagger\Phi_2|^2
+\frac{1}{2}\lambda_5\left[(\Phi_1^\dagger\Phi_2)^2+\text{h.c.}\right], \label{pot_thdm2}
\end{align}
where $\Phi_1$ and $\Phi_2$ are the isospin doublet scalar fields with $Y=1/2$ whose $Z_2$ transformation is given as
$\Phi_1\to +\Phi_1$ and $\Phi_2\to -\Phi_2$. 
The two Higgs doublet fields can be parameterized as 
\begin{align}
\Phi_i=\left[\begin{array}{c}
w_i^+\\
\frac{1}{\sqrt{2}}(h_i+v_i+iz_i)
\end{array}\right],\hspace{3mm}(i=1,2), 
\end{align}
where $v_1$ and $v_2$ are the VEVs of two doublet fields. 
They are related to the Fermi constant $G_F$ by $v^2\equiv v_1^2+v_2^2 = (\sqrt{2}G_F)^{-1}$.
The ratio of the two VEVs is defined as $\tan\beta=v_2/v_1$.  

The mass eigenstates for the scalar bosons are obtained by the following orthogonal transformations as
\begin{align}
\left(\begin{array}{c}
w_1^\pm\\
w_2^\pm
\end{array}\right)&=R(\beta)
\left(\begin{array}{c}
G^\pm\\
H^\pm
\end{array}\right),\quad 
\left(\begin{array}{c}
z_1\\
z_2
\end{array}\right)
=R(\beta)\left(\begin{array}{c}
G^0\\
A
\end{array}\right),\quad
\left(\begin{array}{c}
h_1\\
h_2
\end{array}\right)=R(\alpha)
\left(\begin{array}{c}
H\\
h
\end{array}\right), \notag\\
\text{with}~R(\theta) &= 
\begin{pmatrix} 
\cos \theta & -\sin\theta \\ 
\sin\theta & \cos\theta 
\end{pmatrix},\label{mixing}
\end{align}
where $G^\pm$ and $G^0$ are the Nambu-Goldstone bosons absorbed by the longitudinal component of $W^\pm$ and $Z$, respectively.  
The masses of $H^\pm$ and $A$ are calculated as 
\begin{align}
m_{H^+}^2=M^2-\frac{v^2}{2}(\lambda_4+\lambda_5),\quad m_A^2&=M^2-v^2\lambda_5, \label{mass1}
\end{align}
where $M^2\equiv m_3^2/(\sin\beta\cos\beta)$ describes the soft breaking scale of the $Z_2$ symmetry~\cite{KOSY}. 
The masses for the CP-even Higgs bosons $h$ and $H$,  and the mixing angle $\alpha$ are given by
\begin{align}
&m_H^2=\cos^2(\beta-\alpha)M_{11}^2+\sin^2(\beta-\alpha)M_{22}^2-\sin2(\beta-\alpha)M_{12}^2,\\
&m_h^2=\sin^2(\beta-\alpha)M_{11}^2+\cos^2(\beta-\alpha)M_{22}^2+\sin2(\beta-\alpha)M_{12}^2,\\
&\tan 2(\beta-\alpha)=\frac{2M_{12}^2}{M_{22}^2-M_{11}^2}, \label{Eq:tan2}
\end{align}
where 
\begin{align}
M_{11}^2&=v^2(\lambda_1\cos^4\beta+\lambda_2\sin^4\beta)+\frac{v^2}{2}\bar{\lambda}\sin^22\beta, \label{Eq:bm11} \\
M_{22}^2&=M^2+v^2(\lambda_1+\lambda_2-2\bar{\lambda})\sin^2\beta\cos^2\beta, \label{Eq:bm22}\\
M_{12}^2&=\frac{v^2}{2}(-\lambda_1\cos^2\beta+\lambda_2\sin^2\beta)\sin2\beta+\frac{v^2}{2}\bar{\lambda}\sin2\beta\cos2\beta. \label{Eq:bm12}
\end{align}
with $\bar{\lambda}\equiv \lambda_3+\lambda_4+\lambda_5$. 
We define the range of $\beta-\alpha$ to be $[0,\pi/2]$ or $[\pi/2,\pi]$, in which 
for a given positive value of $\sin(\beta-\alpha)$, $\cos(\beta-\alpha)$ is positive or negative, respectively.
 
\begin{table}[t]
\begin{center}
\begin{tabular}{c||c|c|c|c|c|c}
\hline\hline & $\Phi_1$ & $\Phi_2$ & $u_R^{}$ & $d_R^{}$ & $\ell_R^{}$ &
 $Q_L$, $L_L$ \\  \hline
Type-I  & $+$ & $-$ & $-$ & $-$ & $-$ & $+$ \\
Type-II & $+$ & $-$ & $-$ & $+$ & $+$ & $+$ \\
Type-X  & $+$ & $-$ & $-$ & $-$ & $+$ & $+$ \\
Type-Y  & $+$ & $-$ & $-$ & $+$ & $-$ & $+$ \\
\hline\hline
\end{tabular} 
\end{center}
\caption{Four types of the charge assignment of the $Z_2$ symmetry. } 
\label{Tab:type}
\end{table}

\begin{table}[t]
\begin{center}
\begin{tabular}{c||c|c|c|c|c|c|c|c|c}
\hline 
& $\xi_h^u$ & $\xi_h^d$ & $\xi_h^\ell$
& $\xi_H^u$ & $\xi_H^d$ & $\xi_H^\ell$
& $\xi_A^u$ & $\xi_A^d$ & $\xi_A^\ell$ \\ \hline \hline
Type-I
& $\cos\alpha/\sin\beta$ & $\cos\alpha/\sin\beta$ & $\cos\alpha/\sin\beta$
& $\sin\alpha/\sin\beta$ & $\sin\alpha/\sin\beta$ & $\sin\alpha/\sin\beta$
& $\cot\beta$ & $-\cot\beta$ & $-\cot\beta$ \\
Type-II
& $\cos\alpha/\sin\beta$ & $-\sin\alpha/\cos\beta$ & $-\sin\alpha/\cos\beta$
& $\sin\alpha/\sin\beta$ & $\cos\alpha/\cos\beta$ & $\cos\alpha/\cos\beta$
& $\cot\beta$ & $\tan\beta$ & $\tan\beta$ \\
Type-X
& $\cos\alpha/\sin\beta$ & $\cos\alpha/\sin\beta$ & $-\sin\alpha/\cos\beta$
& $\sin\alpha/\sin\beta$ & $\sin\alpha/\sin\beta$ & $\cos\alpha/\cos\beta$
& $\cot\beta$ & $-\cot\beta$ & $\tan\beta$ \\
Type-Y
& $\cos\alpha/\sin\beta$ & $-\sin\alpha/\cos\beta$ & $\cos\alpha/\sin\beta$
& $\sin\alpha/\sin\beta$ & $\cos\alpha/\cos\beta$ & $\sin\alpha/\sin\beta$
& $\cot\beta$ & $\tan\beta$ & $-\cot\beta$ \\
\hline
\end{tabular}
\end{center}
\caption{The mixing factors in each type of Yukawa interactions in the THDMs~\cite{Ref:AKTY}.} 
\label{Tab:Yukawa_Couplings}
\end{table}

The Yukawa Lagrangian under the $Z_2$ symmetry is given by  
\begin{align}
{\mathcal L}^Y_\text{THDM} =
&-Y_u\overline{Q_L}\widetilde{\Phi}_uu_R^{}
-Y_d\overline{Q_L}\Phi_dd_R^{}
-Y_\ell\overline{L_L}\Phi_\ell \ell_R^{}+\text{h.c.},
\end{align}
where $\Phi_{u,d,\ell}$ are either $\Phi_1$ or $\Phi_2$, and $\widetilde{\Phi}_u=i\tau_2\Phi_u^*$. 
There are four independent ways of the charge assignment of the $Z_2$ symmetry as summarized in TABLE~\ref{Tab:type}, 
which are named as Type-I, Type-II, Type-X and Type-Y Yukawa interactions according to Ref.~\cite{Ref:AKTY}. 
After we specify the types of Yukawa interactions, the Yukawa coupling constants are expressed in the mass eigenstate of the Higgs bosons as
\begin{align}
{\mathcal L}^Y_\text{THDM} 
&=
-\sum_{f=u,d,e}\frac{m_F}{v} \left( \xi_h^f{\overline f}fh+\xi_H^f{\overline f}fH-i\,\xi_A^f{\overline f}\gamma_5fA\right)\notag\\
&+\left[\frac{\sqrt2V_{ud}}{v}\overline{u}
\left(m_u\xi_A^uP_L+m_d\xi_A^dP_R\right)d\,H^+
+\frac{\sqrt2m_\ell\xi_A^e}{v}\overline{\nu^{}}P_Re^{}H^+
+\text{h.c.}\right],\label{yukawa_thdm}
\end{align}
where $P_{L,R}=(1\mp\gamma_5)/2$, and the factors $\xi_\phi^f$ ($\phi=h,~H,$ and $A$) are listed in TABLE~\ref{Tab:Yukawa_Couplings}.
We note that the $\xi_h^f$ and $\xi_H^f$ are rewritten by 
\begin{align}
\xi_h^f =\sin(\beta-\alpha)+2T_3^f\xi_A^f\cos(\beta-\alpha),~
\xi_H^f =\cos(\beta-\alpha)-2T_3^f\xi_A^f\sin(\beta-\alpha), \label{Eq:Yukawa2}
\end{align}
where $T_3^f=1/2~(-1/2)$ for $f=u$ ($d,e$).  

After taking the same rotation of the scalar bosons given in Eq.~(\ref{mixing}), 
the Higgs-Gauge-Gauge type terms are expressed by 
\begin{align}
\mathcal{L}_{\text{kin}}^{\phi VV}=&
\Big[\sin(\beta-\alpha) h+ \cos(\beta-\alpha)H\Big] 
\left(\frac{m_W^2}{v} W^{+\mu} W^-_\mu +\frac12 \frac{m_Z^2}{v} Z^\mu Z_\mu \right).\label{Eq:gauge}
\end{align}

Here, we comment on two important limits; the $SM$-$like~limit$ and the $decoupling~limit$~\cite{Gunion:2002zf} which are
realized by taking $\sin(\beta-\alpha)\to 1$ and $M^2\to\infty$, respectively. 
In the former limit, as seen in Eqs.~(\ref{yukawa_thdm}), (\ref{Eq:Yukawa2}) and (\ref{Eq:gauge}), 
the strength of the Yukawa interaction and the gauge interaction of $h$ become the same as in the SM. 
We thus define $h$ as the SM-like Higgs boson which should be identified as the discovered Higgs boson 
with the mass of around 126 GeV, and all the other Higgs bosons $H^\pm$, $A$ and $H$ are regarded as the additional Higgs bosons. 
On the other hand, in the decoupling limit, 
all the masses of additional Higgs bosons become infinity as long as we take the SM-like limit. 
As a result, only the mass of $h$ remains at the electroweak scale. 

If we consider the case without the SM-like limit,  
we cannot take the decoupling limit. 
This can be seen by looking at Eq.~(\ref{Eq:tan2}) which tells us that in order to keep a fixed non-zero value of 
$\tan2(\beta-\alpha)$, we need sizable contributions from $M_{11}^2$ and $M_{12}^2$ to cancel a large value of $M_{22}^2$ by 
the $M^2$ term in Eq.~(\ref{Eq:bm22}).  
However, as seen in Eqs.~(\ref{Eq:bm11}) and (\ref{Eq:bm12}), 
$M_{11}^2$ and $M_{12}^2$ are given like a form of $\lambda_iv^2$, 
so that too large these terms make $\lambda$ coupling constants too large, which are disfavored by the constraints from perturbative unitarity~\cite{Ref:Uni-2hdm0,Ref:Uni-2hdm}. 
Therefore, there is an upper limit for the mass of additional Higgs bosons when we retain the deviation from the SM-like limit. 

\subsection{Vacuum Stability and Unitarity}

In order to keep a stability of the vacuum, 
the Higgs potential should be bounded from below in any directions with a large value of scalar fields. 
The sufficient condition is given by~\cite{VS_THDM,VS_THDM2}
\begin{align}
\lambda_1>0.\quad \lambda_2>0,\quad \sqrt{\lambda_1\lambda_2}+\lambda_3+\text{MIN}(0,\lambda_4+\lambda_5,\lambda_4-\lambda_5)>0.
\end{align}
In addition, the magnitude of several combinations of the quartic Higgs coupling constants 
are constrained by unitarity. 
When we consider the elastic scatterings of 2 body boson states, 
there are 14 neutral, 8 singly-charged and 3 doubly-charged channels. 
After the diagonalization of the $T$ matrix for the $S$-wave amplitude of these processes, 
we obtain the following 12 independent eigenvalues~\cite{Ref:Uni-2hdm} as  
\begin{align}
x_1^\pm &=  \frac{1}{32\pi}
\left[3(\lambda_1+\lambda_2)\pm\sqrt{9(\lambda_1-\lambda_2)^2+4(2\lambda_3+\lambda_4)^2}\right],\\
x_2^\pm &=
\frac{1}{32\pi}\left[(\lambda_1+\lambda_2)\pm\sqrt{(\lambda_1-\lambda_2)^2+4\lambda_4^2}\right],\\
x_3^\pm &= \frac{1}{32\pi}\left[(\lambda_1+\lambda_2)\pm\sqrt{(\lambda_1-\lambda_2)^2+4\lambda_5^2}
\right],\\
x_4^\pm &= \frac{1}{16\pi}(\lambda_3+2\lambda_4\pm 3\lambda_5),\\
x_5^\pm &= \frac{1}{16\pi}(\lambda_3\pm\lambda_4),\\
x_6^\pm &= \frac{1}{16\pi}(\lambda_3\pm\lambda_5).
\end{align}
For each eigenvalue, we impose the following criterion\footnote{Constraints on the parameter space 
using scale dependent coupling constants have been studied in the THDM in Refs.~\cite{VS_THDM2,THDM_RGE}. }
\begin{align}
|x_i^\pm|\leq\frac{1}{2}. 
\end{align}

\begin{figure}[!t]
\begin{center}
\includegraphics[width=70mm]{ma_kv_p.eps}\hspace{5mm}
\includegraphics[width=70mm]{ma_kv_m.eps}
\caption{
Upper limit of $1-\kappa_V$ as a function of $m_A$ for each value of $\tan\beta$
from the constraints of unitarity and vacuum stability in the case where $M$ is scanned over the range of $m_A\pm 500$ GeV and $m_{H^+}=m_A$.  
The left and right panels respectively show the results with $\cos(\beta-\alpha)>0$ and 
$\cos(\beta-\alpha)<0$.  
The solid curves show the case with $m_H=m_A$, while the dotted curves show the result with $m_H$ to be scanned over the range of  $m_A\pm 500$ GeV.  }
\label{unitarity1}
\end{center}
\end{figure}

\begin{figure}[t]
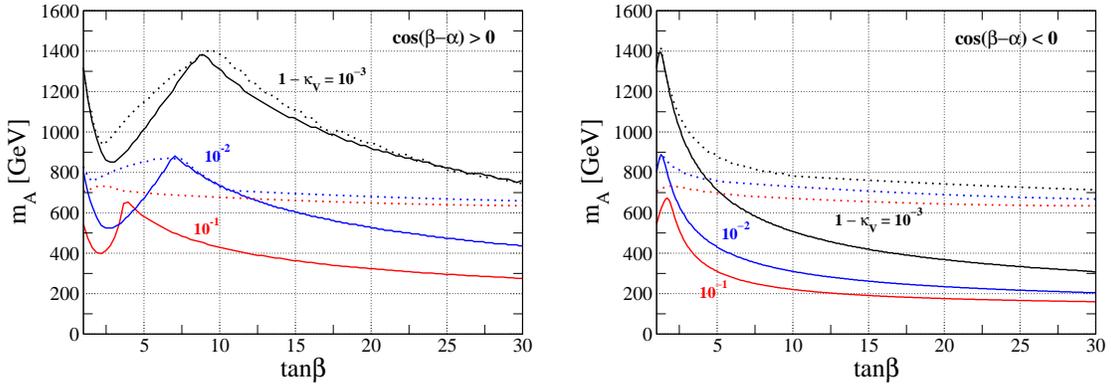

\begin{center}
\includegraphics[width=70mm]{tanb_ma_p.eps}\hspace{5mm}
\includegraphics[width=70mm]{tanb_ma_m.eps}
\caption{
Upper limit of $m_A$ as a function of $\tan\beta$ for each value of $1-\kappa_V$ 
from the constraints of unitarity and vacuum stability in the case where $M$ is scanned over the range of $m_A\pm 500$ GeV and $m_{H^+}=m_A$.  
The left and right panels respectively show the results with $\cos(\beta-\alpha)>0$ and 
$\cos(\beta-\alpha)<0$.  
The solid curves show the result with $m_H=m_A$, while the dotted curves show the result with $m_H$ to be scanned over the range of  $m_A\pm 500$ GeV. 
}
\label{unitarity2}
\end{center}
\end{figure}

As we mentioned in the previous subsection, the unitarity and vacuum stability bounds can be used to obtain the upper limit on 
the mass of additional Higgs bosons when $\sin(\beta-\alpha)$ deviates from unity.  
We introduce the scaling factor $\kappa_V^{}$ defined by the ratio of the $hVV$ coupling constant to the corresponding SM value, which 
coincides with $\sin(\beta-\alpha)$ at the tree level. 
In Fig.~\ref{unitarity1}, we show the upper limit of $m_A$ from the unitarity and vacuum stability bounds
for given values of $1-\kappa_V$ and $\tan\beta$. The value of $M$ is scanned over the range of  $m_A\pm 500$ GeV.  
To avoid the constraint from the rho parameter, we take $m_{H^+}=m_A$ in these plot, so that the one-loop 
corrections to the rho parameter from the additional Higgs boson loops become zero due to the custodial symmetry in the Higgs potential~\cite{
Ref:rho-2hdm,Ref:rho5-2hdm,Gerard1,Gerard2,Ref:rho6-2hdm,Ref:Peskin_Wells,Ref:KOTT}. 
The value of $m_H$
is taken to be the same as $m_A$ (scanned over the range of $m_A\pm 500$ GeV) in the solid (dotted) curves.  
The left and right panels show the cases with $\cos(\beta-\alpha)>0$ and $\cos(\beta-\alpha)<0$, respectively. 
It is seen that the maximal allowed value of $m_A$ is getting larger when the deviation in $\kappa_V$ from unity is getting small. 
Therefore, larger deviations in the $hVV$ coupling constant give a severe upper bound on masses for additional Higgs bosons. 

In Fig.~\ref{unitarity2}, we show the $\tan\beta$ dependence of the upper limit of $m_A$ from the unitarity and vacuum stability bounds for a given value of $1-\kappa_V$. 
The other parameters are taken to be the same as in Fig.~\ref{unitarity1}. 

\subsection{Decay of the Higgs Bosons}

In this subsection, we discuss the decays of Higgs bosons with the four types of Yukawa interaction in the THDM. 
The decay property can be drastically different between the case with $\sin(\beta-\alpha)=1$ and that with $\sin(\beta-\alpha)\neq 1$~\cite{Ref:AKTY}. 
When the SM-like limit is taken, the additional Higgs bosons can dominantly decay into a fermion pair 
whose decay branching ratio strongly depend on the type of Yukawa interactions and $\tan\beta$. 
On the other hand, 
when we take $\sin(\beta-\alpha)\neq 1$, 
$H$ can decay into the gauge boson pairs $W^+W^-$ and $ZZ$, and the SM-like Higgs boson pair $hh$, where these decay rates are proportional to $\cos^2(\beta-\alpha)$. 
At the same time, $H^\pm$ and $A$ can decay into $W^\pm$ and $Z$ associated with the SM-like Higgs boson $h$ whose 
decay amplitudes are also proportional to $\cos(\beta-\alpha)$~\cite{Mukai,Craig:2013hca,Baglio:2014nea,Su}.

In order to calculate the decay rates, we use the following inputs from Particle Data Group~\cite{PDG}
\begin{align}
&m_Z=91.1876~\text{GeV},~m_W=80.385~\text{GeV},~G_F=1.1663787\times 10^{-5}~\text{GeV}^{-2},    \notag\\   
&m_t=173.07~\text{GeV},~\alpha_s(m_Z)=0.1185,~V_{cb}= 0.0409,~V_{ts}= 0.0429. \label{eq:input} 
\end{align}  
The running quark masses at the scale of $m_Z$ are quoted from Ref.~\cite{Fusaoka:1998vc} as
\begin{align}
\bar{m}_b=3.0~\text{GeV},~\bar{m}_c=0.677~\text{GeV},~\bar{m}_s=0.0934~\text{GeV}. 
\end{align} 
The mass of the SM-like Higgs boson $h$ is taken to be 126 GeV in the following calculations. 
 
All the other parameters shown in Eq.~(\ref{eq:input}) are quoted from PDG~\cite{PDG}. 
We note that the effects of Cabibbo-Kobayashi-Maskawa matrix elements $V_{cb}$ and $V_{ts}$ appear in the $H^\pm \to cb$ and $H^\pm \to ts$ decays.
For simplicity, we take all the masses of additional Higgs bosons to be the same; i.e.,  $m_{H^+}=m_A=m_H~(\equiv m_\Phi)$. 
In that case, there are four free parameters in the Higgs potential, which are chosen as $m_\Phi$, $M^2$, $\tan\beta$ and $\sin(\beta-\alpha)$. 

We here comment on the $H^\pm\to W^\pm Z$ and $H^\pm\to W^\pm \gamma$ processes. 
The $H^\pm  W^\mp \gamma$ vertex is obtained at the one-loop level whose magnitude is suppressed due to the $U(1)_{\text{em}}$ gauge invariance. 
This nature does not depend on a model. 
On the other hand, 
in the THDM, 
although the $H^\pm  W^\mp Z$ vertex appears at the one-loop level, 
it can enhance if there is a large violation of the custodial symmetry. 
As we discussed in Introduction, the mass splitting between the top and bottom quarks breaks the custodial symmetry, 
and it gives the $(m_t-m_b)^2$ dependence in the one-loop corrected rho parameter. 
Similar effect appears in the $H^\pm  W^\mp Z$ vertex~\cite{Haber}. 
In addition, when the mass splitting between $H^\pm$ and $A$ is given, which breaks the custodial symmetry in the Higgs potential, 
the $H^\pm  W^\mp Z$ vertex can be enhanced due to the $(m_{H^+}-m_A)^2$ dependence. 
In Ref.~\cite{Kanemura_HWZ}, full one-loop calculation of the $H^\pm W^\mp Z$ vertex have been done. 
It has been shown that the branching ratio of $H^\pm\to W^\pm Z$ can be $\mathcal{O}(10^{-2})$ in the case of $m_{H^+}=300$ GeV 
when the mass splitting between $H^\pm$ and $A$ is taken to be $\mathcal{O}(100)$ GeV. 
In the following calculation, we assume $m_{H^+}=m_A$, so that only the top and bottom quarks loop contribution to the $H^\pm\to W^\pm Z$ vertex is important. 
In this case, typical values of the branching fractions of $H^\pm\to W^\pm Z$ and $H^\pm\to W^\pm \gamma$ are smaller than $\mathcal{O}(10^{-3})$
and $\mathcal{O}(10^{-5})$, respectively~\footnote{If the Higgs sector contains $exotic$ Higgs fields whose isospin is larger than 1/2, 
the $H^\pm  W^\mp Z$ vertex appears at the tree level~\cite{Grifols}. 
The magnitude depends on VEVs from exotic Higgs fields which are usually severely constrained by the rho parameter. 
In the GM model and in the Doublet-Septet model, such a VEV can be taken as $\mathcal{O}(10)$ GeV. 
Therefore, measuring the $H^\pm  W^\mp Z$ vertex can be a probe of exotic Higgs sectors. 
The feasibility study for the measurement of the vertex has been performed in Refs.~\cite{HWZ-LHC} at the LHC and in Ref.~\cite{HWZ-ILC} at the ILC. }. 
We thus safely neglect these modes in the following calculation. 

\begin{figure}[t]
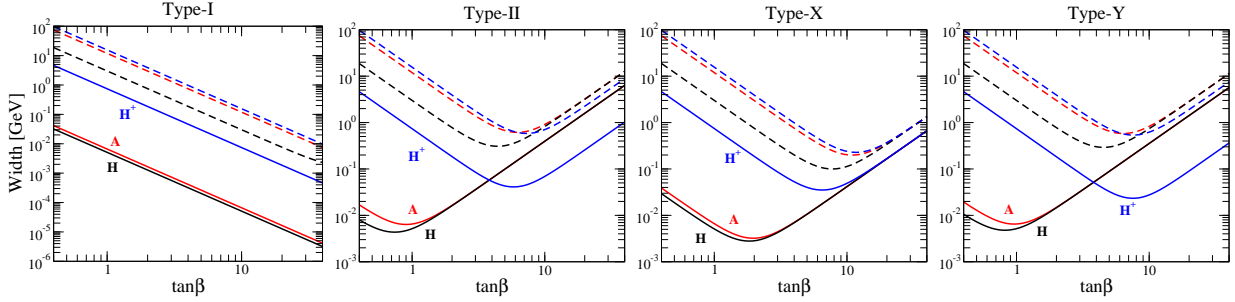

\includegraphics[width=42mm]{width_1.eps}
\includegraphics[width=39mm]{width_2.eps}
\includegraphics[width=39mm]{width_X.eps}
\includegraphics[width=39mm]{width_Y.eps}
\caption{
Total widths for $H$, $A$ and $H^\pm$ as a function of $\tan\beta$ in the case of $\sin(\beta-\alpha)=1$. 
The solid and dashed curves respectively show the results with $m_{\Phi}=M=200$ GeV and $m_{\Phi}=M=400$ GeV. 
}
\label{FIG:width}
\end{figure}

First, we show the total widths for $H$, $A$ and $H^\pm$ in Fig.~\ref{FIG:width} as a function of $\tan\beta$ in the case of $\sin(\beta-\alpha)=1$. 
The solid (dashed) curves show the results with $m_{\Phi}=M=200~(400)$ GeV. 
Except in the Type-I THDM, the widths have a minimum in a certain value of $\tan\beta$, because 
the sum of the decay rates of fermion pair mode are given by terms proportional to $\cot^2\beta$, $\tan^2\beta$ and those without $\tan\beta$ dependence. 
In the Type-I THDM, all the decay rates with the fermion pair final state 
are suppressed by $\cot^2\beta$, so that the $\tan\beta$ dependence of the widths is monotonic decrease. 
In the Type-X THDM, all the widths for $H$, $A$ and $H^\pm$ approach roughly the same value in the high $\tan\beta$ region for a fixed value of $m_\Phi$. 
This can be understood in such a way that the decay rate of $H^\pm\to tb$ mainly deviates the width of $H^\pm$ from that of $H$ and $A$, 
which can be neglected in 
the high $\tan\beta$ region in the Type-X THDM. 
In the Type-I THDM, although the decay rate of $H^\pm\to tb$ is suppressed as in the Type-X THDM, all the other fermion pair decay 
modes are also suppressed at the same time. 
Therefore, the $H^\pm\to tb$ decay is not negligible in the Type-I THDM, and then it deviates the width of $H^\pm$ from that of $H$ and $A$.

\begin{figure}[t]
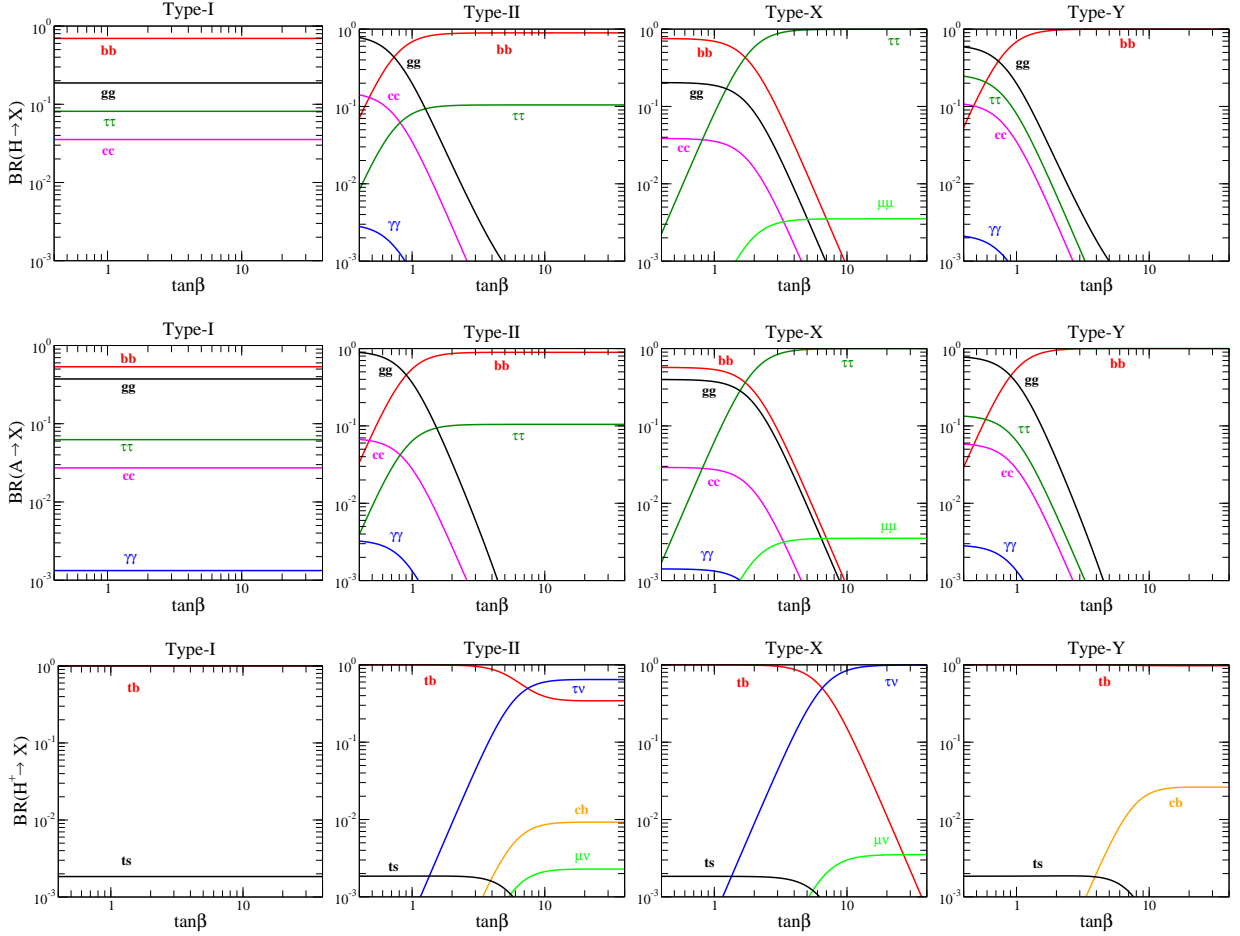

\includegraphics[width=42mm]{H_200_1.eps}\hspace{0mm}
\includegraphics[width=39mm]{H_200_2.eps}\hspace{0mm}
\includegraphics[width=39mm]{H_200_X.eps}\hspace{0mm}
\includegraphics[width=39mm]{H_200_Y.eps}\\ \vspace{3mm}
\includegraphics[width=42mm]{A_200_1.eps}\hspace{0mm}
\includegraphics[width=39mm]{A_200_2.eps}\hspace{0mm}
\includegraphics[width=39mm]{A_200_X.eps}\hspace{0mm}
\includegraphics[width=39mm]{A_200_Y.eps}\\ \vspace{3mm}
\includegraphics[width=42mm]{Hp_200_1.eps}\hspace{0mm}
\includegraphics[width=39mm]{Hp_200_2.eps}\hspace{0mm}
\includegraphics[width=39mm]{Hp_200_X.eps}\hspace{0mm}
\includegraphics[width=39mm]{Hp_200_Y.eps}
\caption{
Decay branching ratios for $H$, $A$ and $H^\pm$ as a function of $\tan\beta$ in the case of $m_{\Phi}=M=200$ GeV and $\sin(\beta-\alpha)=1$. 
}
\label{FIG:BR1}
\end{figure}

In Fig.~\ref{FIG:BR1}, 
we show the decay branching fractions of $H$ (top panels), $A$ (middle panels) and $H^\pm$ (bottom panels) 
as a function of $\tan\beta$ in the case of $\sin(\beta-\alpha)=1$ and $m_{\Phi}=M=200$ GeV. 
It is seen that only in the Type-X THDM, $H$ and $A$ can mainly decay into $\tau^+\tau^-$ in the case of $\tan\beta\gtrsim 3$. 
Besides, $H$ and $A$ can also decay into $\mu^+\mu^-$ with about 0.3\% in the Type-X THDM. 
Regarding the $H^\pm$ decay, 
although the main decay mode is basically $tb$ in all the types, that is replaced by $H^\pm \to \tau^\pm\nu$
in the Type-II and Type-X THDMs with $\tan\beta\gtrsim 7$. 
%

\begin{figure}[t]
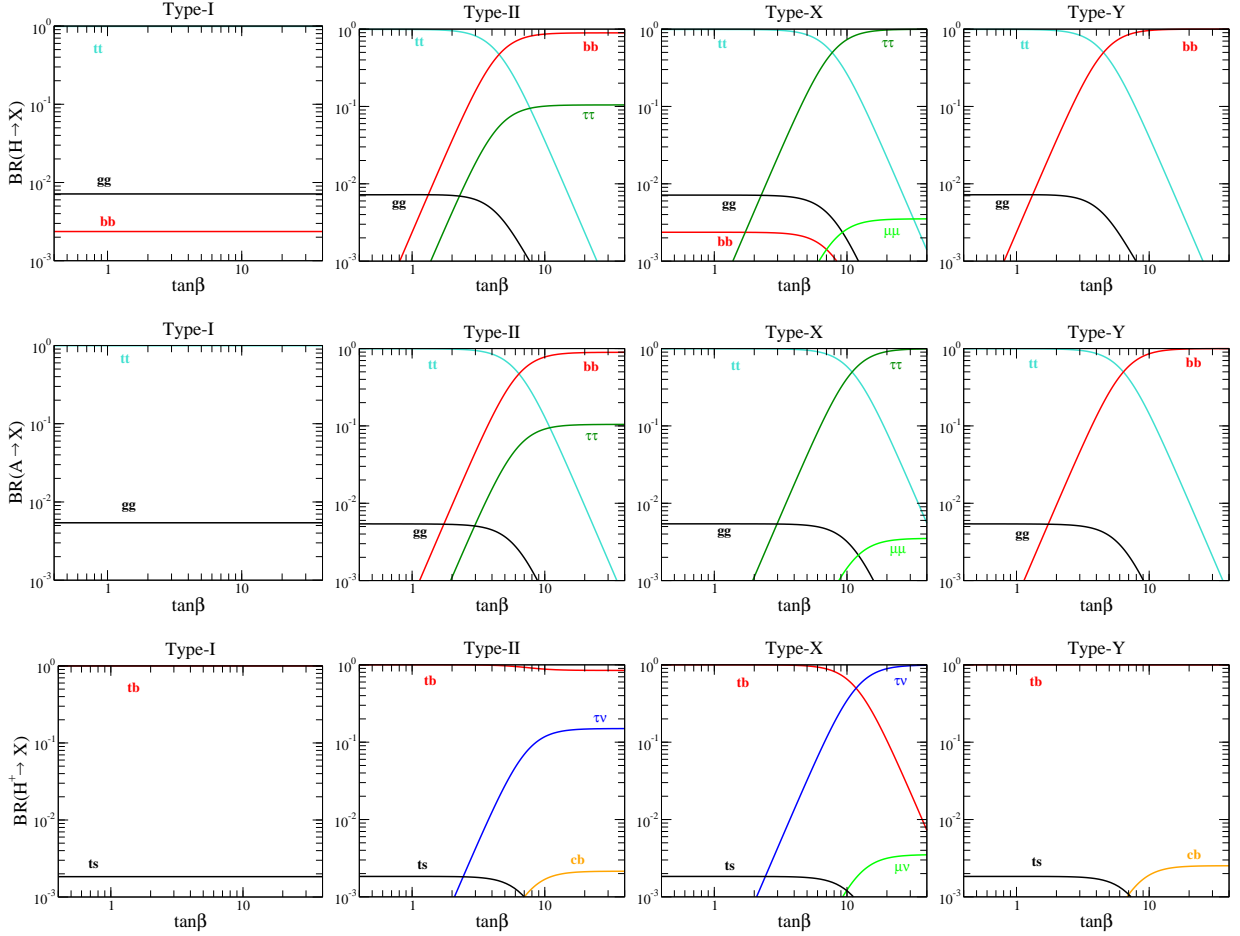

\includegraphics[width=42mm]{H_400_1.eps}\hspace{0mm}
\includegraphics[width=39mm]{H_400_2.eps}\hspace{0mm}
\includegraphics[width=39mm]{H_400_X.eps}\hspace{0mm}
\includegraphics[width=39mm]{H_400_Y.eps}\\ \vspace{3mm}
\includegraphics[width=42mm]{A_400_1.eps}\hspace{0mm}
\includegraphics[width=39mm]{A_400_2.eps}\hspace{0mm}
\includegraphics[width=39mm]{A_400_X.eps}\hspace{0mm}
\includegraphics[width=39mm]{A_400_Y.eps}\\ \vspace{3mm}
\includegraphics[width=42mm]{Hp_400_1.eps}\hspace{0mm}
\includegraphics[width=39mm]{Hp_400_2.eps}\hspace{0mm}
\includegraphics[width=39mm]{Hp_400_X.eps}\hspace{0mm}
\includegraphics[width=39mm]{Hp_400_Y.eps}
\caption{
Decay branching ratios for $H$, $A$ and $H^\pm$ as a function of $\tan\beta$ in the case of $m_{\Phi}=M=400$ GeV and $\sin(\beta-\alpha)=1$. 
}
\label{FIG:BR2}
\end{figure}

Similarly, Fig.~\ref{FIG:BR2} shows the branching fractions of $H$, $A$ and $H^\pm$ in the case of $m_{\Phi}=M=400$ GeV and $\sin(\beta-\alpha)=1$. 
In all the types of THDMs, $H$ and $A$ mainly decay into the top pair in the lower $\tan\beta$ region. 
However, that is replaced by $b\bar{b}$ ($\tau^+\tau^-$) in the Type-II and Type-Y (Type-X) THDMs with $\tan\beta \gtrsim 5$ ($\tan\beta \gtrsim 8$).   
The decay of $H^\pm$ does not change so much from that in the case of $m_{\Phi}=M=200$ GeV. 
Notice here that the magnitude relation between 
the branching fraction of $H/A\to gg$ and that of $H/A\to b\bar{b}$ is flipped compared to the results in Fig.~\ref{FIG:BR1} except 
in the Type-II and Type-Y THDMs with $\tan\beta\gtrsim 1$. 
We note that in the case of $\sin(\beta-\alpha)=1$ and $m_H=m_A$, 
only the difference between the decay rate of $H\to f\bar{f}$ and that of $A\to f\bar{f}$ appears in the power of the phase space factor; 
i.e., that is the cubic (linear) power for $H$ ($A$)~\cite{Ref:HHG}. 
Thus, the decay rate of $A\to f\bar{f}$  is slightly larger than that of $H\to f\bar{f}$. 
Moreover, 
the decay rates of loop induced modes such as the decays into $gg$, $\gamma\gamma$ and $Z\gamma$ are different between $H$ and $A$, 
because of the CP-property.

\begin{figure}[t]
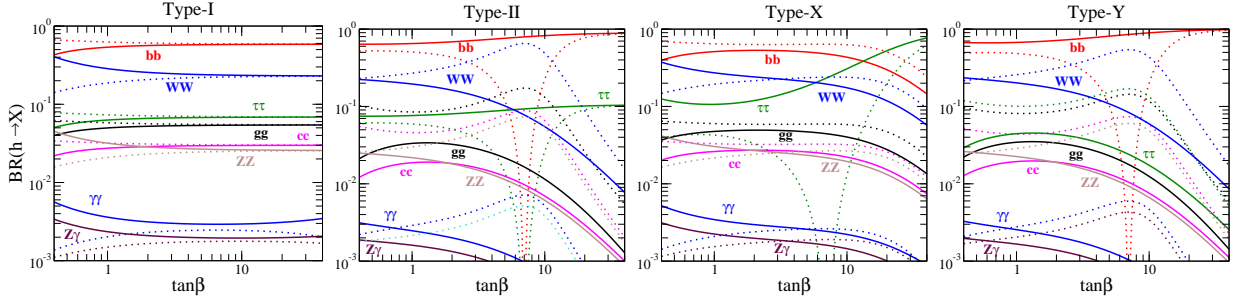

\includegraphics[width=42mm]{h_1_99.eps}
\includegraphics[width=39mm]{h_2_99.eps}
\includegraphics[width=39mm]{h_X_99.eps}
\includegraphics[width=39mm]{h_Y_99.eps}
\caption{
Decay branching ratios for $h$ as a function of $\tan\beta$ in the case of $m_{\Phi}=M=200$ GeV and $\sin(\beta-\alpha)=0.99$. 
The solid and dashed curves respectively show the cases with $\cos(\beta-\alpha)<0$ and $\cos(\beta-\alpha)>0$. 
}
\label{FIG:BR3}
\end{figure}

\begin{figure}[t]
\includegraphics[width=42mm]{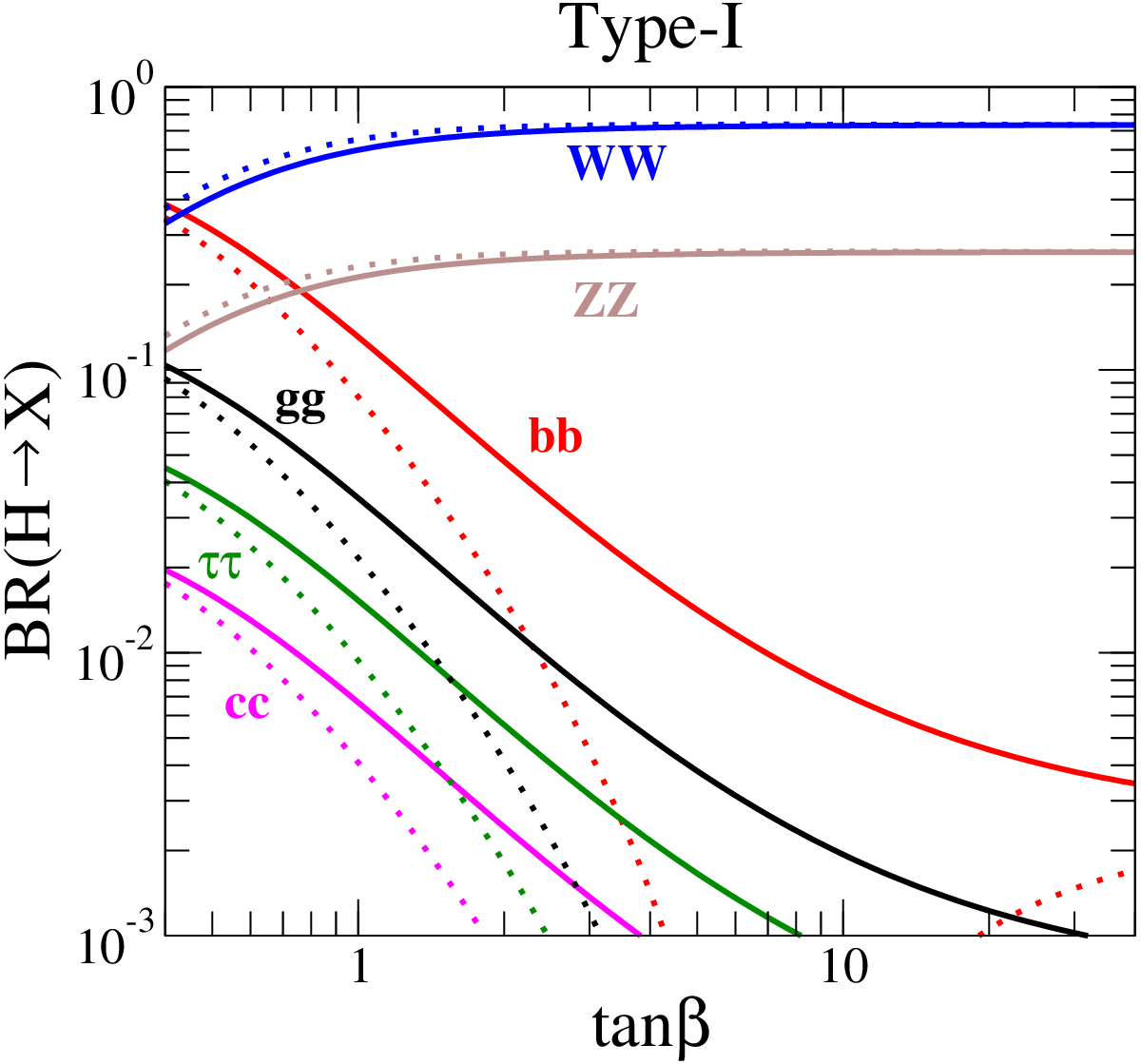}\hspace{0mm}
\includegraphics[width=39mm]{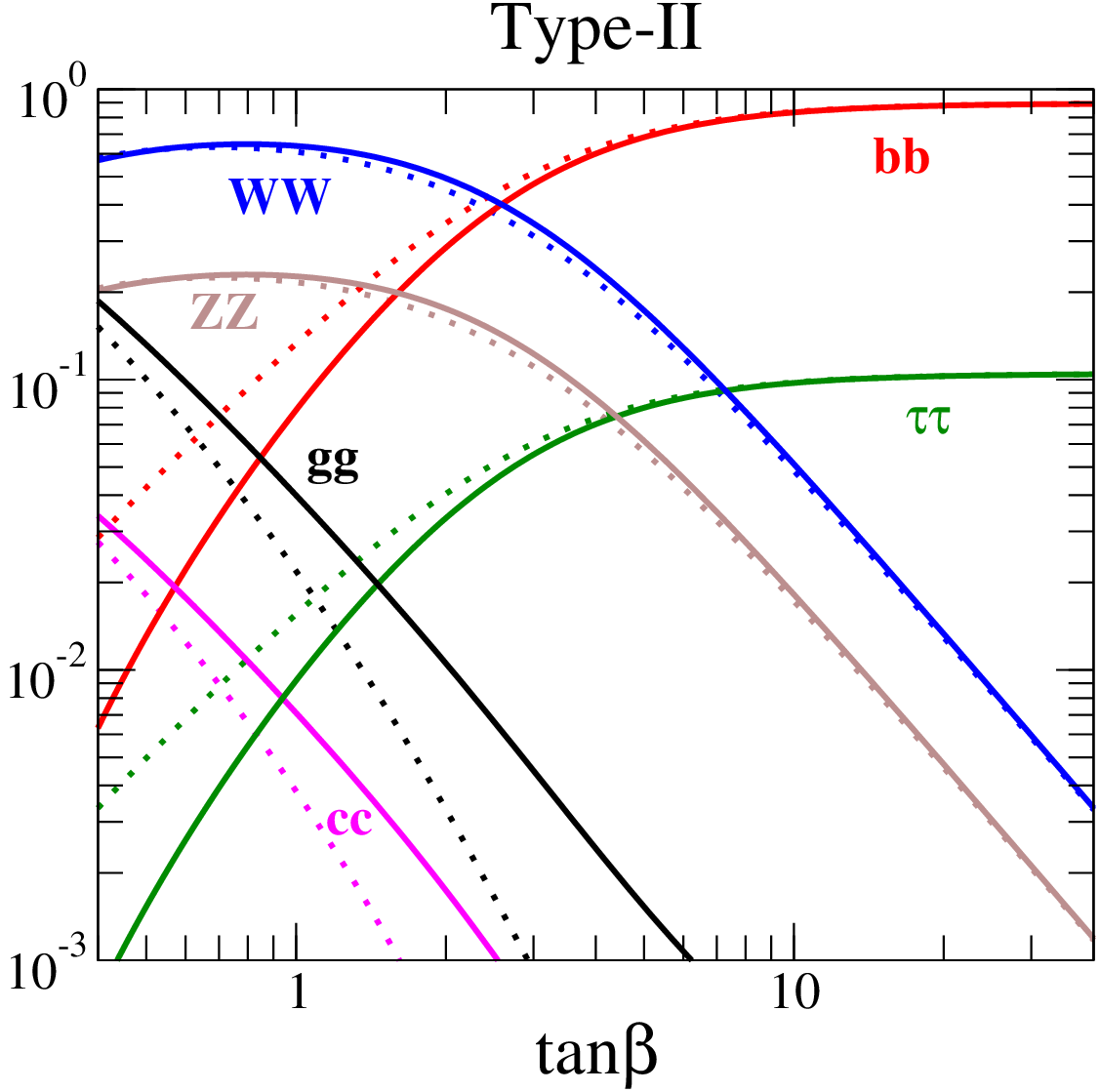}\hspace{0mm}
\includegraphics[width=39mm]{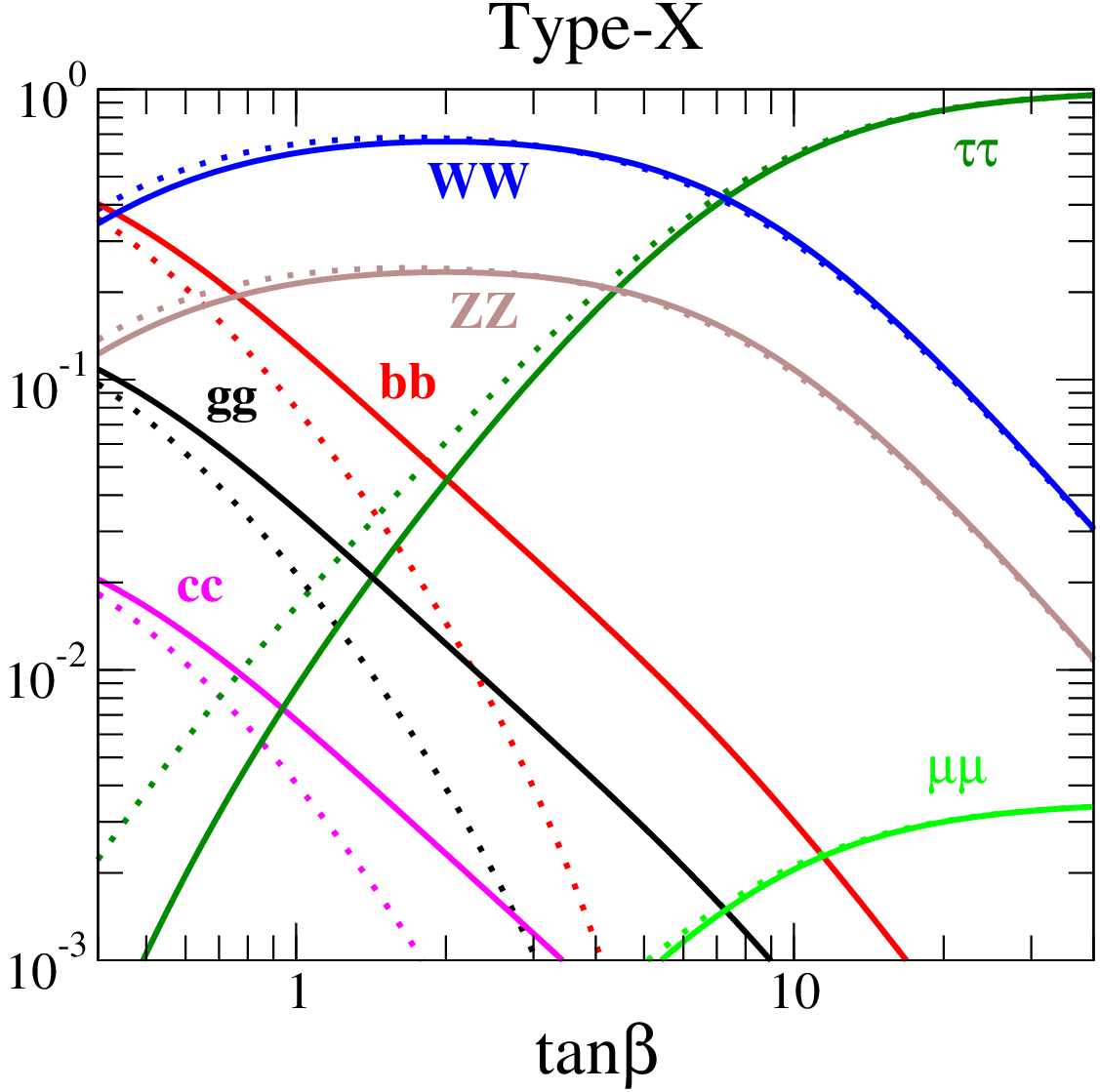}\hspace{0mm}
\includegraphics[width=39mm]{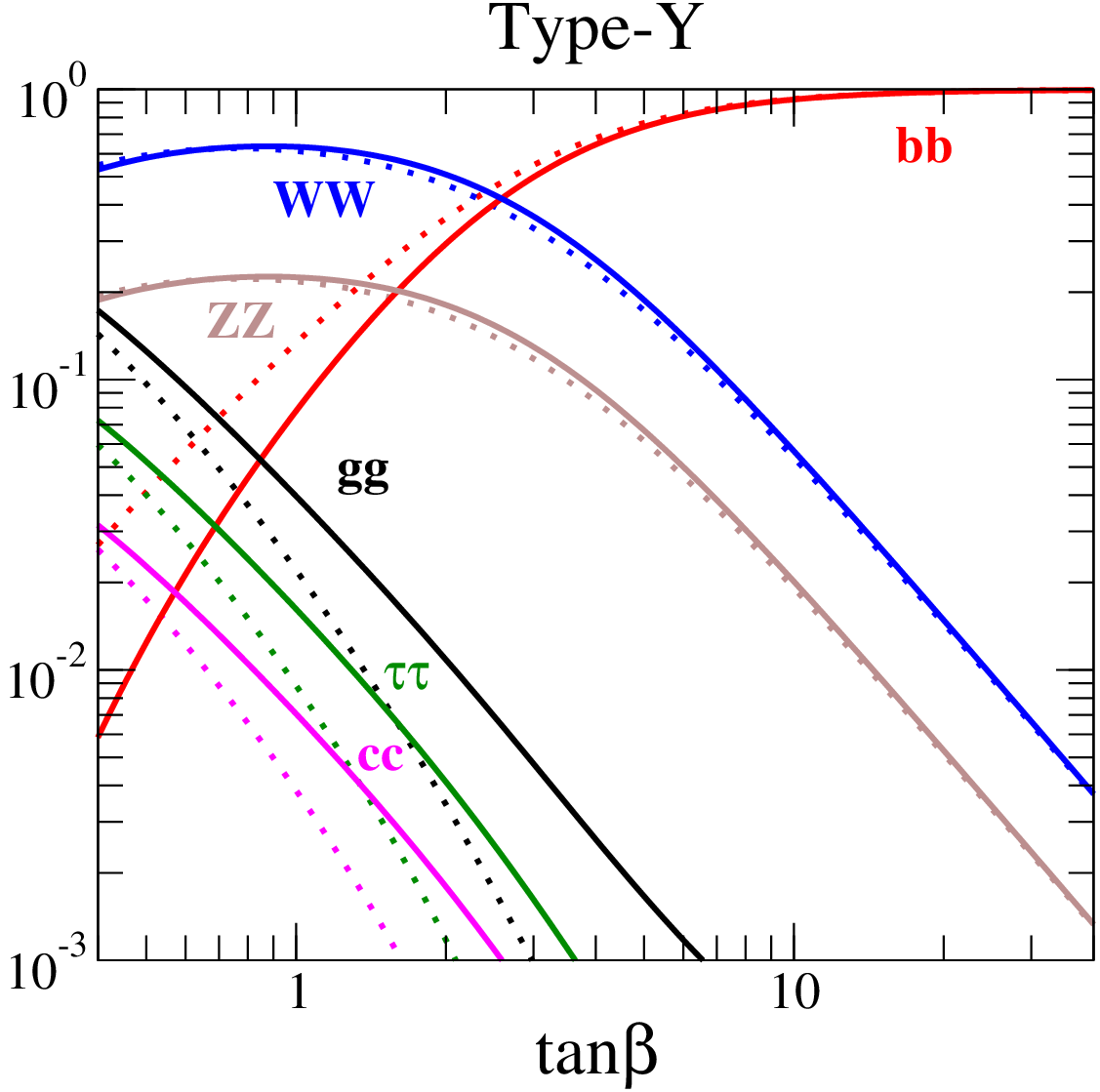}\\ \vspace{3mm}
\includegraphics[width=42mm]{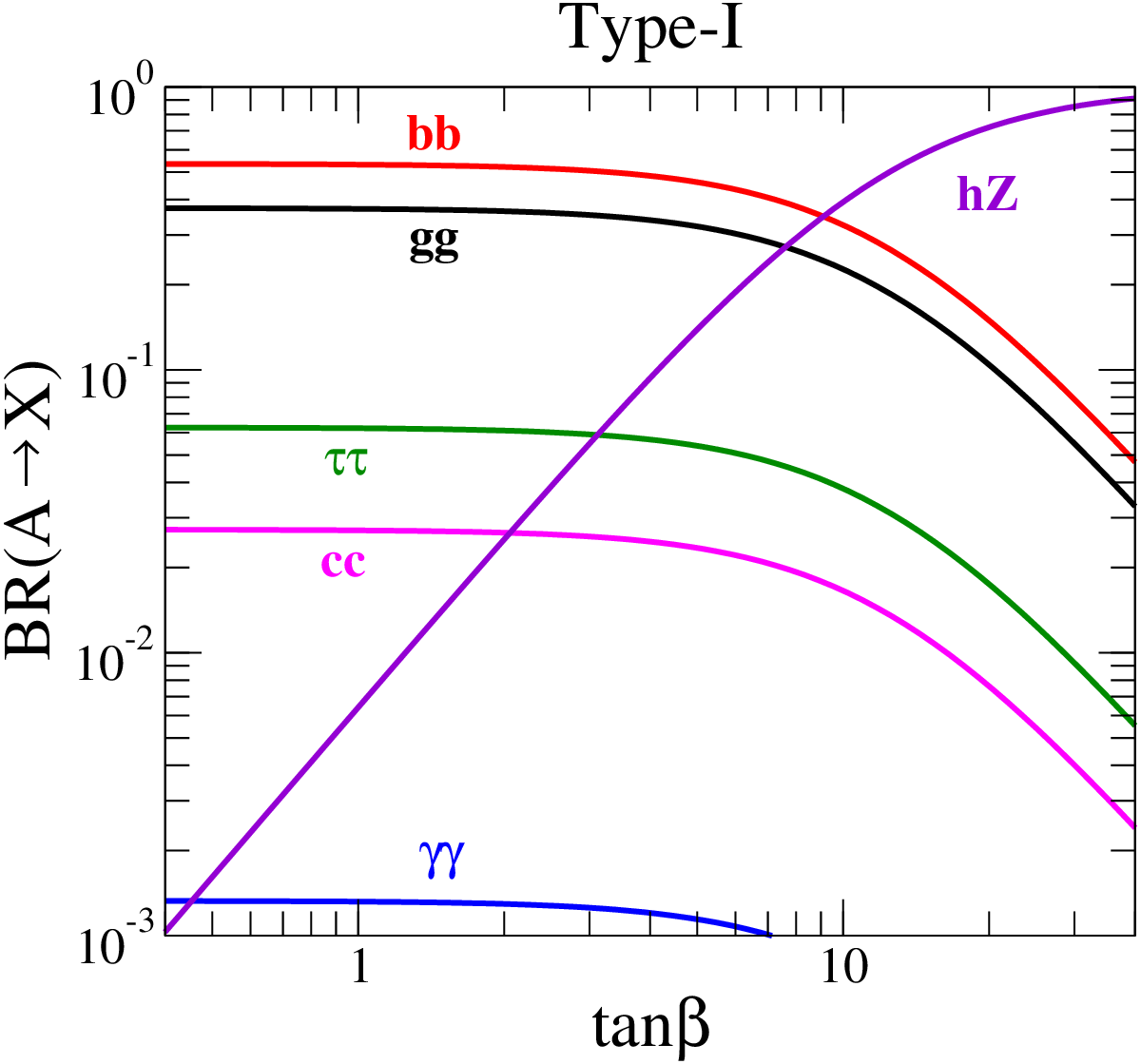}\hspace{0mm}
\includegraphics[width=39mm]{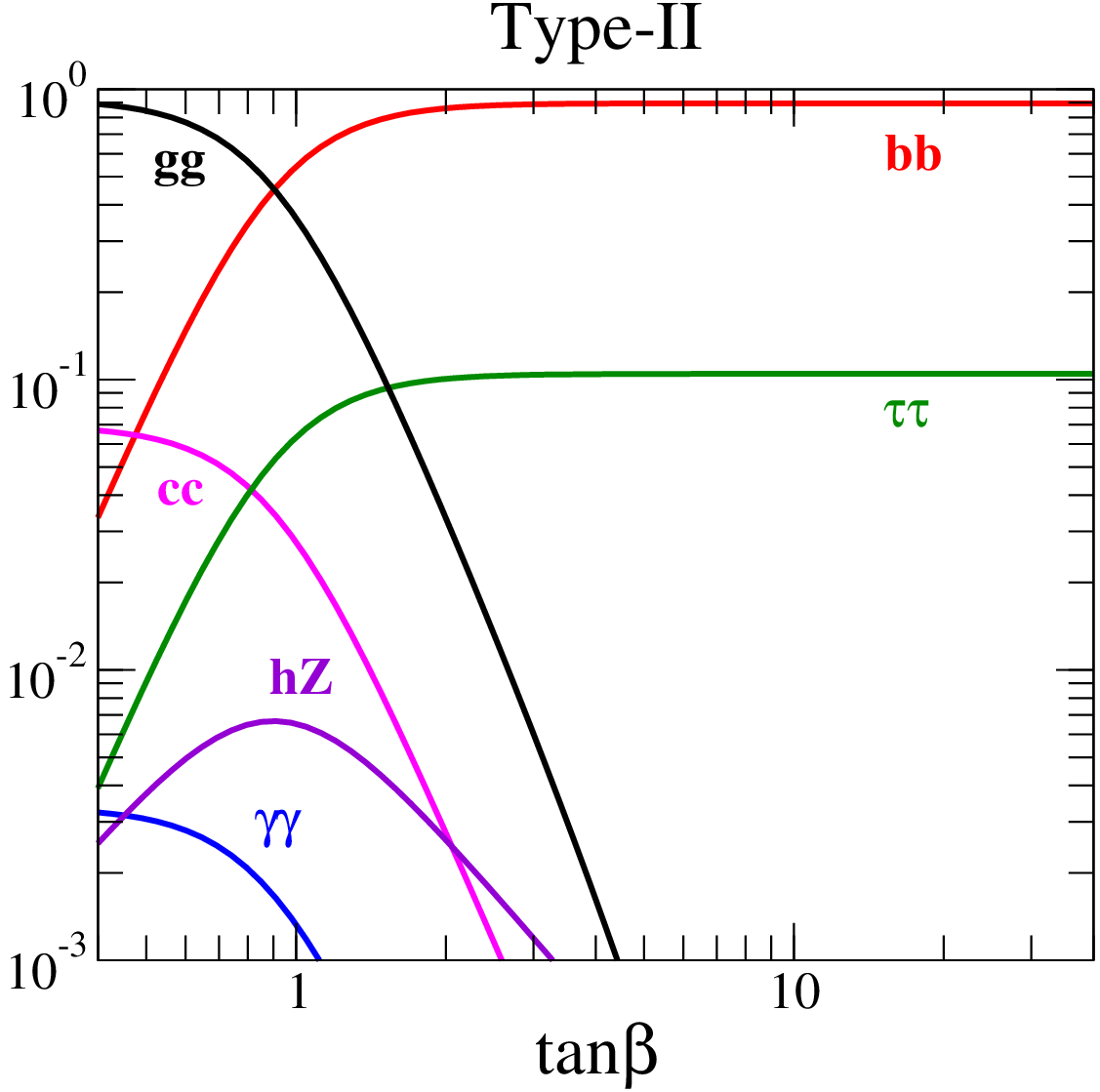}\hspace{0mm}
\includegraphics[width=39mm]{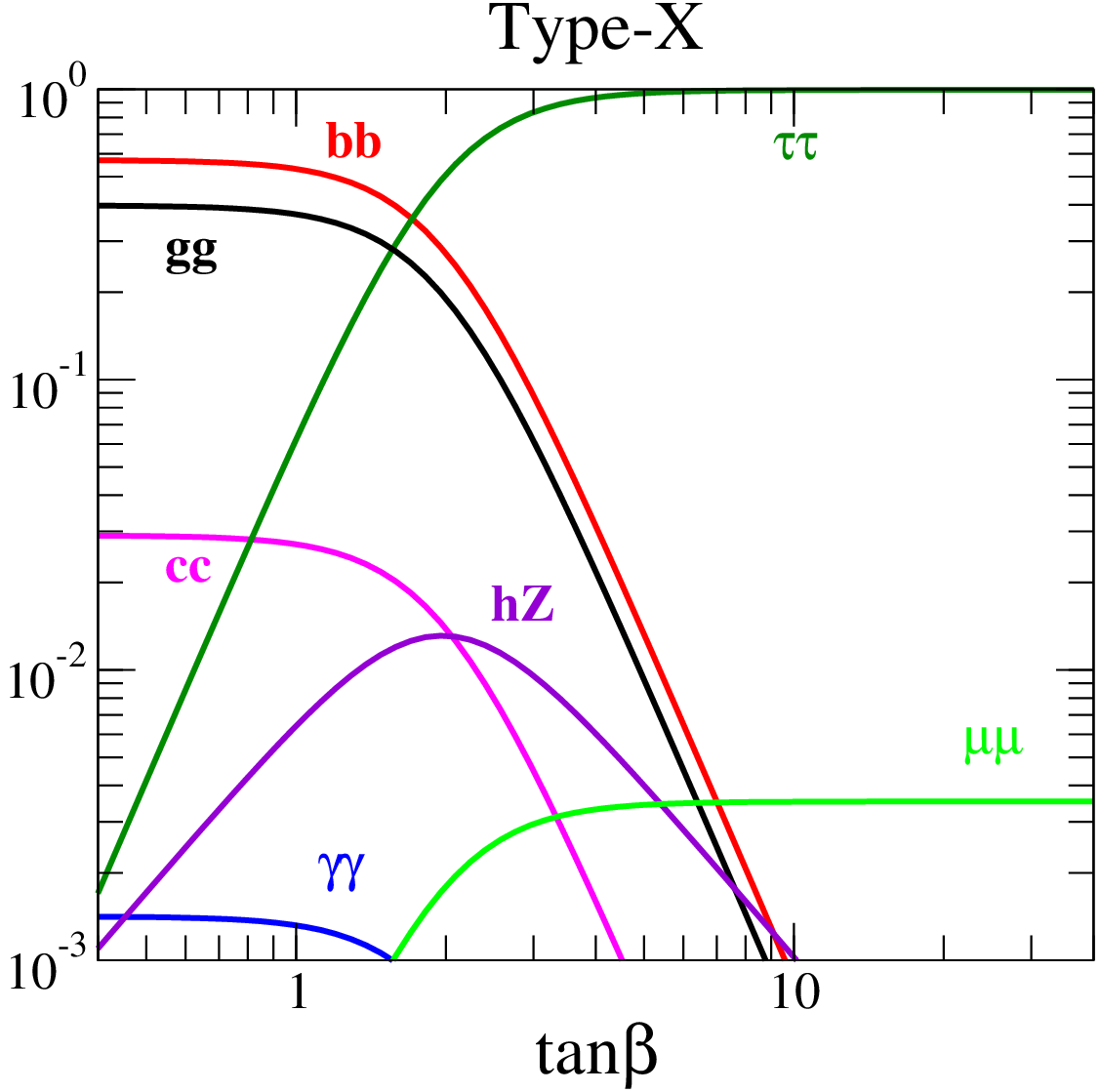}\hspace{0mm}
\includegraphics[width=39mm]{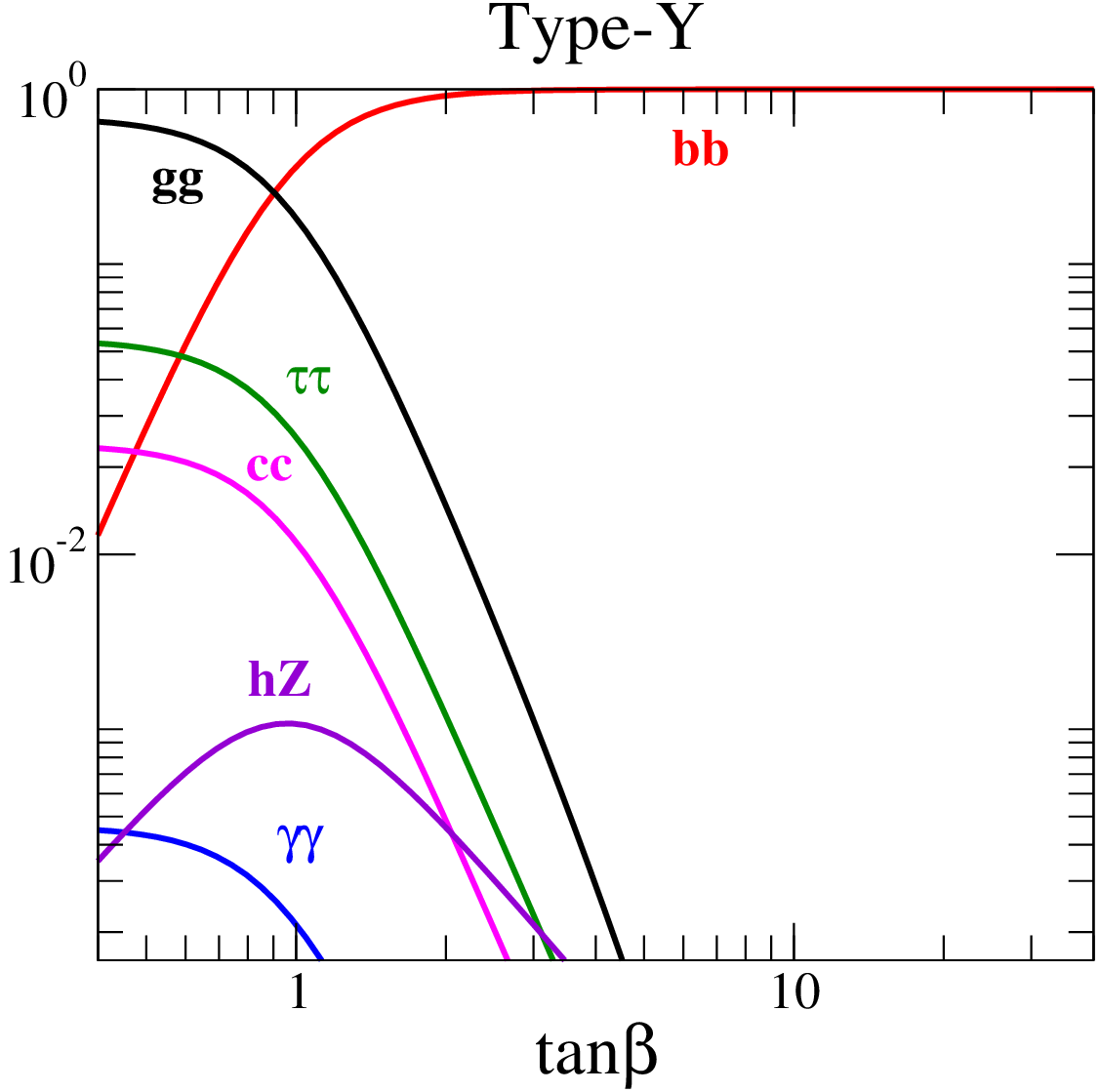}\\ \vspace{3mm}
\includegraphics[width=42mm]{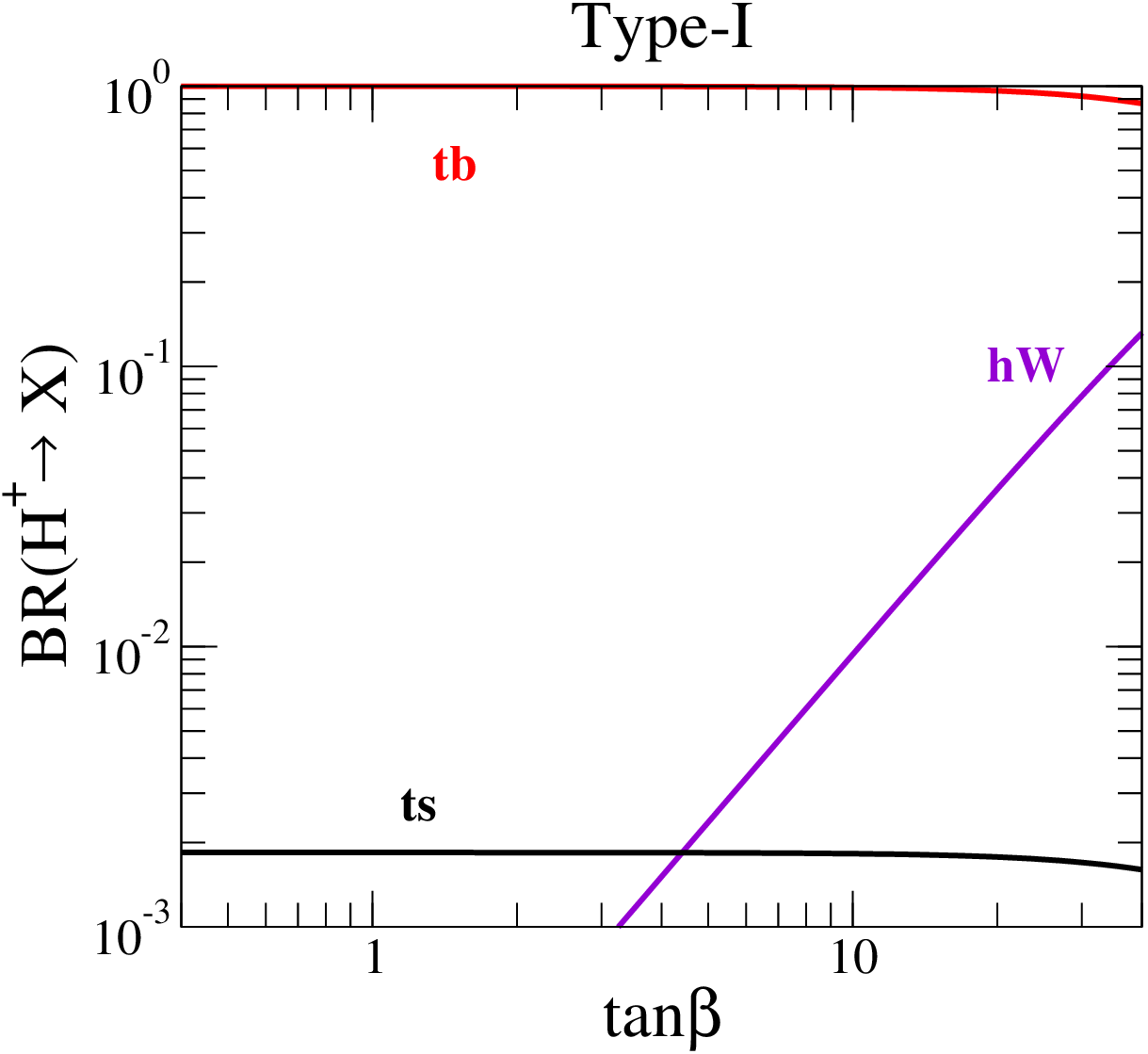}\hspace{0mm}
\includegraphics[width=39mm]{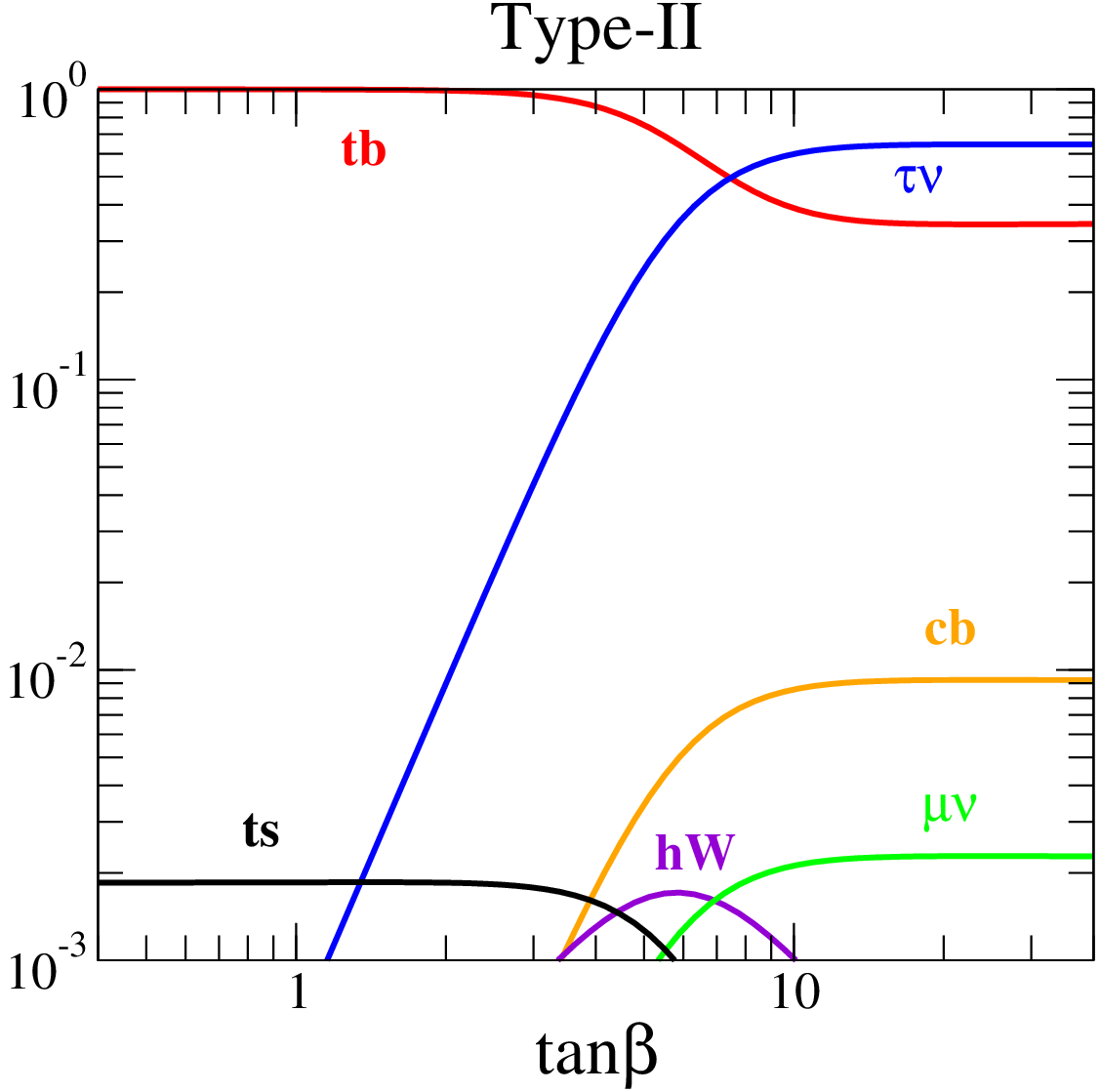}\hspace{0mm}
\includegraphics[width=39mm]{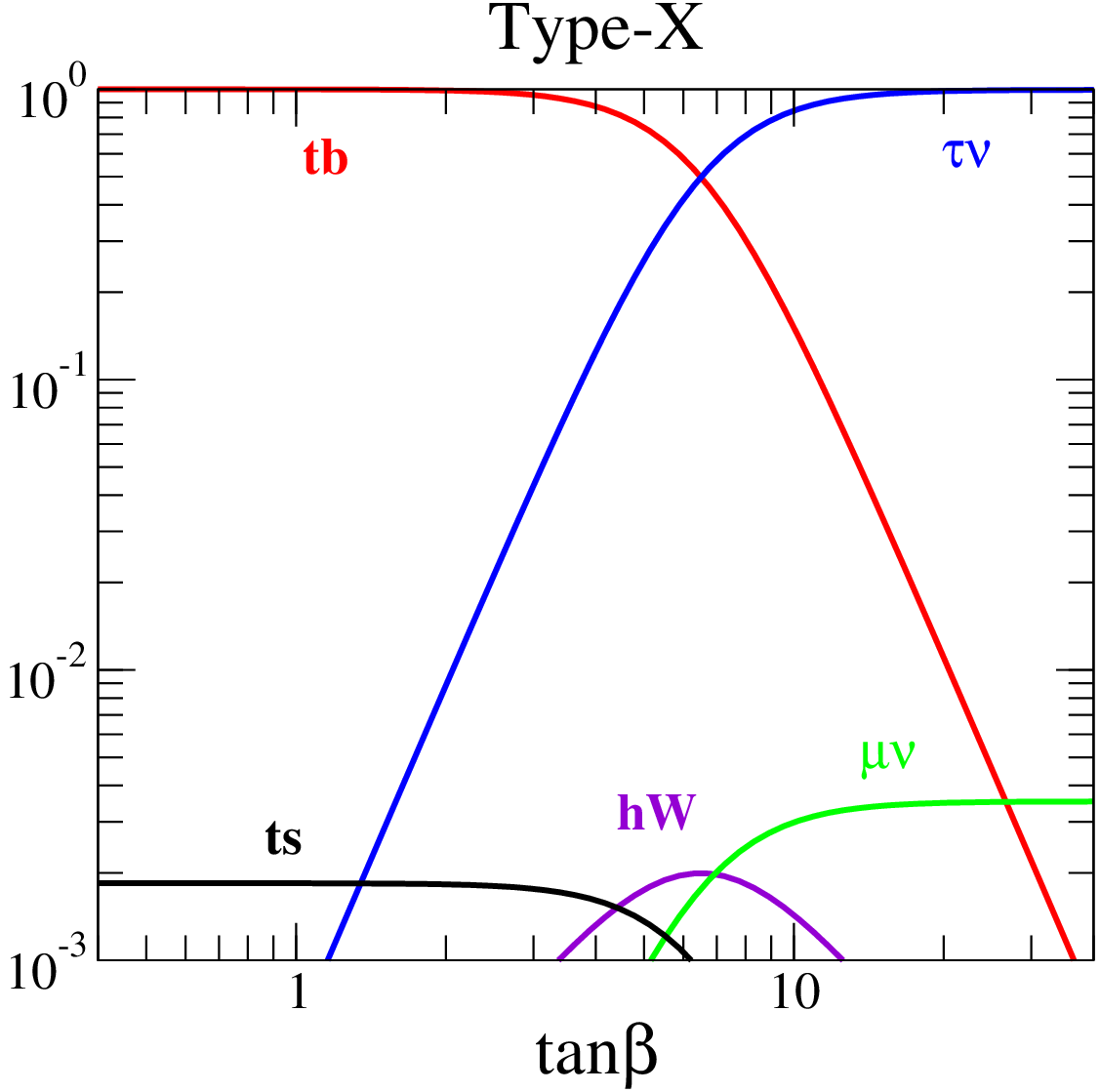}\hspace{0mm}
\includegraphics[width=39mm]{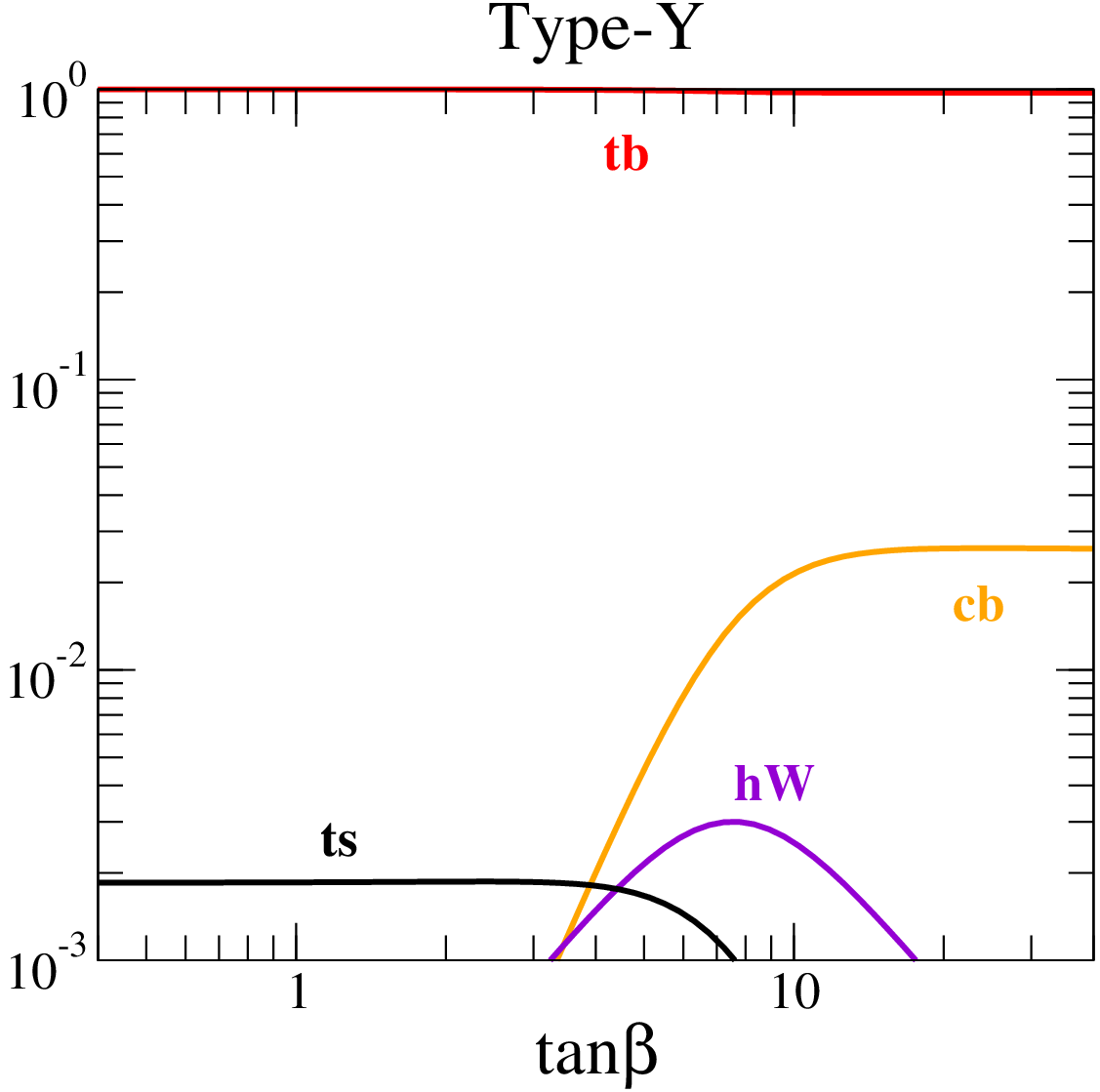}\\ \vspace{3mm}
\caption{
Decay branching ratios for $H$, $A$ and $H^\pm$ as a function of $\tan\beta$ in the case of $m_{\Phi}=M=200$ GeV and $\sin(\beta-\alpha)=0.99$. 
For the $H$ decay, the solid and dashed curves respectively show the cases with $\cos(\beta-\alpha)<0$ and $\cos(\beta-\alpha)>0$. 
}
\label{FIG:BR4}
\end{figure}

\begin{figure}[t]
\includegraphics[width=42mm]{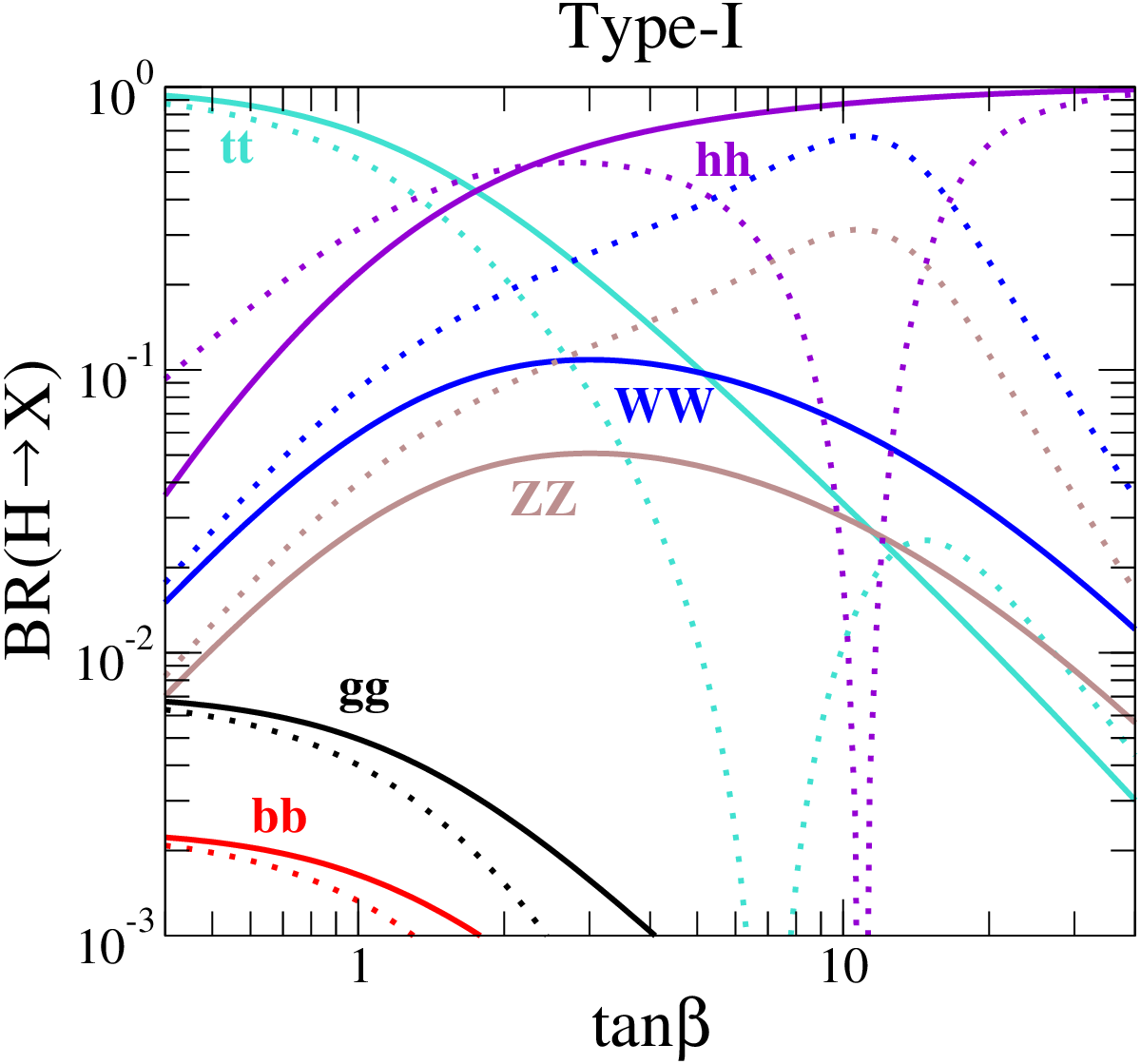}\hspace{0mm}
\includegraphics[width=39mm]{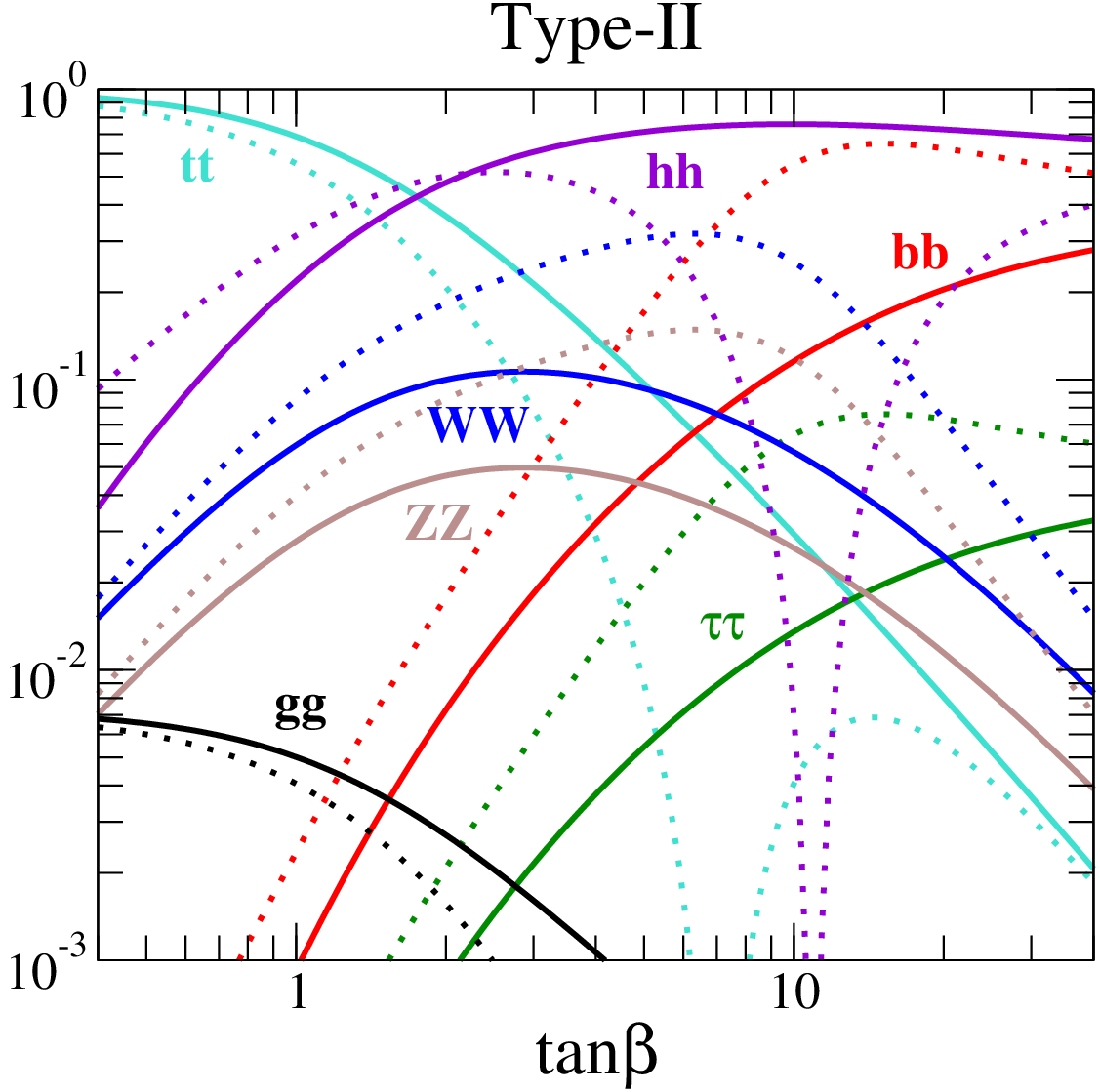}\hspace{0mm}
\includegraphics[width=39mm]{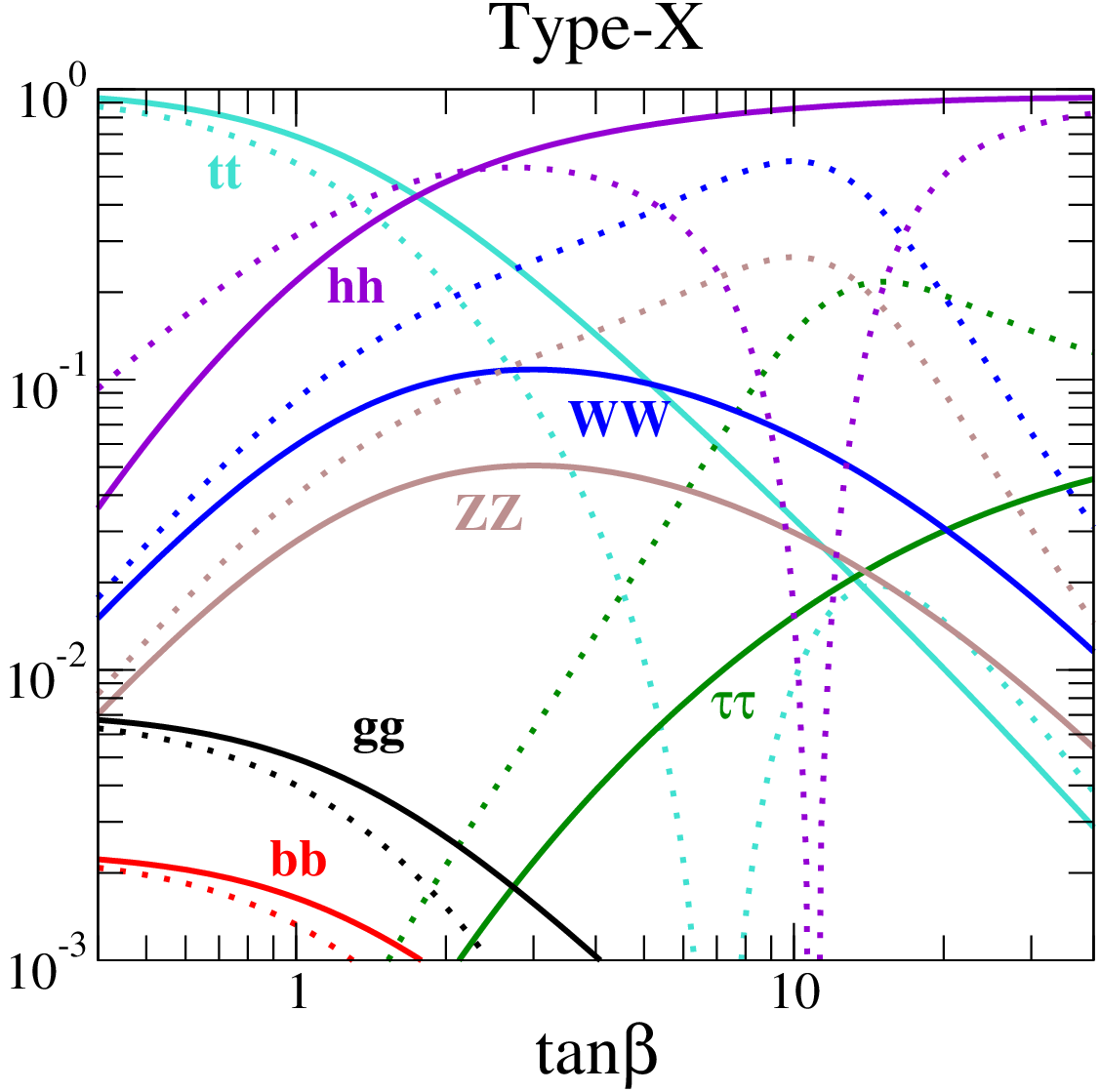}\hspace{0mm}
\includegraphics[width=39mm]{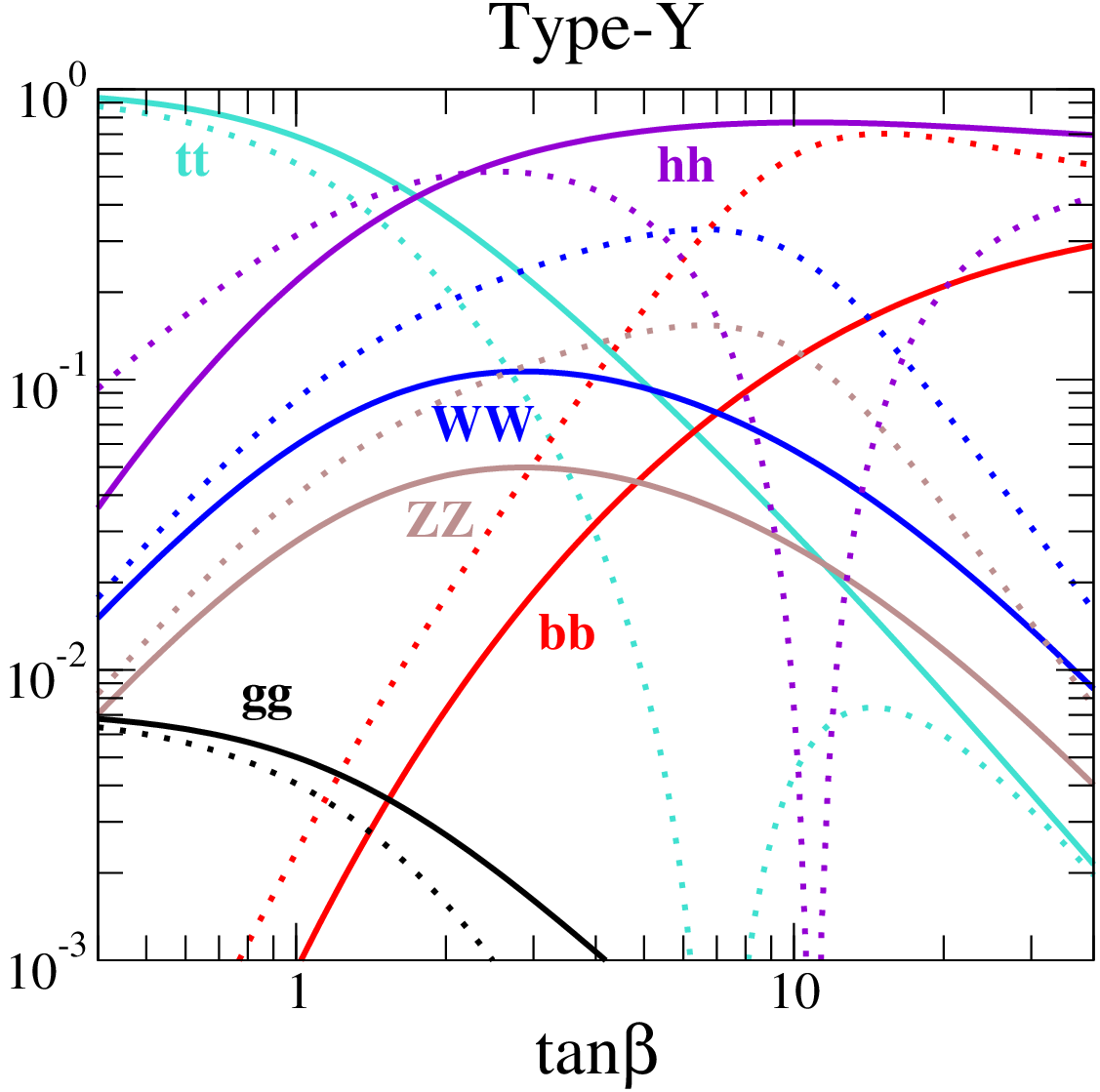}\\ \vspace{3mm}
\includegraphics[width=42mm]{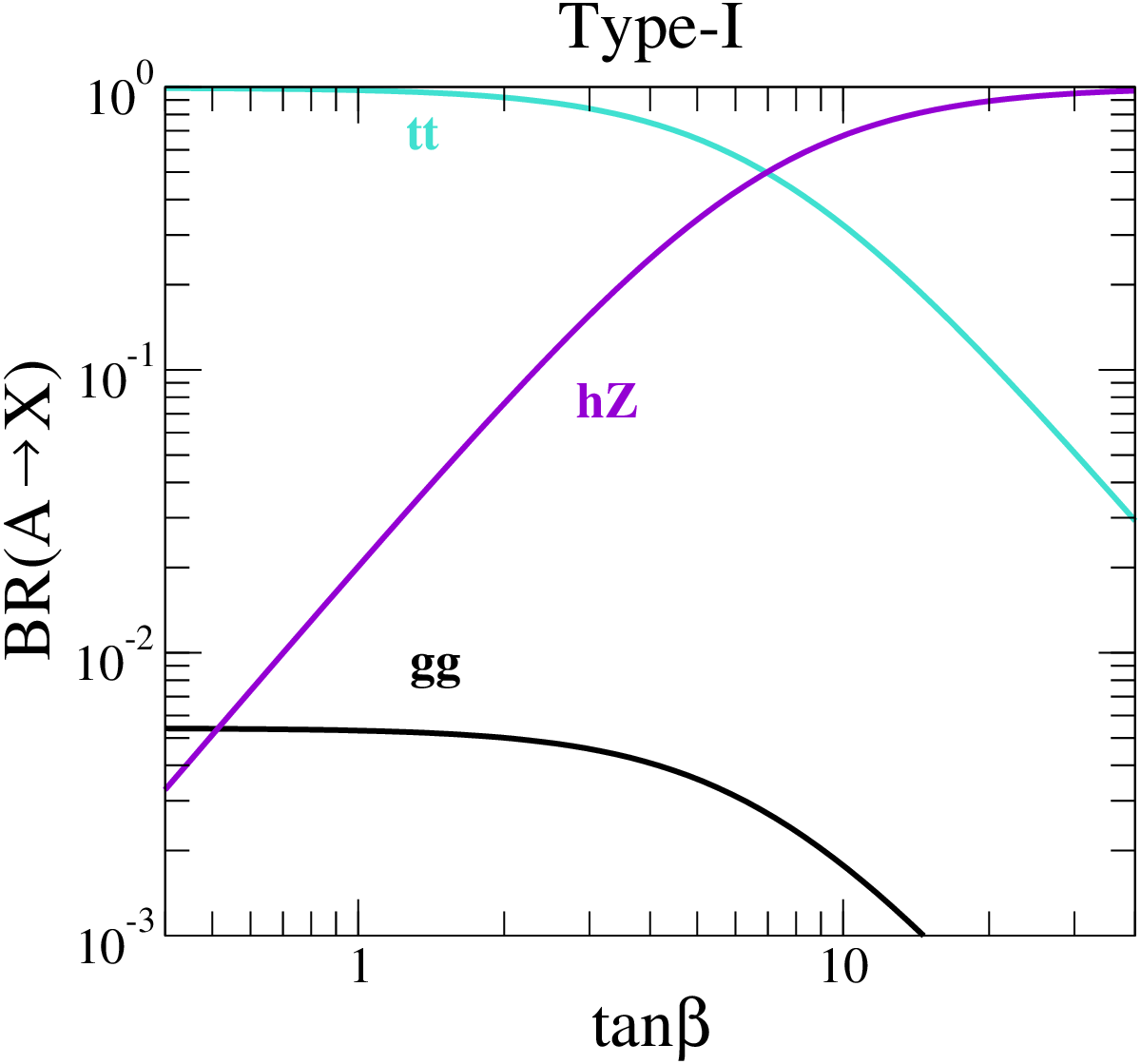}\hspace{0mm}
\includegraphics[width=39mm]{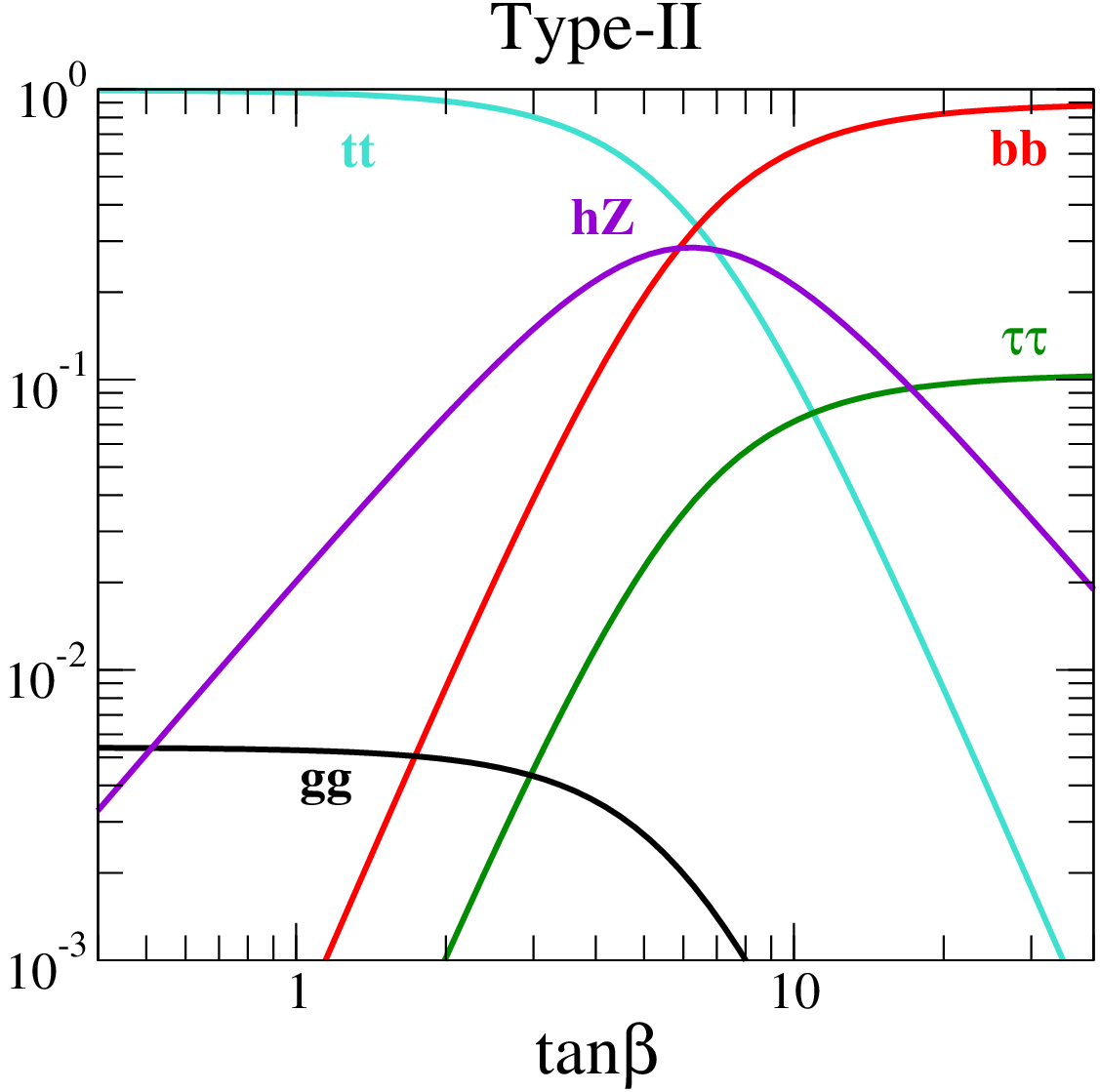}\hspace{0mm}
\includegraphics[width=39mm]{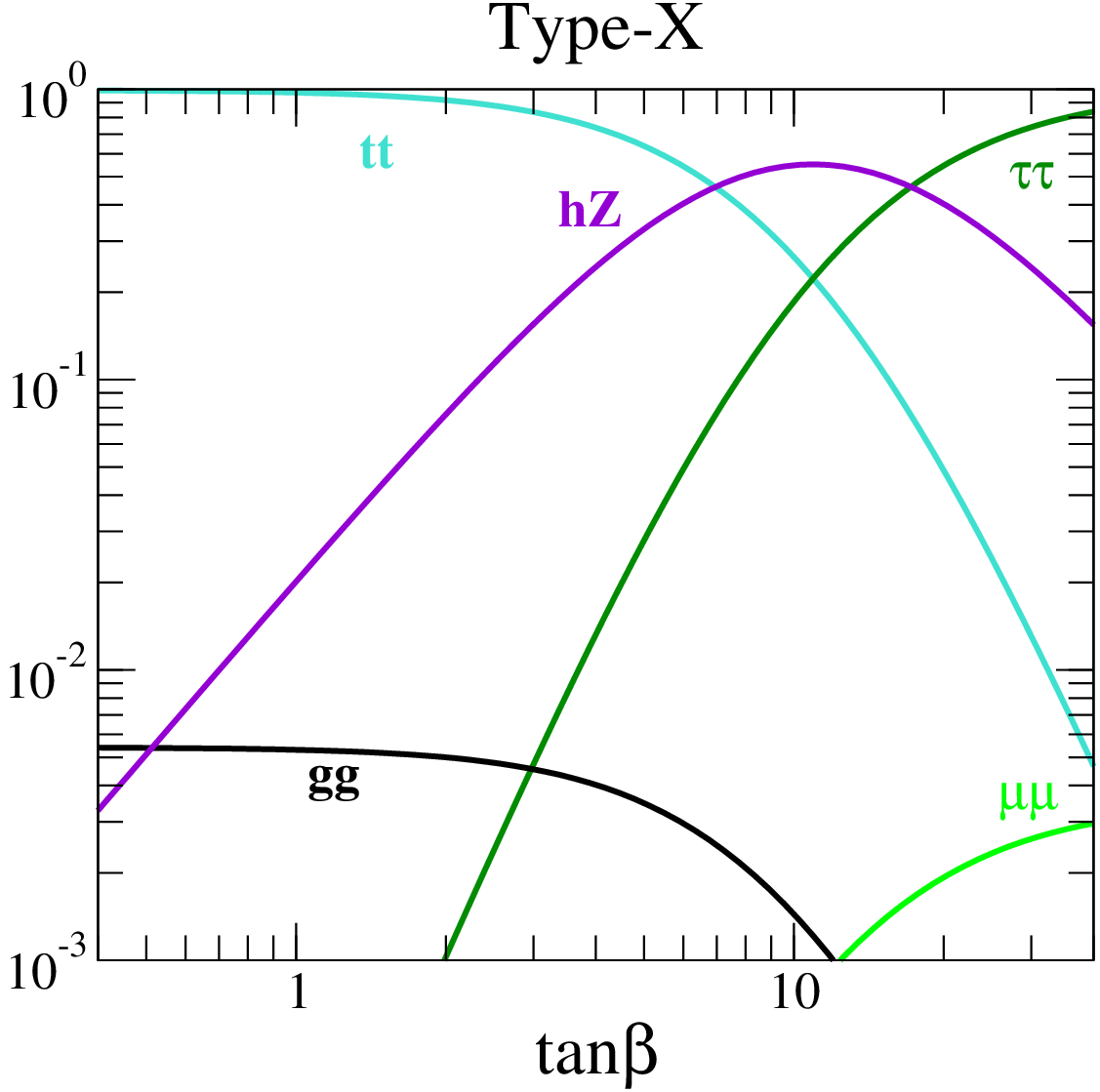}\hspace{0mm}
\includegraphics[width=39mm]{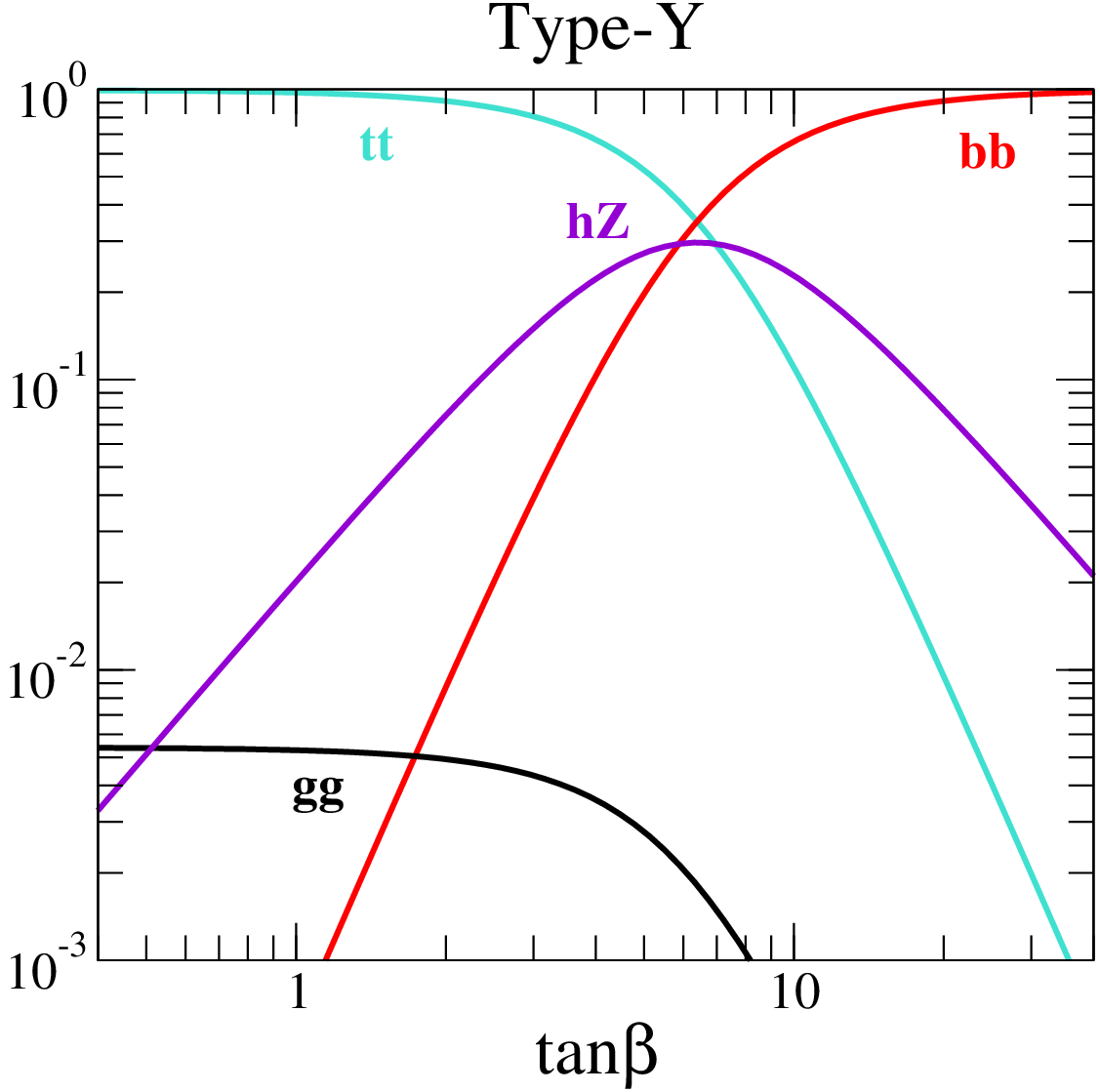}\\ \vspace{3mm}
\includegraphics[width=42mm]{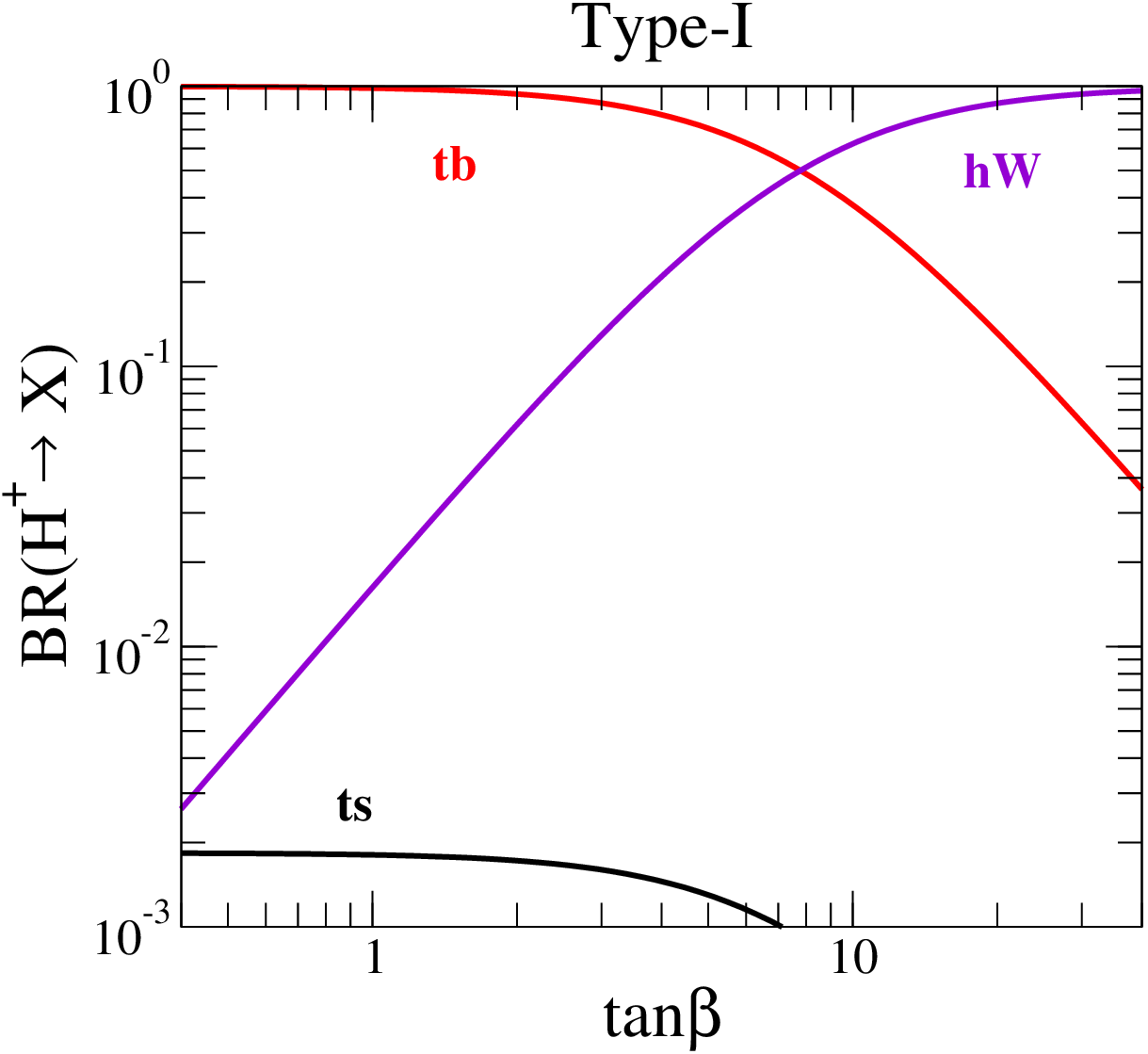}\hspace{0mm}
\includegraphics[width=39mm]{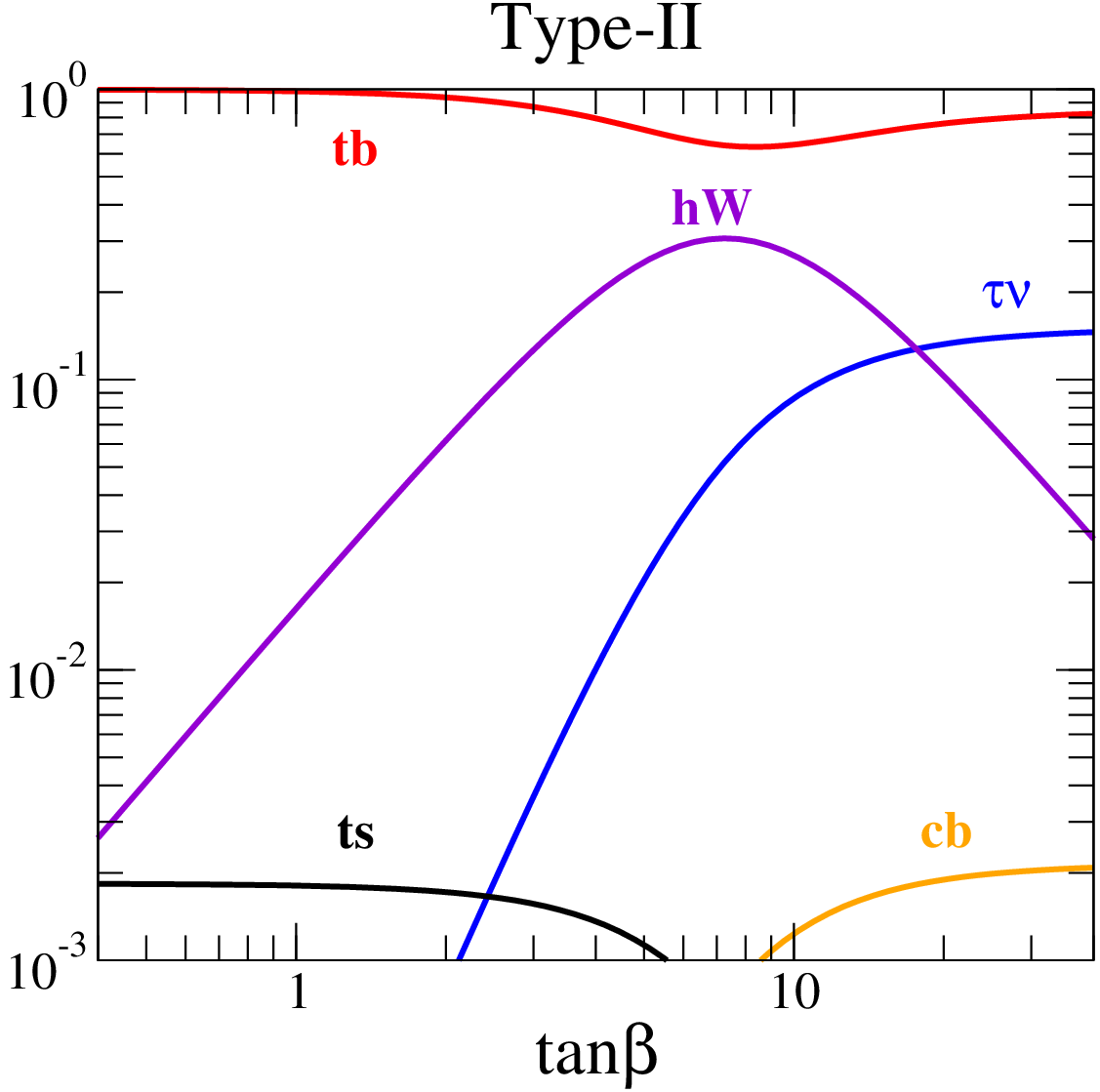}\hspace{0mm}
\includegraphics[width=39mm]{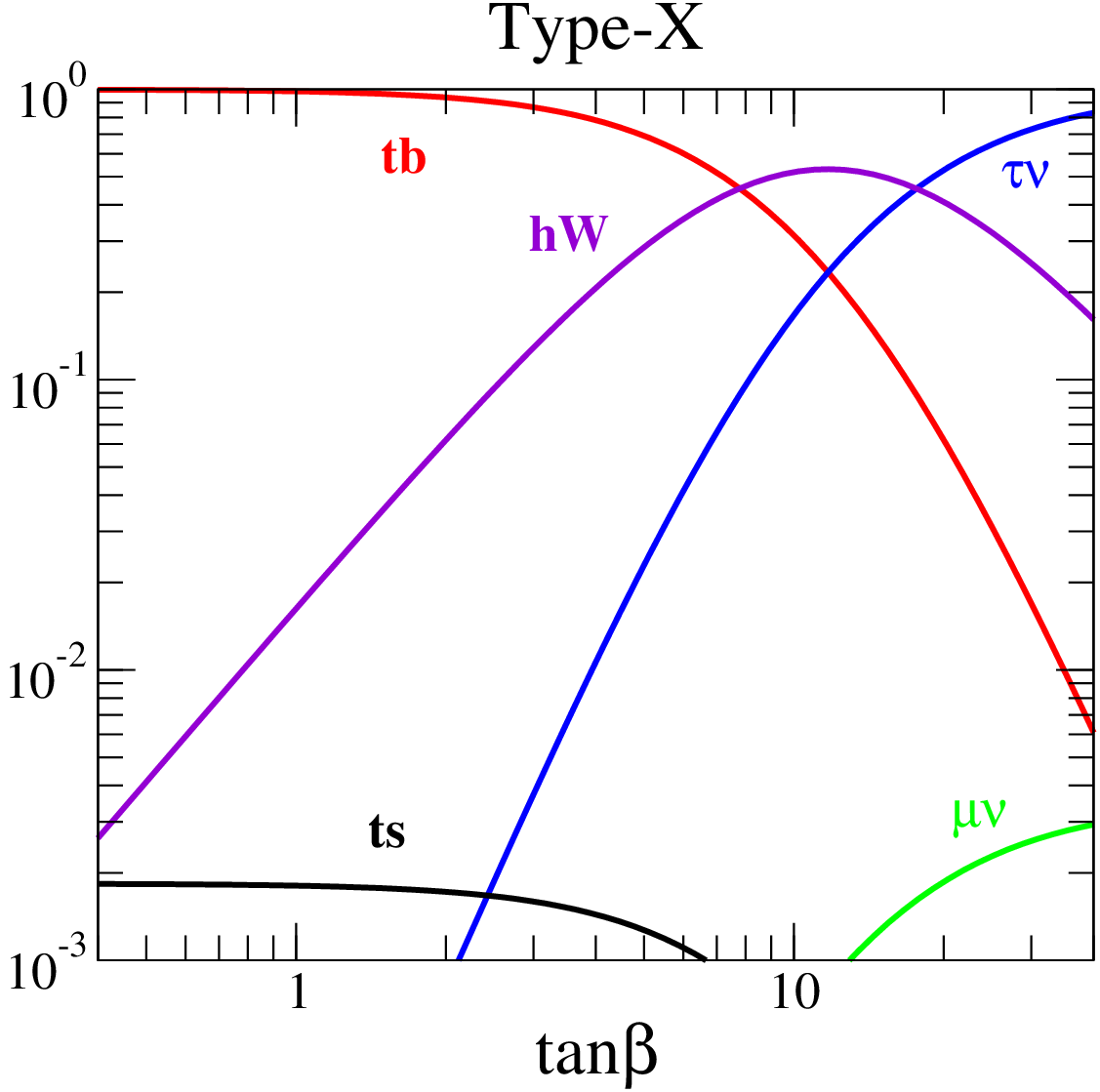}\hspace{0mm}
\includegraphics[width=39mm]{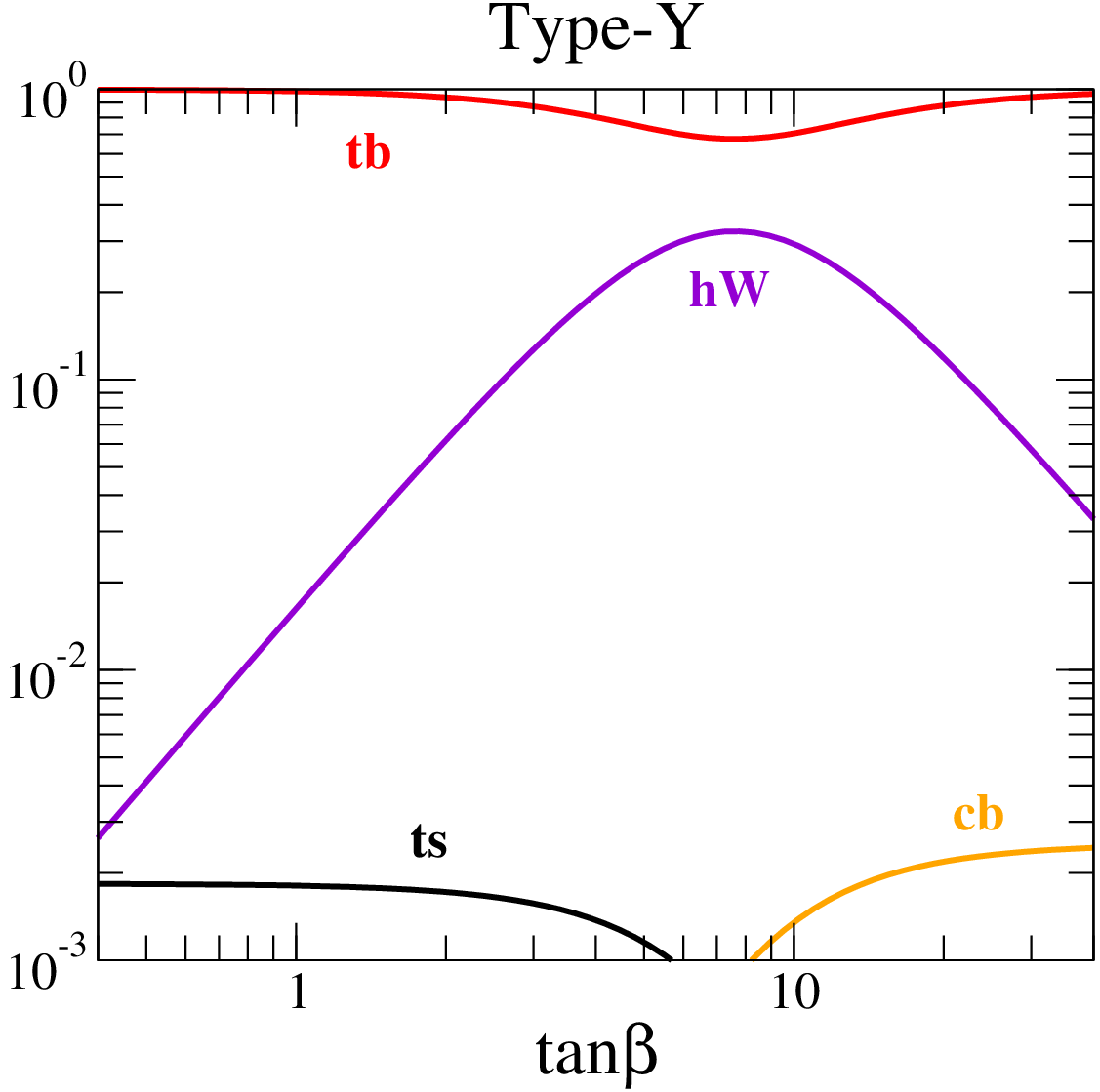}\\ \vspace{3mm}
\caption{
Decay branching ratios for $H$, $A$ and $H^\pm$ as a function of $\tan\beta$ in the case of $m_{\Phi}=M=400$ GeV and $\sin(\beta-\alpha)=0.99$. 
For the $H$ decay, the solid and dashed curves respectively show the cases with $\cos(\beta-\alpha)<0$ and $\cos(\beta-\alpha)>0$. 
}
\label{FIG:BR5}
\end{figure}

Next, we show the branching fractions in the case without taking the SM-like limit; e.g., $\sin(\beta-\alpha)=0.99$. 
In this case, the sign of $\cos(\beta-\alpha)$ can affect decay properties for the CP-even Higgs bosons, 
so that we consider both the cases with $\cos(\beta-\alpha)<0$ and $\cos(\beta-\alpha)>0$. 

In Fig.~\ref{FIG:BR3}, the branching fractions for the SM-like Higgs boson $h$ is shown as a function of 
$\tan\beta$ in the case of $m_{\Phi}=M=200$ GeV. 
For the $h$ decay, the $m_{\Phi}$ and $M$ parameters affect the 
$H^\pm$ loop contribution to the decay rates of $h\to \gamma\gamma$ and $h\to Z\gamma$. 
When we take a larger value of $m_{\Phi}$ keeping $m_{\Phi}=M$, the $H^\pm$ loop contribution vanishes. 
The solid and dashed curves respectively show the cases with $\cos(\beta-\alpha)<0$ and $\cos(\beta-\alpha)>0$. 
We can see that several fermionic decay channels vanish at $\tan\beta\simeq 7$ in the case of $\cos(\beta-\alpha)>0$
in the Type-II, Type-X and Type-Y THDMs. 
Let us explain this behavior by introducing $\delta$ defined by $\sin(\beta-\alpha)=1-\delta$. 
When $\delta \ll 1$,  
the $\xi_h^f$ and $\xi_H^f$ factors in Eq.~(\ref{Eq:Yukawa2}) can be approximately expressed by 
\begin{align}
\xi_h^f \simeq 1+\text{Sign}\left[\cos(\beta-\alpha)\right]2\sqrt{2\delta}T_3^f \, \xi_A^f,\quad
\xi_H^f \simeq \text{Sign}\left[\cos(\beta-\alpha)\right]2\sqrt{2\delta}-2T_3^f \, \xi_A^f. 
\end{align}
From TABLE~\ref{Tab:Yukawa_Couplings}, we can obtain 
$\xi_h^f\simeq 1-\text{Sign}[\cos(\beta-\alpha)]\sqrt{2\delta}\tan\beta$ for 
$f=b$ $(f=\tau)$ in Type-II and Type-Y (Type-II and Type-X) THDMs. 
Thus, when $\cos(\beta-\alpha)$ is positive, and $\delta$ is taken to be 0.01, $\xi_h^f$ becomes zero at around $\tan\beta= 7$.
We note that the $\xi_h^f$ factor can be $-1$ in the case of $\cos(\beta-\alpha)<0$, 
in which the sign of Yukawa coupling constant is opposite compared to the 
SM value. 
Signatures of additional Higgs bosons in the parameter regions with $\xi_h^f\simeq -1$ have been studied in Ref.~\cite{Ref:Chiang-Yagyu}, and the testability of the sign 
of Yukawa couplings has been investigated at a future linear collider in Ref.~\cite{Ref:Ferreira}. 

The branching fractions for the additional Higgs bosons are also shown in Fig.~\ref{FIG:BR4} in the case of $m_{\Phi}=M=200$ GeV and 
those in the case of $m_{\Phi}=M=400$ GeV in Fig.~\ref{FIG:BR5}.  
For the $H$ decay, we use the solid and dashed curves respectively to show the cases with $\cos(\beta-\alpha)<0$ and $\cos(\beta-\alpha)>0$. 
It can be seen that, the gaugephobic nature of $H$ is lost, and the $H\to W^+W^-/ZZ$ modes can be dominate. 
Regarding the $A$ and $H^\pm$ decays, the $A\to hZ$ and $H^\pm \to hW^\pm$ modes are added to the case with $\sin(\beta-\alpha)=1$. 
When we consider heavier case of $H$; $m_{\Phi}=M=400$ GeV, the $H\to hh$ mode is kinematically allowed whose 
decay rate is proportional to $\cos^2(\beta-\alpha)$. 
This can be the main decay mode as we can see in the top panels in Fig~\ref{FIG:BR5}.

We comment on the case without degeneracy in mass of the additional Higgs bosons. 
In that case, heavier additional Higgs bosons can decay into lighter ones associated with a gauge boson even in the SM-like limit. 
For instance, when $m_H>m_A$, the $H\to A Z^{(*)}$ mode is allowed. Recently, 
signatures from $H\to AZ$ and $A\to HZ$ decays have been studied at the LHC in Ref.~\cite{Su}. 

\section{Direct search for additional Higgs bosons at the LHC}

At the LHC with the collision energy to be 7 and 8 TeV, so far, 
there is no report for a discovery of new particles other than a Higgs boson, and only
exclusion bounds for masses of hypothetical particles are obtained.  

First of all, we review the current bounds on parameter space in the THDMs from 7 and 8~TeV data at the LHC. 
The signal of neutral Higgs bosons in the $\tau^+\tau^-$ decay mode
has been searched for in the inclusive production and bottom-quark
associated production processes~\cite{Aad:2012cfr,CMS:2013hja}.
For the Type-II THDM, bounds on $\tan\beta$ have been obtained for given
values of $m_A$, e.g., $\tan\beta\lesssim10$ for $m_A=300$~GeV and
$\tan\beta\lesssim40$ for $m_A=800$~GeV~\cite{CMS:2013hja}.
In addition, the searches for the $b\bar{b}$ decay of neutral Higgs
bosons in the bottom-quark associated process have been
performed~\cite{Ref:A2bb}. 
The $b\bar{b}$ decay mode gives a rather weaker bound on $\tan\beta$ than the $\tau^+\tau^-$ decay mode. 
These bounds can be used to constrain parameter regions in both the Type-II and Type-Y THDMs.
Furthermore, for $\sin(\beta-\alpha)<1$, searches for the $H\to
W^+W^-$ signal has been performed~\cite{ATLAS:2013zla}, and a bound on
the $m_H$-$\cos\alpha$ plane is obtained for given values of $\tan\beta$.
This bound is not sensitive to the type of Yukawa interaction.
In Ref.~\cite{CMS:2013eua}, $H\to hh$ and $A\to Zh$ decays have been searched, and
bounds on the cross section times branching ratio have been obtained.
These can be translated into the exclusion regions in the
$\cos(\beta-\alpha)$-$\tan\beta$ plane for given values of $m_{H/A}$ for
each type of Yukawa interaction.  

In the following, 
we discuss expected excluded regions on the $m_A$-$\tan\beta$ plane at the LHC with the collision energy to be 14 TeV.  
We first focus on the search for $H$ and $A$ by using the tau decay
from the gluon fusion and bottom quark associate production processes as 
\begin{align}
&gg \to \phi^0\to \tau^+\tau^-, \label{gfusion} \\
&gg \to b\bar b \phi^0 \to b\bar b \tau^+\tau^-, \label{bass}
\end{align}
where $\phi^0=H$ or $A$. 
The cross sections for the above processes can be estimated by\footnote{Regarding Eq.~(\ref{bass}), 
the equation for $\phi^0=A$ holds when the bottom quark mass in the phase space function is neglected. }
\begin{align}
\sigma(gg \to \phi^0 )
&=\frac{\Gamma(\phi^0\to gg)}{\Gamma(h_{\text{SM}}\to gg)_
{
m_{\phi^0}
}}\sigma(gg\to h_{\text{SM}})_
{
m_{\phi^0}
},\label{gfusion}\\
\sigma(gg \to b\bar{b}\phi^0 )&=(\xi^{\phi^0}_d)^2\sigma(gg\to b\bar{b}h_{\text{SM}})_
{
m_{\phi^0}
}, \label{bass}
\end{align}
where $h_{\text{SM}}$ is the SM Higgs boson.
In Eq.~(\ref{gfusion}), 
$\Gamma(h_{\text{SM}}\to gg)_{m_{\phi^0}}$ and $\sigma(gg\to h_{\text{SM}})_{m_{\phi^0}}$ 
are respectively the decay rate of $h_{\text{SM}}\to gg$ and 
the cross section of the gluon fusion process by taking 
the mass of $h_{\text{SM}}$ to be replaced by the mass of $\phi^0$ ($m_{\phi^0}$). 
We use the values of gluon fusion cross section in the SM at 14 TeV from Ref.~\cite{gfusion14}. 
In Eq.~(\ref{bass}), $\sigma(gg\to b\bar{b}h_{\text{SM}})_{m_{\phi^0}}$ is the cross section for 
the bottom quark associate production of $h_{\text{SM}}$ with the mass of $h_{\text{SM}}$ to be replaced by $m_{\phi^0}$. 
We calculate $\sigma(gg\to b\bar{b}h_{\text{SM}})_{m_{\phi^0}}$ by using 
{\tt CalcHEP}~\cite{calchep} with {\tt CTEQ6L}~\cite{Ref:CTEQ6L} for the parton distribution functions (PDFs). 

\begin{table}[t]
\begin{center}
\begin{tabular}{c|ccccccc}
\hline\hline 
$m_A$ [GeV] & 150 & 200 & 300 & 400 & 450 & 500 \\  \hline
$\mathcal{S}_{\text{MSSM}}^{\phi^0}$~\cite{RichterWas}& 5.6 & 5.8 & 1.7 &1.1 && 0.2 \\  \hline
$\mathcal{S}_{\text{MSSM}}^{b\bar{b}\phi^0}$~\cite{TDR1}   & 8.0 &  &2.1 & &1.1& \\
\hline\hline
\end{tabular} 
\end{center}
\caption{Significance for the gluon fusion process $\mathcal{S}_{\text{MSSM}}^{\phi^0}$ 
and the bottom quark associated process $\mathcal{S}_{\text{MSSM}}^{b\bar{b}\phi^0}$
in the MSSM with $\tan\beta=10$ at the LHC with the collision energy to be 14 TeV and the 
integrated luminosity to be 30 fb$^{-1}$ quoted from~\cite{TDR1,RichterWas}.} \label{Tab:significance}
\end{table}

\begin{figure}[t]
\begin{center}
\includegraphics[width=80mm]{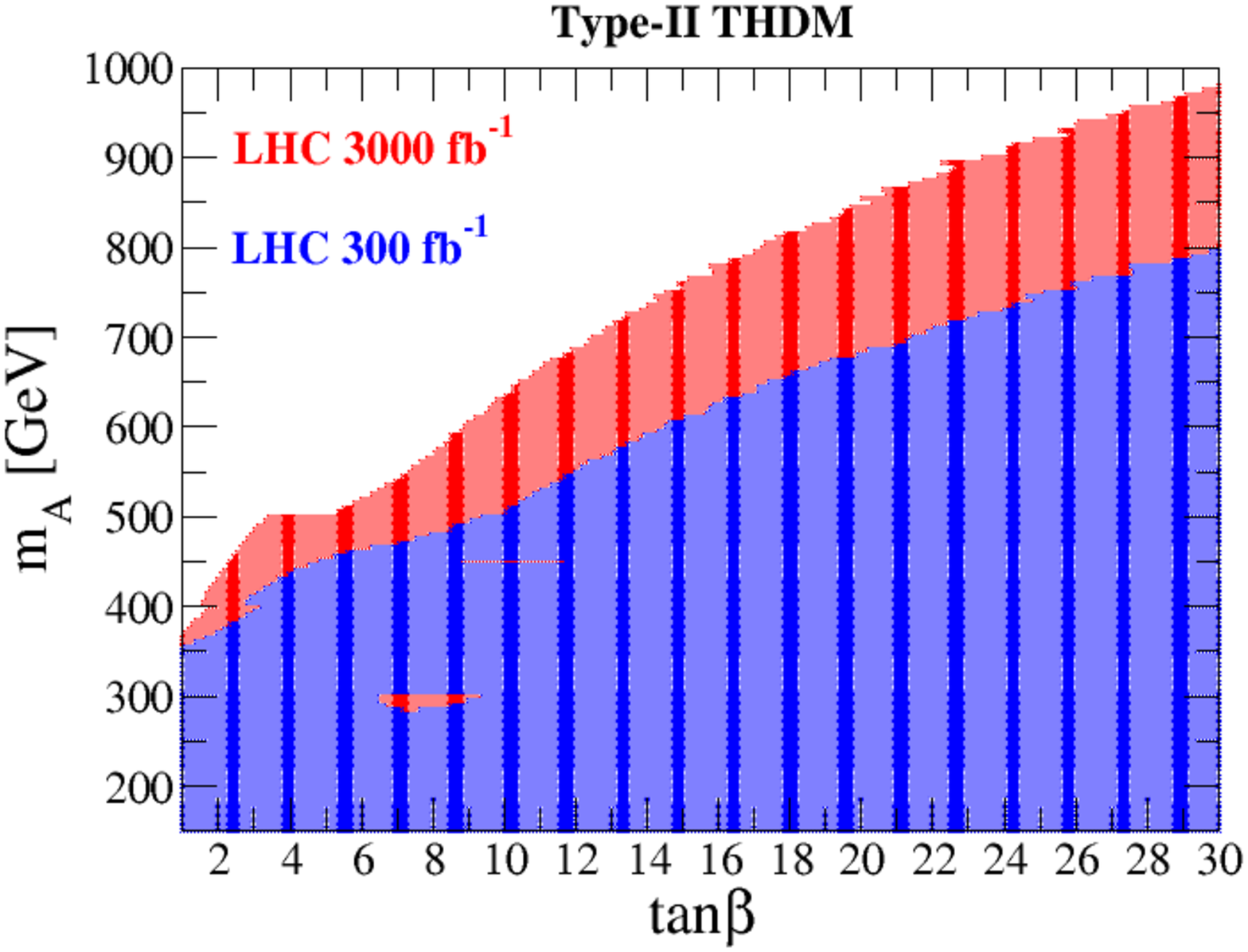}\hspace{1mm}
\includegraphics[width=80mm]{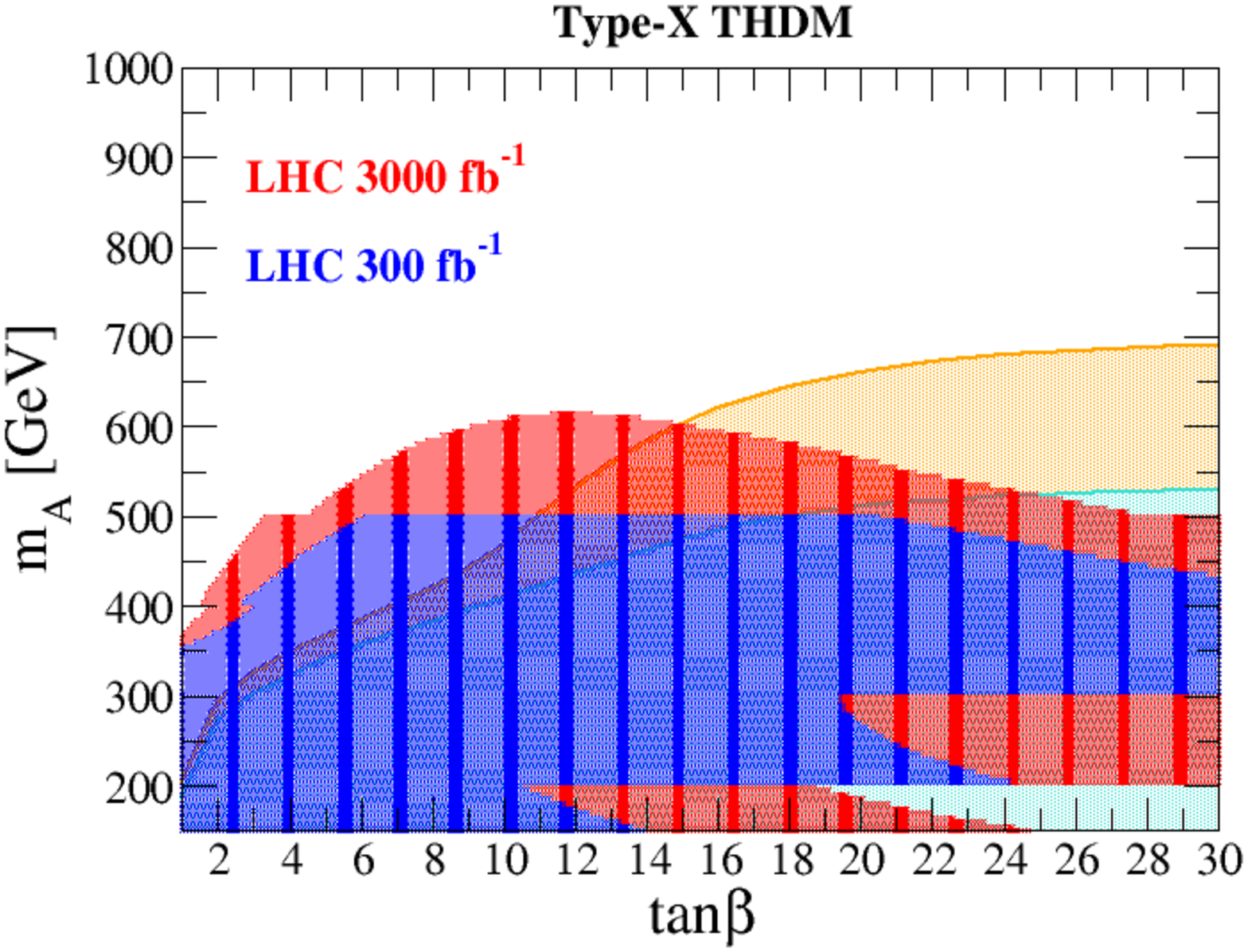}
\caption{
Expected excluded regions on the $\tan\beta$-$m_A$ plane at the 95\% CL from the $gg\to \phi^0 \to \tau^+\tau^-$ and 
$gg\to b\bar{b}\phi^0\to b\bar{b}\tau^+\tau^-$ 
processes by using Eqs.~(\ref{sig_THDM}) and (\ref{comb}) in the case of $m_A=m_H$ and $\sin(\beta-\alpha)=1$. 
The left and right panels show the results in the Type-II and Type-X THDM, respectively. 
The blue (red) shaded regions are excluded regions assuming the 
integrated luminosity to be 300 fb$^{-1}$ (3000 fb$^{-1}$). 
In the right panel, the constraint from the 
$q\bar{q}\to HA\to \tau^+\tau^-\tau^+\tau^-$
 processes is also shown by the light colored regions in the Type-X THDM. 
}
\label{direct_type2_v2}
\end{center}
\end{figure}

The signal and background analysis for these processes have been 
done in the MSSM in Refs.~\cite{RichterWas,TDR1}. 
The signal significances for the processes expressed in Eqs. (\ref{gfusion}) and (\ref{bass}) are 
given in the case of $\tan\beta=10$ and several fixed masses of $A$ 
with the collision energy to be 14 TeV and the integrated luminosity to be 30 fb$^{-1}$.
In TABLE~\ref{Tab:significance}, 
the significance for each fixed value of $m_A$ is listed, where 
$\mathcal{S}_{\text{MSSM}}^{\phi^0}$ and $\mathcal{S}_{\text{MSSM}}^{b\bar{b}\phi^0}$ are respectively
the significances for the gluon fusion process and the bottom quark associated process.  
These significances evaluated in the MSSM can be converted into those in the THDMs by
using the following equations
\begin{align}
\mathcal{S}_{\text{THDM}}^{\phi^0}
&=\mathcal{S}_{\text{MSSM}}^{\phi^0}\times\frac{\sum_{\phi^0=H,A}\sigma(gg \to \phi^0)\times\mathcal{B}(\phi^0\to\tau^+\tau^-)|_{\text{THDM}}}{\sum_{\phi^0=H,A}\sigma(gg \to \phi^0)\times\mathcal{B}(\phi^0 \to \tau^+\tau^-)|_{\text{MSSM}}}
\times \sqrt{\frac{\mathcal{L}}{30~\text{fb}^{-1}}},\\
\mathcal{S}_{\text{THDM}}^{b\bar{b}\phi^0}
&=\mathcal{S}_{\text{MSSM}}^{b\bar{b}\phi^0}\times 
\frac{\sum_{\phi^0=H,A}\sigma(gg \to b\bar{b}\phi^0 )\times\mathcal{B}(\phi^0 \to\tau^+\tau^-)|_{\text{THDM}}}{\sum_{\phi^0=H,A}\sigma(gg \to b\bar{b}\phi^0)\times\mathcal{B}(\phi^0 \to \tau^+\tau^-)|_{\text{MSSM}}}
\times \sqrt{\frac{\mathcal{L}}{30~\text{fb}^{-1}}}, \label{sig_THDM}
\end{align}
where $\mathcal{L}$ is the assumed integrated luminosity. 
In the above expression, when $m_{\phi^0}$ 
is taken in the range of $X\leq m_{\phi^0} \leq Y$, where 
$X$ and $Y$ are values of $m_A$ listed in TABLE~\ref{Tab:significance},  
we use the value of $\mathcal{S}_{\text{MSSM}}^{\phi^0}$ and $\mathcal{S}_{\text{MSSM}}^{b\bar{b}\phi^0}$ given in the case with $m_A=X$.
The combined significance is calculated by 
\begin{align}
\mathcal{S}_{\text{comb}} = \sqrt{(\mathcal{S}_{\text{THDM}}^{\phi^0})^2+(\mathcal{S}_{\text{THDM}}^{b\bar{b}\phi^0})^2  }, \label{comb}
\end{align}
and the expected excluded region with the 95\% confidence level (CL)
is obtained by requiring $\mathcal{S}_{\text{comb}}\geq 2$. 
In the following analysis, we assume $m_A=m_H$ and $\sin(\beta-\alpha)=1$, 
and we sum over the processes in Eqs.~(\ref{gfusion}) and (\ref{bass}) 
mediated by $H$ and $A$.
These assumptions are valid as long as we consider the case with $m_A\gtrsim 150$ GeV, because 
in the MSSM $m_H\simeq m_A$ and $\sin(\beta-\alpha)\simeq 1$ are the good approximation in that case.

In Fig.~\ref{direct_type2_v2}, we show
the expected excluded regions by using Eq.~(\ref{comb}) in the Type-II (left panel) and Type-X (right panel) 
THDMs. 
The blue and red shaded regions are respectively the excluded regions assuming $\mathcal{L}$ to be 300 fb$^{-1}$ and 3000 fb$^{-1}$. 
In the Type-II THDM, exclusion reach of $m_A$ increases when a larger value of $\tan\beta$ is taken, 
because the cross sections of the bottom quark associated processes are enhanced due to 
the coefficient $\xi_{H}^b=\xi_A^b=\tan\beta$, 
and the branching fraction of $\phi^0\to \tau^+\tau^-$ is approaching to be 10\% in high $\tan\beta$ regions as shown in Figs.~\ref{FIG:BR1} and \ref{FIG:BR2}. 
On the other hand, in the Type-X THDM, both the gluon fusion and the bottom quark associated production cross sections are suppressed by $\cot^2\beta$
as $\tan\beta$ is getting larger, while
the branching fraction of $\phi^0\to \tau^+\tau^-$ increases. 
Consequently, the cross section times branching ratio takes maximal obtained at $\tan\beta\simeq 12$, and then
$m_A\simeq 600$ GeV can be excluded assuming $\mathcal{L}=3000$ fb$^{-1}$. 
When $\mathcal{L}=300$ fb$^{-1}$ is assumed, the excluded reach is settled to be 500 GeV in the region of $6\lesssim\tan\beta\lesssim 20$ in spite of
the fact that the cross section has the maximal value at around $\tan\beta=12$. 
This can be understood in such a way that the quoted significance $\mathcal{S}_{\text{MSSM}}^{\phi^0}$ given in TABLE~\ref{Tab:significance} is changed at $m_A=500$ GeV, and the combined significance $\mathcal{S}_{\text{comb}}$ defined in Eq.~(\ref{comb}) cannot exceed 2 even in the case with $\tan\beta\simeq 12$. This behavior should vanish by the detailed background analysis with smaller intervals of $m_A$. 

This result is the updated version of Fig.~1.20 in the ILC Higgs White Paper~\cite{Ref:ILC_White}. 
In the previous figure, the excluded regions have been derived by using only one value of the significance for 
the gluon fusion and bottom quark associated processes with $m_A=150$ GeV from~\cite{TDR1}. 
In the current version, we use several values of the significance as shown in TABLE~\ref{Tab:significance}. 

If we take $\sin(\beta-\alpha)\neq 1$, the contribution from $H$ $(A)$ can drastically decrease, because the 
branching fraction of the $H(A)\to \tau^+\tau^-$ mode significantly decreases due to the $H\to VV$ and $H\to hh$ ($A\to hZ$) modes as seen in Fig.~\ref{FIG:BR3}. 
In such a case, $H \to ZZ \to 4\ell$ and $A \to h Z\to b\bar{b}\ell\ell$ channel 
can be important instead of the $\tau^+\tau^-$ mode. 
In fact, these searches have been studied with the LHC data~\cite{Ref:H2ZZ,CMS:2013eua}.
The performance of High Luminosity (HL)-LHC has also been evaluated in
Refs.~\cite{Ref:A2hZ-HL-LHC}.
These results show that masses of $\sim 1$ TeV could be explored for
$1-\kappa_V^{} \gtrsim 10^{-2}$ with low $\tan\beta \lesssim
3$~\cite{Ref:A2hZ-HL-LHC}.
Thus, the parameter space allowed by theoretical consistencies can be
fully probed
by future LHC data for $1-\kappa_V^{} \gtrsim 10^{-2}$.

Next, we consider the Drell-Yan production; 
\begin{align}
pp\to Z^* \to HA. 
\end{align}
For given values of the masses for $H$ and $A$, 
this cross section is purely determined by the gauge coupling constant, so that the cross section does not 
depend on the type of Yukawa interactions. 
When both $H$ and $A$ decay into the tau pairs, the 4-$\tau$ final state is obtained. 
The cross section of the 4$\tau$ process can be large in the Type-X THDM as compared to the other three types of THDMs due to 
the enhancement of the branching fraction of $H/A\to \tau^+\tau^-$ for large $\tan\beta$~\cite{Ref:AKTY}. 
We thus focus on the 4$\tau$ signature from the $HA$ production to test the Type-X THDM in the following. 
Analyses on the $pp\to HH^\pm$ and $AH^\pm$ resulting the $3\tau$ signature 
have been studied in Ref~\cite{Ref:KTY}, where the same order bounds on $m_{H^\pm}$ can be obtained. 

We estimate the cross-section by using the leading order expression with the {\tt CTEQ6L} PDFs~\cite{Ref:CTEQ6L}, 
where the scale of them are set to $\mu=m_H$.
The event rates of the $HA\to4\tau$ signal are obtained by multiplying
the production cross-section by the branching ratios of $H$ and $A$
into $\tau^+\tau^-$. 
Furthermore, by using the kinematical distributions of the decay
products of $\tau$'s which are calculated by {\tt
PYTHIA}~\cite{Ref:PYTHIA} and {\tt TAUOLA}~\cite{Ref:tauola}, we
estimate the efficiency of detecting the signal 
events after the acceptance and kinematical cuts given in
Ref.~\cite{Ref:KTY} for all the final-states lead from the decays of
the four $\tau$'s, such as four $\tau$-jets, three $\tau$-jets plus one
lepton, etc. 
The significance for detecting the $HA\to 4\tau$ process is estimated
for a given value of the integrated luminosity, by combining the
significance of all the channels where each significance is evaluated as
$S=\sqrt{2[(s+b)\ln{(1+s/b)}-s]}$ with $s$ and $b$ being the expected
numbers of the signal and background events after the cuts, respectively.

In the right panel of Fig.~\ref{direct_type2_v2}, the expected exclusion regions are shown on the $\tan\beta$-$m_A$ plane in the
Type-X THDM from the $pp\to HA\to 4\tau$ process. 
The cyan and orange shaded regions are excluded at the 95\% CL assuming the integrated luminosity to be 
300 fb$^{-1}$ and 3000 fb$^{-1}$, respectively. 
The search potential is significantly improved for the large $\tan\beta$
regions due to the enhancement of the decay branching ratios of 
$H$ and $A$ into the $\tau^+\tau^-$ final state. 
For $\tan\beta\gtrsim20$, the discovery regions arrive at around
$m_A=500$~GeV for 300~fb$^{-1}$, while those arrive at around 700~GeV for
3000~fb$^{-1}$.

We note that the $\tau$-jet tagging efficiency shall worsen at the high
luminosity run of the LHC, due to the participation of many hadrons in
an event which prevent the isolation requirement in the $\tau$-jet
tagging procedure~\cite{Ref:taujet}. 
Therefore, the expected significance may be reduced for the channels
with high $\tau$-jet multiplicity, since the $\tau$-jet tagging
efficiency used in our analysis is based on the {\tt PYTHIA}
simulation; see Ref.~\cite{Ref:KTY}. 
Although we have not studied this issue seriously, it can be important
at the high luminosity run of the LHC. 

In the end of this section, we would like to mention the direct search potential for $H$ and $A$ in 
the Type-I and Type-Y THDMs. 
In Type-I THDM, the Yukawa interactions for the additional Higgs bosons are getting weak for large $\tan\beta$,
so it is difficult to generate new bosons via the Yukawa interaction.
In Type-Y THDM, only the down-type quark Yukawa interactions are
enhanced by $\tan\beta$.
Since the process $pp \to HX, AH (H, A \to b\bar b)$ are enhanced for
large $\tan\beta$,
the cross section times the branching ratio are constrained~\cite{Ref:A2bb}.
The bounds are much weaker than those from $H/A \to \tau^+\tau^-$ decay
channels in Type-II
and Type-X THDMs. The analysis with data for high luminosity running will push
these bounds substantially.

Finally, we comment on the constraint from flavour experiments. 
It is well known that the mass of $H^\pm$ in the Type-II THDM is severely constrained
by the precise measurements of the $b\to s\gamma$ process~\cite{4types_barger,4types_grossman,bsg1,Misiak}, 
where the $H^\pm$ loops contribute to this process in addition to 
the $W$ boson loop contribution. 
A lower bound has been found to be $m_{H^+} \gtrsim 380$ GeV (95\% CL) in the Type-II THDM at the next-to-next-to-leading order~\cite{Misiak}.
In the Type-I THDM, the bound from $b\to s\gamma$ is important only in the case with low $\tan\beta$; namely, 
the bound on $m_{H^+}$ is stronger than the LEP bound of around 80 GeV~\cite{PDG} when $\tan\beta <2.5$ is taken~\cite{Misiak}. 
The Type-Y and Type-X THDMs are received similar constraints as in the Type-II and Type-I THDMs, respectively, 
because of the same structure of quark Yukawa interactions. 
Bounds from the other observables such as 
$B\to \tau\nu$~\cite{Btaunu,Maria_Btaunu}, $\tau\to \mu\nu\bar{\nu}$~\cite{Maria_Btaunu,Maria_Tau} 
and the muon anomalous magnetic moment~\cite{Haber_g2,Maria_g2} have been discussed in the Type-II THDM. 
In Ref.~\cite{Stal}, constraints from various flavour experiments have been studied in the four types of Yukawa interactions of the THDM. 
Excluded parameter regions are shown on the $m_{H^+}$-$\tan\beta$ plane. 
The bound on $m_{H^+}$ can be 
converted into that on the masses of neutral Higgs bosons from the electroweak precision data.
Although such a constraint can be stronger than that from the direct search as shown in Fig.~\ref{direct_type2_v2}, 
it is important to search for additional Higgs bosons independently on the flavour experiments.

\section{Precision Measurements for the Higgs boson Couplings and Fingerprinting Extended Higgs Models}

In this section, we discuss the deviation in the SM-like Higgs boson couplings in the THDMs and also in the other models with universal Yukawa couplings. 
In a model with extended Higgs sectors, the Higgs boson couplings can deviate from the SM values as 
we already have discussed in Section.~II in the THDMs as an example. 
Therefore, extended Higgs sectors can be indirectly tested by measuring the deviation 
of various Higgs boson couplings. 
Furthermore, 
the pattern of the deviation strongly depends on the structure of the Higgs sector, so that 
we can discriminate various Higgs sectors by comparing the predicted pattern of the deviations with the 
measured one. 

We here define the scaling factors by normalizing 
the coupling constant of the SM Higgs boson which will be precisely determined by future collier experiments; 
\begin{align}
{\mathcal L} 
=&
\kappa_V h
\Big(\frac{m_W^2}{v} W^{+\mu} W^-_\mu +\frac12 \frac{m_Z^2}{v} Z^\mu Z_\mu \Big)
-\sum_f\kappa_f h\frac{m_f}{v}\bar{f} f. 
\end{align}
These measured values should be compared with corresponding values in extended Higgs models. 
In the THDM, $\kappa$ factors are given at the tree level by 
\begin{align}
\kappa_f = \xi_h^f,\quad \kappa_V = \sin(\beta-\alpha), 
\end{align}
where $\xi_h^f$ are listed in TABLE~\ref{Tab:Yukawa_Couplings}. 
We also discuss the other extended Higgs sectors with universal Yukawa coupling constants; i.e., 
$\kappa_f$ for any fermion $f$ are modified as the same way in the end of this section.

\begin{table}
\centering
\begin{tabular}{lcccccc}\hline\hline
Facility                     &   LHC            &   HL-LHC         &   ILC500  & ILC500-up & ILC1000 &   ILC1000-up \\
$\sqrt{s}$ (GeV)                 &   14,000         &   14,000         &   250/500 & 250/500 & 250/500/1000 & 250/500/1000 \\
$\int{\cal L}dt$ (fb$^{-1}$) & 300/expt         & 3000/expt        &   250+500 & 1150+1600 & 250+500+1000 & 1150+1600+2500  \\
\hline 
 $\kappa_{\gamma}$ &  $5-7$\%      & $2-5$\%       & 8.3\% & 4.4\% & 3.8\%  & 2.3\%            \\
 $\kappa_g$           &  $6-8$\%       & $3-5$\%       & 2.0\% & 1.1\% & 1.1\%  & 0.67\%            \\
 $\kappa_W$           & $4-6$\% & $2-5$\% & 0.39\% & 0.21\% & 0.21\% & 0.2\%         \\
 $\kappa_Z$           & $4-6$\% & $2-4$\% & 0.49\% & 0.24\% & 0.50\% & 0.3\%        \\

 $\kappa_{\ell}$      & $6-8$\%       &  $2-5$\%      & 1.9\%  & 0.98\% & 1.3\%  & 0.72\%            \\
 $\kappa_d =\kappa_b$ & $10-13$\%        &  $4-7$\%       & 0.93\% & 0.60\% & 0.51\% & 0.4\%           \\
 $\kappa_u =\kappa_t$ & $14-15$\%        &  $7-10$\%      & 2.5\%  &  1.3\% & 1.3\%   & 0.9\%            \\
\hline\hline
 \end{tabular}
 \caption{Expected precisions on the Higgs boson couplings and total width
   from a constrained 7-parameter fit quoted from Table 1-20 in Ref.~\cite{Ref:Snowmass}.  }
 \label{Tab:Sensitivity}
\end{table}

The scaling factors will be measured accurately at future collider experiments such as 
the high luminosity running of the LHC (HL-LHC) and the ILC.
In TABLE~\ref{Tab:Sensitivity}, we give a brief summary of expected sensitivities 
on the (SM-like) Higgs boson coupling constant at various future experiments.  
The ranges shown for LHC and HL-LHC represent the conservative and
 aggressive scenarios for systematic and theory uncertainties.  ILC
 numbers assume $(e^-,e^+)$ polarizations of $(-0.8,0.3)$ at 250 and 
 500~GeV and $(-0.8,0.2)$ at 1000~GeV, plus a 0.5\% theory uncertainty.  


\subsection{Higgs boson couplings in the THDMs}

\begin{figure}[t]
\centering
\includegraphics[width=7cm]{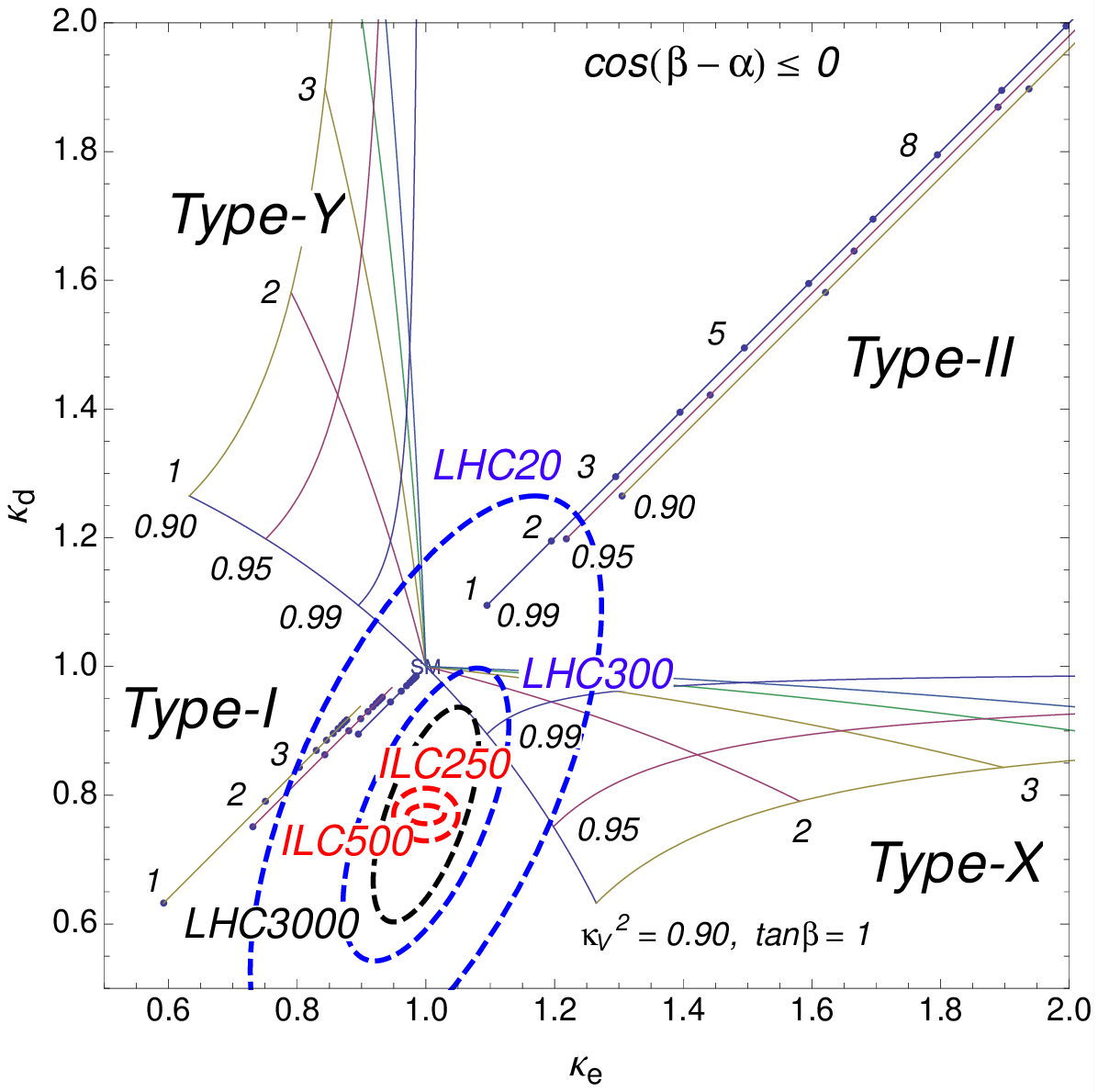}	\hspace{5mm}
\includegraphics[width=7cm]{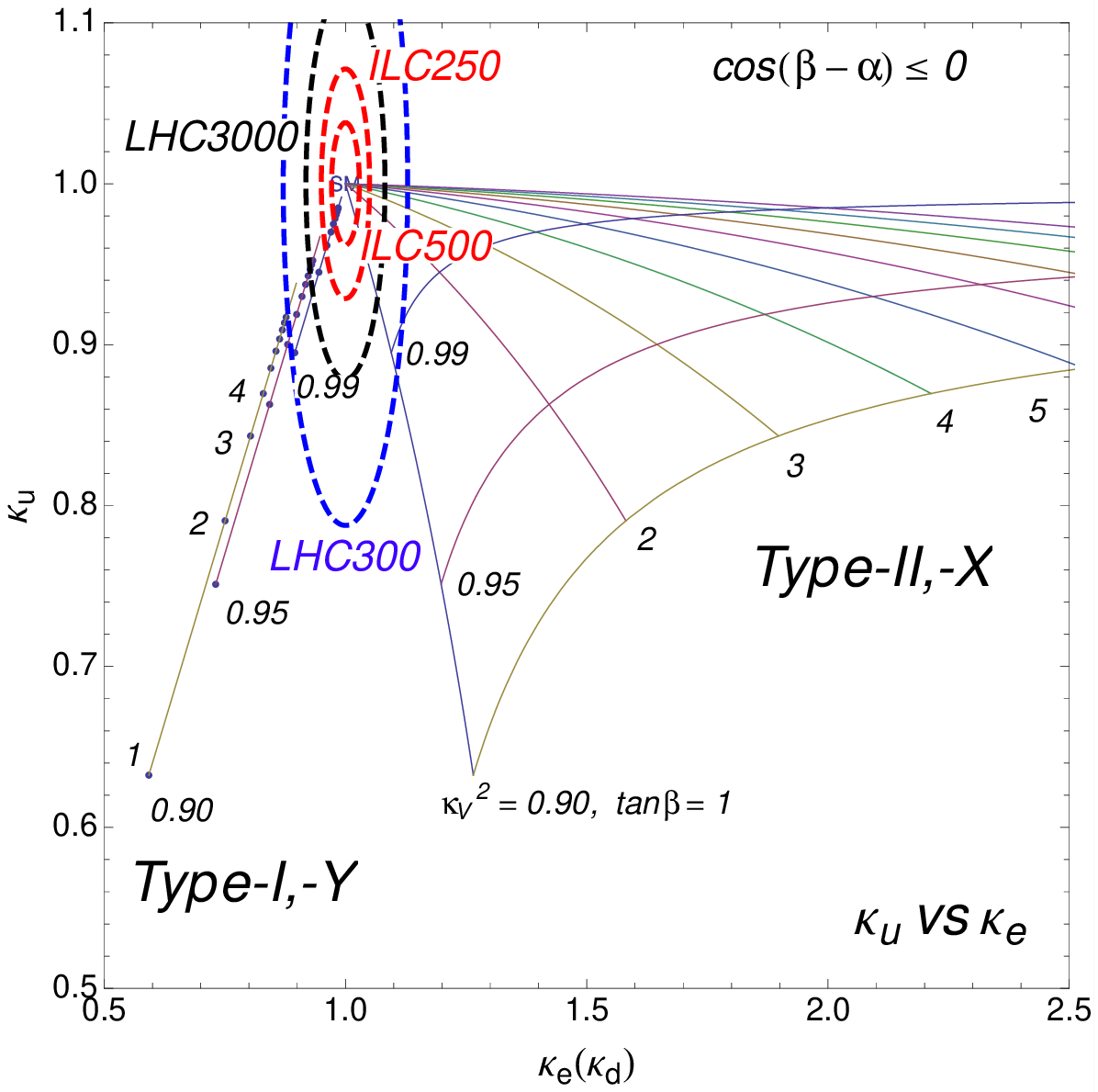}  
\caption{The scaling factors for the Yukawa interaction of the SM-like Higgs boson in THDMs in the case of $\cos(\beta-\alpha)<0$.}
\label{Fig:KFKFm}
\end{figure}

\begin{figure}[t]
\centering
\includegraphics[width=7cm]{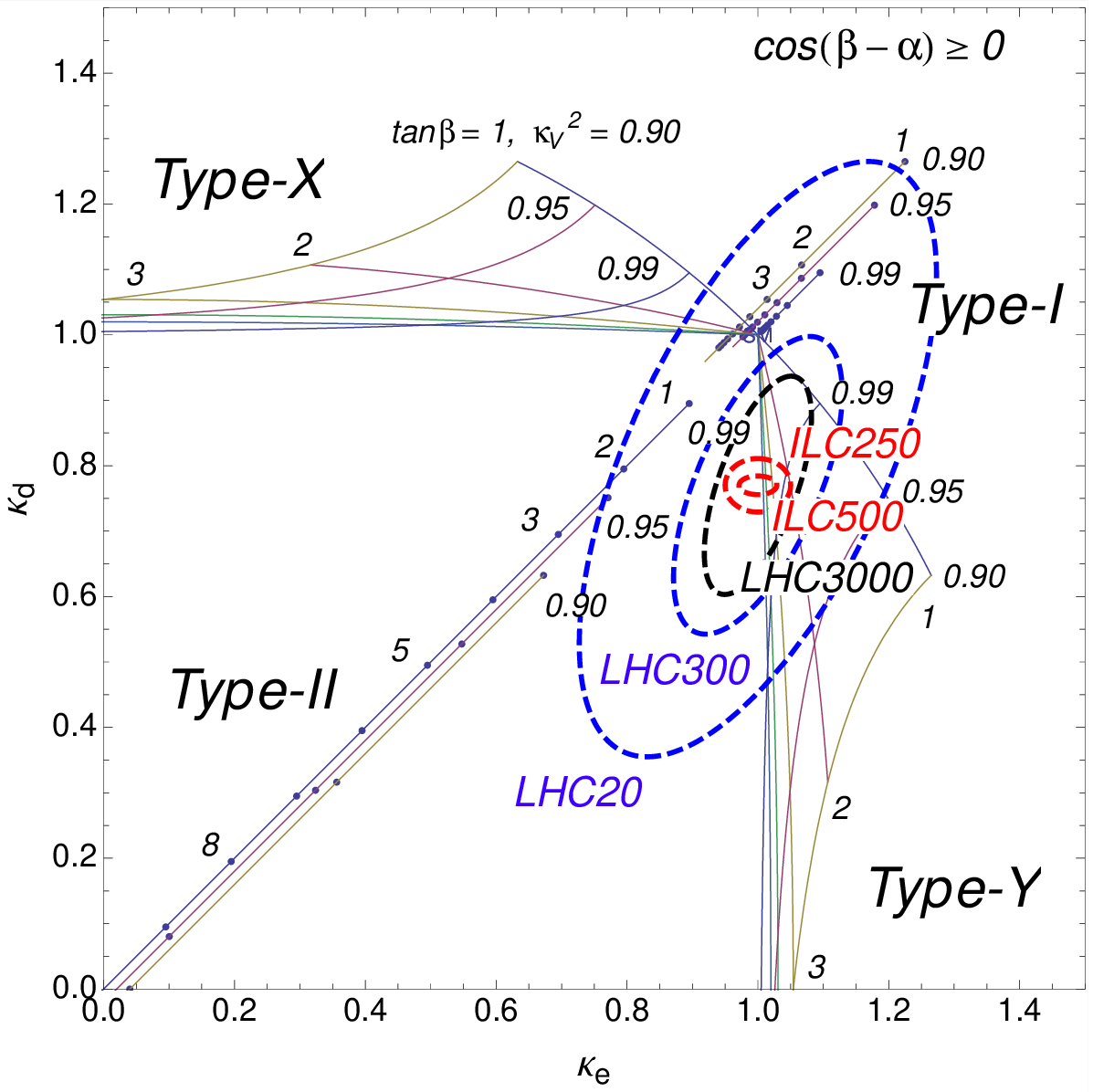}     \hspace{5mm}
\includegraphics[width=7cm]{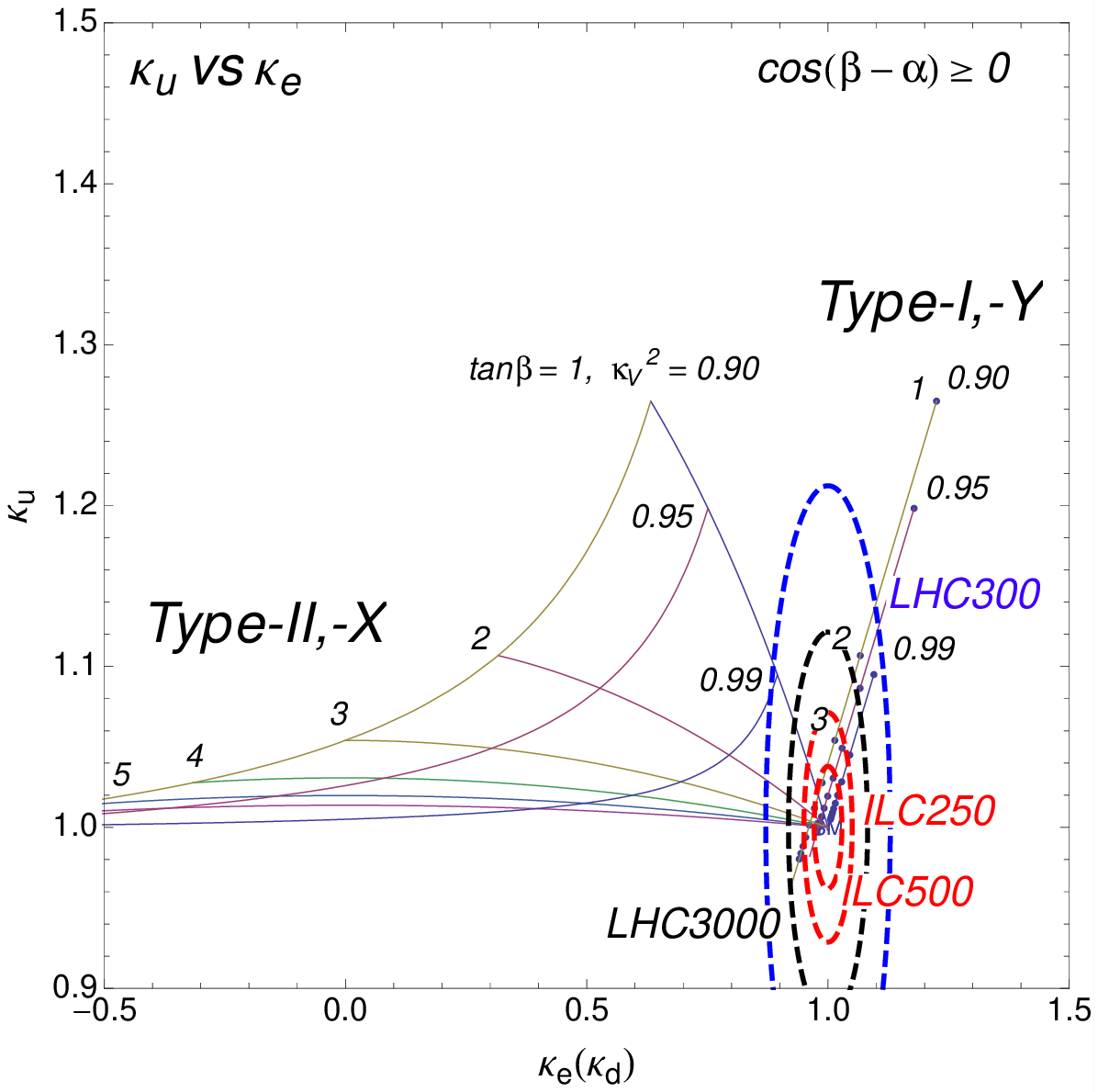}
\caption{The scaling factors for the Yukawa interaction of the SM-like Higgs boson in THDMs in the case of $\cos(\beta-\alpha)>0$.}
\label{Fig:KFKFp}
\end{figure}

We first consider the deviations in the Higgs boson coupling constants in the THDMs.
From TABLE~\ref{Tab:Yukawa_Couplings}, it can be seen that 
all the four types of Yukawa interaction have different combinations of $\xi_h^f$ for $f=u,d$ and $e$ when $\sin(\beta-\alpha)\neq 1$. 
Therefore, the direction and magnitude of modifications for $\kappa_f$ are different in four types of Yukawa interaction. 

In FIGs.~\ref{Fig:KFKFm} and~\ref{Fig:KFKFp}, the scaling factors are shown for each type of Yukawa interaction in the THDMs 
as functions of $\kappa_V^2$ and $\tan\beta$. 
When $\kappa_V^2$ is determined, there still has a sign ambiguity for $\cos(\beta-\alpha)$. 
Thus, we separately plot model predictions for $\cos(\beta-\alpha) < 0$ in FIG.~\ref{Fig:KFKFm} and for $\cos(\beta-\alpha) > 0$ in FIG.~\ref{Fig:KFKFp}. 
Note that the Higgs sector in the MSSM predicts a negative value of $\cos(\beta-\alpha)$. 
In the left (right) panels, the scaling factors of THDMs are given 
in the $\kappa_d$--$\kappa_\ell$ ($\kappa_u$--$\kappa_\ell$) plane. 
Because of the simple scaling in TABLE~\ref{Tab:Yukawa_Couplings}, 
the predictions in the $\kappa_u$--$\kappa_d$ plane are obtained by interchanging
the Type-X and Type-Y THDMs in the right panels. 
For the illustration purpose only, we slightly shift lines along with $\kappa_x=\kappa_y$ 
in order to show $\tan\beta$ dependence for fixed $\kappa_V^2$ to avoid confusions. 
The largest contour (LHC20) denotes the current LHC bound at the 68 \%CL, where the central values and the correlations are taken from Ref.~\cite{Strumia}.  
We also present the projection at the HL-LHC ($\sqrt{s}$=14TeV) with an integrated luminosity of 300 fb$^{-1}$ (LHC300) and 3000 fb$^{-1}$ (LHC3000), 
where the same central values and the correlations are adopted. 
The ILC prospects are also shown for ILC250 and ILC500, where 
the collision energy is 250 GeV and 500 GeV, and the integrated luminosity is 250 fb$^{-1}$ and 500 fb$^{-1}$, respectively. 
Each of the THDMs predicts quite a different region, which can be discriminated by the precision 
measurement of the SM-like Higgs boson coupling constants. 

We note that through the precision measurement of the branching ratios of the SM-like Higgs boson,
not only the discrimination of the type of Yukawa interaction
but also determination of $\tan\beta$ in an indirect way can be accomplished~\cite{Ref:TanBeta_ILC}.
The later complements the determination of $\tan\beta$ by using additional Higgs boson production
directly~\cite{Ref:TanBeta}.

\subsection{Models with Universal Yukawa Couplings}

\begin{table}[t]
\begin{center}
\begin{tabular}{c||l|l|l}\hline\hline 
& $\tan\beta$ & $\kappa_f$ &$\kappa_V^{}$   \\ \hline 
Doublet-Singlet Model & --- & $\cos\alpha$ & $\cos\alpha$ \\ \hline 
Type-I THDM  &$v_0/v_\text{ext}^{}$ &$\cos\alpha/\sin\beta=\sin(\beta-\alpha)+\cot\beta\cos(\beta-\alpha)$ &$\sin(\beta-\alpha)$  \\ \hline 
GM Model &$v_0/(2\sqrt{2}v_\text{ext}^{})$& $\cos\alpha/\sin\beta$ &   $\sin\beta \cos\alpha -\tfrac{2\sqrt6}3 \cos\beta \sin\alpha$ \\  \hline 
Doublet-Septet Model &$v_0/(4v_\text{ext}^{})$&$\cos\alpha/\sin\beta$&$\sin\beta \cos\alpha 
-4 \cos\beta \sin\alpha$\\\hline\hline
\end{tabular}
\end{center}
\caption{The fraction of the VEVs $\tan\beta$ and the scaling factors $\kappa_f$ and $\kappa_V$ in the extended Higgs sectors with 
universal Yukawa couplings. }
\label{Tab:ScalingFactor}
\end{table}

We consider Higgs sectors with a universal shift in the Yukawa coupling constants. 
Such a situation can be realized in a Higgs sector composed of only one doublet field; e.g., 
a model with a scalar doublet plus singlets, triplets and higher isospin multiplets, or 
in a Higgs sector with multi-doublet fields but only one of them giving all the fermion masses; e.g., the Type-I THDM. 
In the following, we first discuss 
the Doublet-Singlet model, the Type-I THDM and the Doublet-Septet model, and then we consider the GM model as models with $\rho_\text{tree}=1$. 
We note that these extended Higgs sectors can predict larger deviations in the $hVV$ couplings as compared to those in models with $\rho_\text{tree}\neq 1$, 
because an amount of the deviation depends on an additional VEV whose magnitude is constrained by the rho parameter if it causes $\rho_\text{tree}\neq 1$. 

In the Doublet-Singlet model and the Doublet-Septet model, 
an isospin singlet field with $Y=0$ and an isospin septet field with $Y=2$ are contained, respectively, in addition to the 
doublet scalar field $\Phi$ with $Y=1/2$. 
The Type-I THDM was already defined in Section II. 
From Eq.~(\ref{Eq:rho_tree}), a VEV from the additional scalar multiplet does not change $\rho_\text{tree}$ from the SM value. 

Except the VEV of the singlet scalar field, 
all the VEVs  from the additional Higgs multiplet $v_\text{ext}$
contribute to the electroweak symmetry breaking.  
They satisfy $v^2= v_0^2 + (\eta_\text{ext}\, v_\text{ext})^2$, where 
$v_0$ is the VEV of $\Phi$ and $\eta_{\text{ext}}$ = 1 and 4 in the Type-I THDM and the Doublet-Septet model, respectively. 
It is convenient to define the ratio of the VEVs as $\tan\beta = v_0/(\eta_\text{ext}\, v_\text{ext})$. 

There are two CP-even scalar states in these three models, and they are mixed with the angle $\alpha$ as
\begin{align}
\begin{pmatrix}
 h_\text{ext}^{}\\
 h_0
 \end{pmatrix}
=
R(\alpha)
\begin{pmatrix} 
H \\ 
h \end{pmatrix}, 
\end{align}
where $h_0$ and $h_\text{ext}$ denote the CP-even scalar components from $\Phi$ and an additional scalar multiplet, respectively. 
The $h$ and $H$ fields are the mass eigenstates, and we assume that $h$ is the observed Higgs boson with the mass of about 126 GeV.

Next, we discuss the GM model whose Higgs sector is composed of a real ($Y=0$) and a complex ($Y=1$) triplet scalar fields in addition to $\Phi$.
When the VEVs of two triplet fields are aligned to be the same ($=v_\text{ext}$), $\rho_\text{tree}=1$ is satisfied, where 
the contributions to the deviation in $\rho_\text{tree}$ from unity by the triplet VEVs are cancelled with each other. 
The value of $\eta_{\text{ext}}$ defined in the above is given as $2\sqrt{2}$. 

In the GM model, there are three CP-even scalar states from $\Phi$ and two triplets.  
They are mixed with each other in the following way~\cite{GVW}
\begin{align}
\left(
\begin{array}{c}
\xi_r\\
h_0\\
\chi_r
\end{array}\right)=\left(
\begin{array}{ccc}
0 & \frac{1}{\sqrt{3}} & -\sqrt{\frac{2}{3}}  \\
1 & 0 & 0\\
0 & \sqrt{\frac{2}{3}} & \frac{1}{\sqrt{3}}
\end{array}\right)
\left(
\begin{array}{ccc}
 \cos\alpha & -\sin\alpha &0\\
 \sin\alpha & \cos\alpha &0\\
0 & 0 & 1 
\end{array}\right)
\left(
\begin{array}{c}
H\\
h\\
H_5
\end{array}\right), 
\end{align}
where $\xi_r$ and $\chi_r$ are respectively the CP-even scalar components in the $Y=0$ and $Y=1$ triplet Higgs fields, and 
$H_5$ is the neutral component of the custodial SU(2) 5-plet Higgs boson.

In TABLE~\ref{Tab:ScalingFactor}, we list the scaling factors $\kappa_f$ and $\kappa_V$ 
in terms of $\alpha$ and $\beta$ in the four models. 
In the Doublet-Singlet model, $\kappa_f$ and $\kappa_V$ have the same expression $\cos\alpha$, because 
both the Yukawa interaction and the gauge interaction are originated from the doublet Higgs field, and they are 
suppressed by the same origin; i.e., the mixing between doublet and singlet fields. 

In the Type-I THDM, both the Yukawa couplings and the gauge couplings are suppressed by $\kappa_f$ and $\kappa_V$, respectively. 
However, $\kappa_f\neq\kappa_V$ is generally allowed unlike the Doublet-Singlet model. 
We have already mentioned in Subsection~II A that we can take the SM-like limit by $\sin(\beta-\alpha)\to 1$, where 
both $\kappa_f$ and $\kappa_V$ become unity. 
Similar limit can be defined in the Doublet-Singlet model by taking $\alpha\to 0$.

In the GM model and the Doublet-Septet model, 
the VEV of the additional multiplet affects the electroweak symmetry breaking 
in a different way from that by the doublet Higgs field; i.e., $\eta_\text{ext}$ in the GM model and the Doublet-Septet model are different 
in the Type-I THDM. 
As a result, $\kappa_V$ can be larger than 1 (see TABLE~\ref{Tab:ScalingFactor}). 
This is a unique feature to identify these models.  
Furthermore, the limit of $\kappa_f\to 1$ and $\kappa_V\to 1$ is taken 
by setting $\beta=0$ and $\alpha=-\pi/2$ which corresponds to the special case in the Type-I THDM.

\begin{figure}
\centering
\includegraphics[width=12cm]{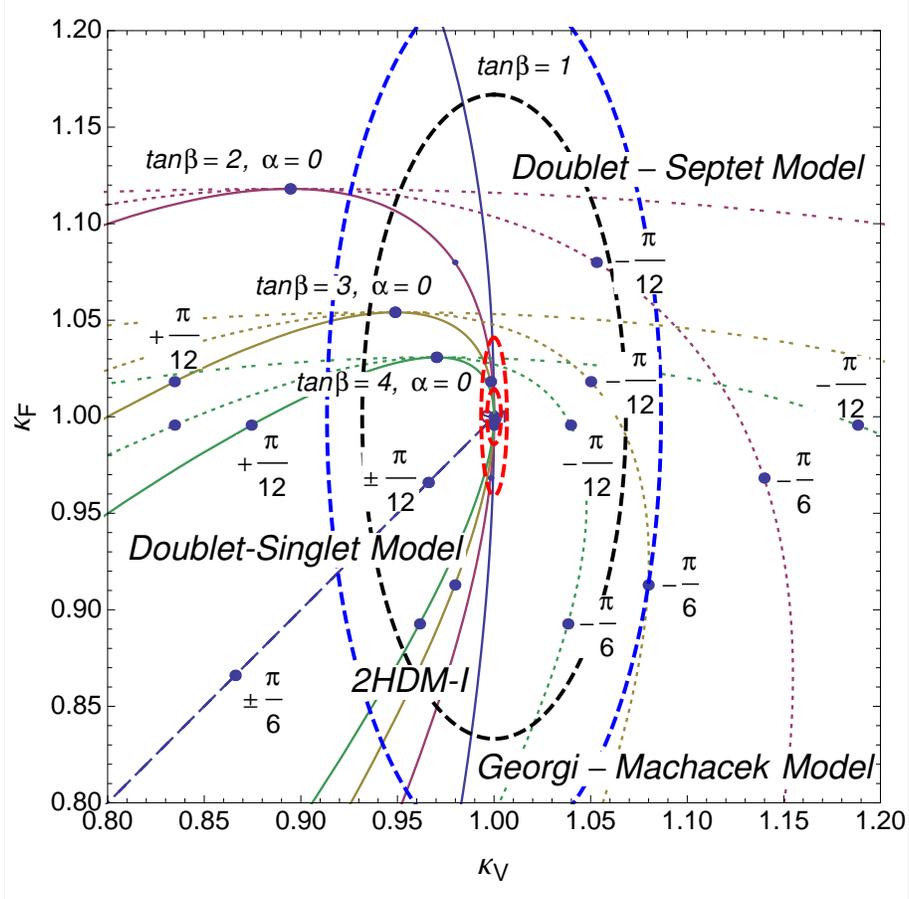}
\caption{The scaling factors $\kappa_f$ and $\kappa_V$ in models with universal Yukawa coupling constants.}
\label{Fig:KVKF}
\end{figure}

In FIG.~\ref{Fig:KVKF}, we show predictions of the scaling factors $\kappa_f$ and $\kappa_V$ for each value of 
$\alpha$ and $\beta$ in the models with universally modified Yukawa couplings. 
If we vary $\alpha$ and $\beta$, a model dependent area (line) is drawn, 
which is a distinctive prediction of the models. 
Note that predictions are the same at $\alpha=0$ in the Type-I THDM, the GM model, 
and the Doublet-Septet model. 
From the current LHC data, the scaling factors are obtained about 20\% accuracy 
at $1\sigma$. It is not sufficient to distinguish these models at this moment. 
Improvements of the (SM-like) Higgs boson coupling measurements 
at the HL-LHC and also at the ILC may resolve model predictions.

\section{Discussions}

We here discuss complementarity of precision measurements of the
coupling constants of the discovered Higgs boson $h$ and direct
searches of additional Higgs bosons at the LHC. 
In addition, we also discuss the importance of direct searches of
additional Higgs bosons at the ILC. 
A key role is taken by the deviation in the coupling constant of $h$ to
weak gauge bosons from the SM prediction, $\delta\kappa_V=1-\kappa_V$. 
When non-zero $\delta\kappa_V$ is found at future colliders, that is identified as an evidence of non-standard effects mainly due
to additional Higgs bosons. 
By combining the theoretical constraints from perturbative unitarity and
vacuum stability, we obtain the upper limit of the energy scale where an
evidence of non-standard Higgs sectors should appear. 
We first discuss the complementarity in the THDMs, and then in the other models later. 

For $\delta\kappa_V\gtrsim 5\%$, which is the expected accuracy at the
LHC with 300~fb$^{-1}$~\cite{Ref:ILC_TDR,Ref:ILC_White,Ref:Snowmass}, 
$m_A$ should be less than about 700~GeV from the conditions
of perturbative unitarity and vacuum stability under the assumptions of $m_{H^+}=m_A$ with varying $M$ and $m_H$ in the $m_A\pm 500$~GeV range.  
In such a case, it is expected that the LHC direct search can find an
evidence of additional Higgs bosons simultaneously. 
For $m_A\lesssim500$~GeV, direct production at the ILC experiment with
$\sqrt{s}=1$~TeV will also be useful to explore the properties of
additional Higgs bosons~\cite{KYZ}. 
On top of above, the precision measurement of the couplings of $h$ at
the ILC will be the most powerful tool to discriminate types of Yukawa interaction as shown in Figs.~\ref{Fig:KFKFm} and \ref{Fig:KFKFp}. 

For $\delta\kappa_V\gtrsim 0.4\%$, which is the expected accuracy at
the ILC with $\sqrt{s}=500$~GeV and
$\mathcal{L}=500$~fb$^{-1}$~\cite{Ref:ILC_TDR,Ref:ILC_White,Ref:Snowmass}, $m_A$
should be less than 1~TeV from the conditions of
perturbative unitarity and vacuum stability under the assumptions of $m_{H^+}=m_A$ with varying $M$ and $m_H$ in the $m_A\pm 500$~GeV range. 
In such a case, there is a possibility that the direct search at the LHC
cannot find any evidence of additional Higgs bosons. 
In other words, the LHC direct search combined with the constraints
from perturbative unitarity and vacuum stability cannot exclude the
extended Higgs sector which predicts $\delta\kappa_V\lesssim 0.4\%$.
At the ILC, at least the precision measurement of the couplings of $h$
can indicate an evidence of the extended Higgs sector. 
Even in such a situation, as we have shown in the last section, the
model discrimination and parameter determination will be still possible
by utilizing only the fingerprinting of the deviation of the couplings
of $h$. 
Furthermore, an upper limit of the mass scale of additional Higgs bosons
can be set by the constraints from perturbative unitarity and vacuum
stability, while the lower limit is given by the direct search at the
LHC. 
Therefore, we could conclude the existence of the non-standard Higgs
sector at a certain energy scale. 
This energy scale will be a crucial information to design next
generation future colliders. 

The accuracy of $\delta\kappa_V$ measurement can be improved at the ILC with 1~TeV and 1~ab$^{-1}$,
and the indirect upper limit of the mass scale can be slightly extended accordingly. 
For $\delta\kappa_V\lesssim 0.2\%$, which is beyond the accuracy of
the coupling measurement of $h$ at the ILC with $\sqrt{s}=1$~TeV and
$\mathcal{L}=1$~ab$^{-1}$~\cite{Ref:ILC_TDR,Ref:ILC_White,Ref:Snowmass}, the upper limit of the mass
scale cannot be obtained from the conditions of perturbative unitarity
and vacuum stability. 
In this case, we cannot separate the extended Higgs sector from the
SM from the coupling measurements of $h$. 
Therefore, the decoupling limit of the extended Higgs sector cannot be
excluded. 
There are possibilities that the additional Higgs bosons can be
discovered at the LHC or the ILC, since the small deviation in
$\kappa_V$ does not necessarily mean the large mass of additional Higgs
bosons in the extended Higgs sector. 
We note that the direct production of additional Higgs bosons at the LHC
and the ILC also have a power to discriminate the models of extended
Higgs sectors, such as the type of Yukawa sector in the THDMs~\cite{KYZ} etc. 

In order to compare the precisely measured values of the Higgs boson couplings, 
precise calculations in each given model are essentially important. 
One-loop corrections to the $hVV$ coupling constants have been calculated in Ref.~\cite{KOSY}, and  
those to $hf\bar{f}$ coupling constants have been calculated in Ref.~\cite{THDM_rad} in the THDM. 
Magnitudes of these corrections due to the additional Higgs boson loops are  
respectively given to be maximally about $1\%$ and $5\%$ for the $hVV$ and $hf\bar{f}$ 
couplings under the constraint from perturbative unitarity and vacuum stability.
Therefore, the pattern of the deviations in the $hf\bar{f}$ shown in Figs.~\ref{Fig:KFKFm} and \ref{Fig:KFKFp} does not 
change even including radiative corrections.  
However, if the $hf\bar{f}$ couplings are determined with an order of 1\% accuracy under the situation where 
the deviation in $hVV$ couplings are also found, 
we may be able to determine not only the type of Yukawa interactions but also some inner parameters
such as $M^2$ in the THDMs. 
In addition to the $hVV$ and $hf\bar{f}$ couplings, 
one-loop corrections to the $hhh$ coupling 
is also important whose amount can be significant due to non-decoupling effect of the additional Higgs bosons. 
In Ref.~\cite{KOS,KOSY}, it has been shown that the size of correction can be $\mathcal{O}(100)\%$ 
under the constraint from perturbative unitarity~\cite{Ref:Uni-2hdm} and vacuum stability~\cite{VS_THDM,VS_THDM2}. 
By studying the correlation among the deviations in the $hhh$~\cite{KOS,KOSY}, 
$h\gamma\gamma$~\cite{septet_ellis,Shifman,THDM_gam,THDM_gam2} and $hZ\gamma$~\cite{THDM_gam2,Ref:hZgam} 
couplings from the SM predictions, 
we can extract properties of additional Higgs bosons running in the loop such as the electric charge, the isospin and the 
non-decoupling nature. 

Finally, we mention models other than the THDMs. 
In the Doublet-Singlet model~\cite{Ref:singlet}, both the $hVV$ and $hf\bar{f}$ coupling constants are suppressed by the same factor. 
Therefore, $\kappa_V^{}=\kappa_f^{}<1$ can be an indirect evidence for this model. 
Detection of an additional CP-even scalar boson, whose Yukawa and gauge interactions are 
given only from the mixing with the doublet Higgs field, can be a direct search for the model. 
The GM model~\cite{Ref:GM} and the Doublet-Septet model~\cite{septet_ellis,Ref:septet,KKY} 
have a unique pattern of the deviation in the Higgs boson couplings; namely, 
$\kappa_V^{}$ can be larger than unity~\cite{KKY}, which is a crucial property to identify these models. 
In addition, multi-charged; e.g., doubly-charged, Higgs bosons  
can significantly contribute to the deviation in the loop induced 
$h\gamma\gamma$ and $hZ\gamma$ couplings. 
When multi-charged scalar bosons are discovered, it can be a direct test of these models. 
Phenomenology of such additional scalar bosons has been discussed in the GM model~\cite{Chiang:2012cn,An-Li,Re} and in the Doublet-Septet model~\cite{Ref:septet_pheno} at the LHC. 
Measuring the $H^\pm W^\mp Z$ vertex~\cite{HWZ-LHC,HWZ-ILC} is also an important probe as discussed in Subsection II C.

In this paper, we concentrate on the models with $\rho_{\text{tree}}=1$. 
However, we here shortly comment on the Higgs Triplet Model (HTM) as an important example for models with $\rho_{\text{tree}}\neq 1$, 
because it is deduced from the type-II seesaw mechanism~\cite{typeII}.
In the HTM, although deviations in the $hVV$ and $hf\bar{f}$ couplings 
cannot be so large due to the constraint from the rho parameter, 
those in the loop induced $h\gamma\gamma$~\cite{HTM_gamgam,HTM_gamgam2,HTM_gamgam3,HTM_gamgam4} 
and $hZ\gamma$~\cite{HTM_gamgam2,HTM_gamgam3,HTM_gamgam4} couplings 
can be significant by the doubly-charged Higgs boson $H^{\pm\pm}$ loop. 
The one-loop corrections to the $hhh$
coupling can also be large as calculated in Refs.~\cite{AKKY_lett,AKKY} due to the non-decoupling effect of additional Higgs bosons similarly to the THDM.
The correlation among the deviations in the decay rate of $h\to \gamma\gamma$ and the $hhh$ coupling constants\footnote{The deviation in the $hhh$ coupling
at the tree level is much suppressed by the triplet VEV similar to the $hVV$ and $hf\bar{f}$ as mentioned in the above. } from the SM values 
have been investigated in Ref.~\cite{AKKY}. 
Direct search for $H^{\pm\pm}$ can be an important clue to test the model with the $Y=1$ triplet field, which 
can decay into the same-sign dilepton~\cite{SS-dilepton1,SS-dilepton2,SS-dilepton3,SS-dilepton4,SS-dilepton5,SS-dilepton6} 
and the same-sign diboson~\cite{SS-diboson1,SS-diboson2} depending on the magnitude of 
the triplet VEV\footnote{If there is a mass difference between $H^{\pm\pm}$ and the singly-charged scalar components, 
the cascade decay of $H^{\pm\pm}$ associated with the $W$ boson is possible~\cite{Cascade}. }.

\section{Conclusions}

We have discussed the determination of the extended Higgs sector by combining the 
direct and indirect searches for additional Higgs bosons at future collider experiments.
Direct searches of the additional Higgs bosons provides the clear evidence for extended Higgs sectors. 
Focusing on the THDM with the softly-broken $Z_2$ symmetry, 
we have studied the expected exclusion regions in the $m_A$-$\tan\beta$ plane
at the LHC with 14~TeV run with 300 fb$^{-1}$ and 3000 fb$^{-1}$ data.
For the neutral Higgs boson searches, 
we have shown that the mass scale up to several hundreds GeV to TeV can be explored at the LHC, 
depending on the type of Yukawa interaction and parameters such as $\tan\beta$ and $\sin(\beta-\alpha)$. 
For the indirect searches of additional Higgs bosons via coupling constants of the SM-like Higgs bosons, 
we have considered various models for the extended Higgs sector, such as 
the THDMs with four types of Yukawa interactions, the Doublet-Singlet model, the Doublet-Septet model, 
and the GM model, as typical models which predict $\rho_{\text{tree}}=1$.
We have demonstrated that  
there exists a variety of patterns in the deviations in the SM-like Higgs boson couplings 
to the gauge bosons and fermions from the SM prediction depending on the structure of the Higgs sector.
Therefore, we can fingerprint the non-minimal Higgs sector 
by detecting the pattern of deviations in an excellent precision at future colliders.

Taking into account the theoretical constraints on the model, such as perturbative unitarity and vacuum stability, 
the complementarity between the direct searches and the indirect searches can be understood to identify the non-minimal Higgs sector.
Observation of the deviation in the coupling constant of the SM-like Higgs boson to the weak gauge bosons 
plays a key role, which also affects the strategy of the direct search of additional Higgs bosons at colliders.
First of all, we have to keep in mind that 
there exists a decoupling limit in extended Higgs sectors in the limit of $\delta\kappa_V\to0$, 
where the SM is a good description as a low energy effective theory up to much higher scales than the electroweak scale.
On the other hand, if a relatively large deviation of $\delta\kappa_V$ is observed, 
the mass scale of the additional Higgs bosons is bounded from the above 
by using the argument of perturbative unitarity and vacuum stability, 
so that the direct discovery of them can be highly expected.
If a small deviation is observed at the ILC,
the direct discovery of the additional Higgs boson can be difficult.
Even in such a situation, 
the fingerprinting of the SM-like Higgs boson couplings 
can be a solid and powerful tool to explore the extended Higgs sector. 

\acknowledgments
S.K. was supported in part by Grant-in-Aid for Scientific Research
from Japan Society for the Promotion of Science (JSPS), Nos.\ 22244031 and 24340046, and 
from Ministry of Education, Culture, Sports, Science and Technology (MEXT), Japan, No.\ 23104006.
K.T. was supported in part by MEXT, Nos.\ 26104704 and 23104011.
K.Y. was supported in part by the National Science Council of R.O.C. under Grant No. NSC-101-2811-M-008-014.
The work of H.Y. was supported in part by Grant-in-Aid for Scientific Research, No. 24340046 and
the Sasakawa Scientific Research Grant from the Japan Science Society.


\end{document}